\documentclass[manuscript,screen]{acmart}

\usepackage{csquotes}
\usepackage{array}
\usepackage{natbib}
\usepackage{booktabs}
\usepackage{threeparttable}
\usepackage{xcolor}
\usepackage{booktabs,tabularx}
\usepackage{multirow}

\newcolumntype{P}[1]{>{\centering\arraybackslash}p{#1}}

\newcolumntype{C}[1]{>{\centering\arraybackslash}p{#1}}

\AtBeginDocument{%
  }

\setcopyright{acmlicensed}
\copyrightyear{2025}
\acmYear{2025}
\acmDOI{XXXXXXX.XXXXXXX}
\acmConference[]{ACM Trans. Comput.-
Hum. Interact}{June 03--05,
  2018}{Woodstock, NY}
\acmISBN{978-1-4503-XXXX-X/2018/06}

\begin{document}

\title{Beyond Awareness: 
Investigating
How AI and Psychological Factors Shape Human Self-Confidence Calibration}


\author{Federico Maria Cau}
\affiliation{%
  \institution{University of Cagliari}
  \city{Cagliari}
  \country{Italy}}
\email{federicom.cau@unica.it}
\orcid{0000-0002-8261-3200}

\author{Lucio Davide Spano}
\affiliation{%
  \institution{University of Cagliari}
  \city{Cagliari}
  \country{Italy}}
\email{davide.spano@unica.it}
\orcid{0000-0001-7106-0463}

\renewcommand{\shortauthors}{F. M. Cau and L. D. Spano}

\begin{abstract}
Human-AI collaboration outcomes depend strongly on human self-confidence calibration, which drives reliance or resistance toward AI's suggestions. This work presents two studies examining whether calibration of self-confidence before decision tasks, low versus high levels of Need for Cognition (NFC), and Actively Open-Minded Thinking (AOT), leads to differences in decision accuracy, self-confidence appropriateness during the tasks, and metacognitive perceptions (global and affective). The first study presents strategies to identify well-calibrated users, also comparing decision accuracy and the appropriateness of self-confidence across NFC and AOT levels. The second study investigates the effects of calibrated self-confidence in AI-assisted decision-making (no AI, two-stage AI, and personalized AI), also considering different NFC and AOT levels. 
Our results show the importance of human self-confidence calibration and psychological traits when designing AI-assisted decision systems. We further propose design recommendations to address the challenge of calibrating self-confidence and supporting tailored, user-centric AI that accounts for individual traits.

\end{abstract}





\maketitle

\section{Introduction}
\label{sec:introduction}
Given that Artificial Intelligence (AI) is not perfect when delivering assistance, it becomes necessary to inform individuals when suggestions may be flawed so they can accept correct recommendations and ignore erroneous ones, fostering appropriate reliance on AI \cite{Bansal2021FeatureBased,Zhao2023AIUncertaintyVisualization,He2023DunningKruger,Cao2024UncertaintyPresentation,Vaccaro2024HumanAISynergy}. 
An approach to achieve this is communicating the AI's correctness likelihood by providing confidence estimates for individual decisions \cite{Bertrand2022BiasesInXAIMitigation,Prabhudesai2023UncertantyHCI,Lai2023SurveyHumanAIDecisionMaking}. However, past research has shown that displaying confidence estimates does not always guarantee improved human-AI team performance or appropriate reliance on AI \cite{Zhang2020ConfidenceExplanationsAccuracyTrust,Vodrahalli2022HumanTrustAIUserConfidence,zhang2024confidenceAccuracyCorrelation,li2025confidencealigns}. In addition, these confidence estimates need to be calibrated (i.e., reflecting prediction correctness probabilities, so a confidence score of 0.7 implies a 70\% chance of a correct prediction) to improve human-AI decision-making and mitigate the risks associated with misleading/inflated AI confidence
\cite{Rechkemmer2022AIPerformanceAndConfidenceEffectsOnUsers,Cao2024UncertaintyPresentation, Ma24UserConfidenceCalibration,Marusich2024AIUncertaintyQuantification,li2024overconfidentunconfidentaihinderARXIV,li2025confidencealigns}.
Building on this, recent studies in Human-Computer Interaction (HCI) have highlighted the importance of calibrating human self-confidence alongside AI, as it significantly influences how individuals interact with AI assistance \cite{Ma2023CorrectnessLikelihoodAIUsersIncome,Ma24UserConfidenceCalibration,Cau2025CuriosityTraits}. 
For example, overconfident individuals show inflated self-knowledge in their judgements when they are wrong, as well as the tendency to engage in risky behaviors and potentially dismiss valuable external advice (e.g., AI) \cite{Frankenberger1997decision,Moore2008OverconfidenceOverprecision,Razmdoost2015UnderconfidentOverconfident,Pescetelli2021GroupDecisionMetacognition,Soll2022OverconfidenceAdvice,Binnendyk2024IndividualDifferencesOverconfidence,li2025confidencealigns,LiHaleMoore_2025OverconfidenceIndividualDifference}.
Conversely, underconfident individuals tend to have underestimated self-knowledge in their judgments, often seeking external advice (e.g., AI) even when their assessments are accurate \cite{Pallier2002_individual_differences,Razmdoost2015UnderconfidentOverconfident,Mandel2018UnderconfidenceExpertJudgements, Pescetelli2021GroupDecisionMetacognition,CHONG2022HumanConfidence}.
Although self-confidence calibration mechanisms do not fully resolve the issue of users' overreliance on AI, they can improve the appropriateness of confidence and human-AI team performance compared to uncalibrated baselines. 
Furthermore, previous studies underscore that self-confidence is strictly related to metacognition \cite{KEREN1991_calibration,EFKLIDES2006MetacognitionAndAffect,Ayanna2022thinkingAboutThinking,Yeung2012ConfidenceMetacognition,Lee2024MetacognitionOfEmotions,Cushing2024Metacognitionasawindowintosubjectiveaffectiveexperience,KATYAL2024_metacognition}, emphasizing the importance of collecting measures from individuals related to monitoring self-confidence during tasks (such as local and global metacognition \cite{KLEITMAN2007SelfConfidenceMetacognitivePRocesses,Rouault2019GlobalEstimatesFromLocalSelfPerformance,Fleming2024_metacognition,Katyal2025AnxietyDepressionLocalGlobalMetacognition}).\footnote{Local metacognition refers to judgments of
performance on individual tasks or instances of a task. 
In contrast, global metacognition refers to a more global level judgment like how individuals would rank, e.g.,  their driving or intellectual abilities \cite{KATYAL2024_metacognition}.} However, these studies also highlight a lack of integration and measurement of meta-affective processes (for example, metacognitive experiences which do not have an obvious ground truth and related to certainty of feeling specific emotions \cite{Smith2016MemoryAcuteStress,Tankelevitch&Kewenig2024MetacognitionGenAI,KATYAL2024_metacognition,Mladen2024AffectiveAndMotivationalStatesMetacognition,Cushing2024Metacognitionasawindowintosubjectiveaffectiveexperience}). 
Consequently, how people regulate and benefit from their meta-affective states remains underexamined, even though evidence suggests that affective and metacognitive processes are deeply intertwined in guiding decision-making and learning processes. 

Self-confidence calibration is also influenced by a phenomenon called confidence alignment, where individual members’ confidence tends to align with that of their peers, either human or AI \cite{Bang2014Doesinteractionmatter,Bang2017ConfidenceGroupDecisionMaking,Pescetelli2021GroupDecisionMetacognition,CHONG2022HumanConfidence,Pescetelli2022Benefitsofspontaneousconfidence}. 
Specifically, a recent study by Li et al. \cite{li2025confidencealigns} confirmed that displaying AI confidence influences individuals' self-confidence calibration and elicits confidence alignment, which persists even after AI involvement ends when users resume making decisions independently. Furthermore, the authors found that the impact of confidence alignment on human self-confidence calibration also depends on the type of AI assistance provided (i.e., advisor\footnote{The term \enquote{advisor} refers to the AI two-stage decision-making paradigm, where individuals initially make decisions on their own and may subsequently update them after receiving AI assistance.} \cite{He2023AnalogyExplanations,He2023DunningKruger,Agudo2024HumanError,Cao2024UncertaintyPresentation,Morrison2024ImperfectXAI,Salimzadeh2024PrognosticVsDiagnosticTasks} and peer collaborator\footnote{The term \enquote{peer collaborator} refers to the Maximum Confidence Slating (MCS) algorithm, in which the outcome of a task is determined by selecting the decision of the more confident member within the human-AI dyad.}   \cite{Bang2014Doesinteractionmatter,Koriat2015MCS,Nguyen2025DiadDecisions,li2025confidencealigns}) and the individuals' self-confidence level compared to AI confidence level (e.g., overconfident or underconfident).
Nevertheless, previous work has mainly exposed users to fixed types of AI assistance, without exploring the potential of tailoring AI support according to users' self-confidence status resulting from a calibration phase. 
While limited research has investigated how to detect overreliant or non-overreliant users and adapt AI assistance accordingly to improve human-AI team decision outcomes \cite{Swaroop2024NFCTimePressureOverreliance, Swaroop2025AIPersonalizedOverrelianceRate}, personalizing AI support based on individuals' self-confidence remains an underexplored area.

In addition, prior studies have highlighted that individual differences can also influence self-confidence calibration and decision-making in general \cite{Pallier2002_individual_differences,BEELER2022IndividualFactorsDisclosure,Khadar2025UserCharacteristicsXAI}.
Past work on psychological factors such as the Need for Cognition (NFC)\footnote{NFC is a stable personality trait that reflects an individual’s preference for cognitively
demanding activities.}  \cite{Cacioppo1984NFC} has shown that high-NFC individuals might perform better in specific domains (e.g., music recommendation, AI-assisted nutrition decisions, and intelligent tutoring systems), and especially when interacting with AI explanations \cite{Millecamp2019NFC,Millecamp2020NFC,Conati2021NFC,Bucinca2021NFC,Gajos2022NFC,BahelConati2024NFC}. 
Yet, they are more prone to be overconfident in their decisions \cite{Vogt2022NonAbilityConfidence,Zerna2024NFCReview,Cau2025CuriosityTraits}. 
Moreover, individuals with high Actively Open-minded Thinking (AOT)\footnote{AOT is a trait which includes the tendency to
weigh new evidence against a favored belief, to spend
sufficient time on a problem before giving up, and to consider
carefully the opinions of others in forming one’s
own.} \cite{Baron_1985_AOT,Ottati2023AOT,Stanovich2023AOTMeasurement} are generally related to a higher accuracy and lower overconfidence \cite{Haran2013AOTinAccuracyAndCalibration}, good self-confidence calibration \cite{Martin2024CalibrationFeedbackAOT}, and more appropriate reliance on AI \cite{Swaroop2025AIPersonalizedOverrelianceRate}. 
Still, how users with diverse individual traits respond to self‑confidence calibration mechanisms, and how these mechanisms impact the accuracy and appropriateness of their self‑confidence, remain largely understudied topics \cite{Raees2025PersuasivepersonalizedAIAssistance}.

Building on this foundation, our study aims to address how AI assistance and individual differences, such as NFC and AOT, influence decision accuracy, the appropriateness of human self-confidence, and both metacognitive and meta-affective measures throughout the decision-making process.

To investigate these aspects, we conducted two user studies\footnote{For more details on data processing, model training, user study results, and statistical analysis, please refer to this link.\label{open_repo}}  in which participants completed a series of income prediction tasks \cite{Ribeiro_Singh_Guestrin_2018income,Zhang2020ConfidenceExplanationsAccuracyTrust,Ghai2021income,Ma2023CorrectnessLikelihoodAIUsersIncome,Ma24UserConfidenceCalibration,li2025confidencealigns,Khadar2025UserCharacteristicsXAI}.
In the first study (N = 128), we aimed to define (i) what constitutes a \textit{well-calibrated} user in terms of self-confidence by establishing a threshold $\delta$ to distinguish between overconfident, underconfident, and well-calibrated individuals, and (ii) whether low and high levels of the Need for Cognition (NFC) and Actively Open-minded Thinking (AOT) individual traits impact decision accuracy and appropriateness of self-confidence\footnote{We measured appropriateness of self-confidence using the Expected Calibration Error (ECE) metric.} already in the calibration phase.
Participants first completed the NFC and AOT questionnaires, followed by 20 income prediction tasks without AI support, receiving feedback on the correctness of their answers and their confidence status (overconfident, underconfident, or well-calibrated).
This phase followed the \enquote{Calibration Status Feedback}  method by Ma et al. \cite{Ma24UserConfidenceCalibration}, with the key difference that we did not provide post-hoc calibration feedback, as our focus was on deriving a general classification rule for self-confidence across all tasks.
The results indicate that a value of $\delta = 5$ serves as a suitable threshold for categorising users' self-confidence into Over-confident, Under-confident, and Well-calibrated groups within our study settings. Additionally, most users were overconfident (90.6\%) while underconfident users were rare (0.8\%). We observed no differences in decision accuracy and appropriateness of self-confidence (ECE) between low and high NFC and AOT groups in the calibration phase.

In the second study (N = 159, between-subjects), our goal was to understand how different types of AI assistance (no AI, two-stage AI, and personalized AI) and individual psychological trait levels (low or high) of NFC and AOT impact users' accuracy, appropriateness of self-confidence, and meta-perceptions (global and affective metacognition).
Participants underwent the same self-confidence calibration as in Study 1, where we also provided post-hoc calibration feedback and status (i.e., over-confident, under-confident, or well-calibrated) determined by the $\delta$ threshold. Then, participants completed 20 additional income prediction tasks under one of the AI assistance conditions, also reporting 
their global and affective metacognition throughout the study, as well as their subjective experience during the main task.
The findings indicate that AI assistance significantly improved performance: both two-stage and personalized AI outperformed no-AI on accuracy, while only the two-stage system led to better confidence appropriateness (ECE) and global metacognition compared to no AI. 
High-NFC participants generally reported greater global and affective metacognition in both task phases, whereas high-AOT participants achieved higher accuracy in the main task. 
Post hoc analyses highlighted three main points: i) the increases in accuracy and confidence appropriateness (ECE) from the calibration phase to the main task were driven by the introduction of AI assistance, for which the two-stage AI seemed to promote excessive agreement on AI and confidence alignment behaviors, ii) high-AOT participants showed greater agreement and less reliance on AI advice, as well as higher perceived decision-making autonomy in the main task, and iii) higher AI literacy predicted greater global and affective metacognition across both task phases, although the affective gain occurred only for high-NFC participants.

In this work, we make the following contributions: 

\begin{itemize}

    \item We propose a classification of users into overconfident, underconfident, or well-calibrated, considering a low-stakes, easy task. We show that most individuals are \textit{overconfident}, with a low percentage of \textit{well-calibrated} ones. Instead, \textit{underconfident} individuals are rare.
    
    \item We show that, although informing people about their self-confidence calibration status, this alone is insufficient to enhance their decision-making accuracy or confidence appropriateness (ECE) in subsequent decisions without AI assistance. In fact, improvements in decision-making across calibration and main tasks only occur when AI assistance is provided. Two-stage and personalised AI significantly boost accuracy compared to no AI, while only the two-stage AI enhances self-confidence appropriateness (ECE) and global metacognition. However, these improvements are due to people's excessive alignment with the AI in both predictions and confidence scores, rather than critically assessing AI advice.

    \item We show that high-NFC individuals tend to have inflated perceptions of their skills (global metacognition) and positive sentiment (affective metacognition), which does not translate into actual improved decision-making accuracy. Instead, high-AOT individuals achieved greater decision-making accuracy, agreement with AI, and reduced under-reliance, further reporting higher perceived autonomy in the main task. 

    \item  We outline practical implications for decision-support design and offer guidance on improving personalized AI assistance while accommodating users with different levels of individual traits.
 
\end{itemize}

\section{Related Work}
\label{sec:related}
In this section, we provide an overview of past research on the importance of calibrating human self-confidence and AI confidence, as well as the impact of various AI paradigms on decision-making within human-AI teams. 
Then, we discuss the significance of metacognition in humans' cognitive and emotional self-monitoring processes, further examining how individual traits influence self-confidence calibration and AI-assisted decision-making.

\subsection{Human Self-confidence, AI confidence, and Calibration in Human-AI Team Decisions}
\label{sec:related_confidence}
Humans' self-confidence can be described as the likelihood of being correct given the evidence and the decision made \cite{Fleming2017HumanConfidence,Pescetelli2021DecisionConfidenceTrust}. It significantly moderates acceptance of external advice, either in human group decisions or when following AI assistance \cite{Koriat2019ConfidenceJudgmentsMetacognition,Pescetelli2021GroupDecisionMetacognition,Lu2021HumanRelianceAIUserConfidence,Pescetelli2022Benefitsofspontaneousconfidence,CHONG2022HumanConfidence,Vodrahalli2022HumanTrustAIUserConfidence,He2023DunningKruger,Snijders2023IndividualContextualConfidence,li2025confidencealigns,Papantonis2025UserConfidenceExplanationsAI}. 
Previous work emphasized that human self-confidence relates to \textit{metacognition}, which involves individuals' assessment of their abilities, knowledge, and understanding of task-relevant factors, with their self-confidence seen as an outcome of that self-monitoring process
\cite{KEREN1991_calibration,Yeung2012ConfidenceMetacognition,KLEITMAN2007SelfConfidenceMetacognitivePRocesses,moore2015OverprecisionInJudgement,Rouault2019GlobalEstimatesFromLocalSelfPerformance,Tankelevitch&Kewenig2024MetacognitionGenAI,KATYAL2024_metacognition,Fleming2024_metacognition,Rahnev2025_metacognition} (see Section \ref{sec:related_metacognition}).
However, humans' expression of self-confidence can be dangerous in decision-making when it is not predictive of accuracy (i.e., confidence is miscalibrated), in both experts and lay people  \cite{Arkes2001Overconfidence,BIER2004OverconfidenceExpertise,Miller2015OVerconfidenceUnderconfidence,Pescetelli2021DecisionConfidenceTrust,Ma2023CorrectnessLikelihoodAIUsersIncome,FosterAndRenie2024HumanConfidenceCalibrationStudents,Ma24UserConfidenceCalibration,Cau2025CuriosityTraits,li2025confidencealigns}. 
For example, prior work suggests that users are generally \textit{overconfident} in their decisions \cite{Frankenberger1997decision,Razmdoost2015UnderconfidentOverconfident,Soll2022OverconfidenceAdvice,Binnendyk2024IndividualDifferencesOverconfidence,li2025confidencealigns,LiHaleMoore_2025OverconfidenceIndividualDifference}, thus demonstrating inflated self-knowledge in their judgments or risky behaviors, which might lead to dismissing potentially fruitful external advice \cite{Moore2008OverconfidenceOverprecision,Pescetelli2021GroupDecisionMetacognition, Cau2025CuriosityTraits}.
Conversely, \textit{underconfident} users tend to have a deflated self-knowledge \cite{Pallier2002_individual_differences,Razmdoost2015UnderconfidentOverconfident,Mandel2018UnderconfidenceExpertJudgements, Pescetelli2021GroupDecisionMetacognition}, with a tendency to seek external advice even when correct about their assessments \cite{Pescetelli2021GroupDecisionMetacognition,CHONG2022HumanConfidence}. 
The causes of miscalibration (i.e., overconfidence or underconfidence) and consequent effects are numerous and stem from people's characteristics, such as cognitive abilities and biases \cite{Pallier2002_individual_differences,Haran2013AOTinAccuracyAndCalibration,Rastogi2022CognitiveBiasAIConfidence,Martin2024CalibrationFeedbackAOT,Soleimanof2025OverUnderConfidentAI,Nattapat2025CognitiveBiases}, personality traits \cite{Bucinca2021NFC,Vogt2022NonAbilityConfidence,Zerna2024NFCReview,Swaroop2025AIPersonalizedOverrelianceRate,Cau2025CuriosityTraits}, and mental health issues \cite{ROUAULT2018443,Seow2020-ko,Benwell2022-yj,Hoven2023Psychopathology,KATYAL2024_metacognition,Katyal2025AnxietyDepressionLocalGlobalMetacognition}, to name a few.\footnote{We will further examine the role of human factors in calibrating human self-confidence in Section \ref{sec:related_human_factors}.}
Overall, this highlights the importance of \textit{self-confidence calibration}, which leads to better decision-making considering dyadic collaborations with other humans or AI \cite{Papantonis2025UserConfidenceExplanationsAI,Ma2023CorrectnessLikelihoodAIUsersIncome,Ma24UserConfidenceCalibration,Cau2025CuriosityTraits,li2025confidencealigns}. 

Similarly, AI can also deliver confidence estimates about the correctness of its outcomes in multiple formats, such as numerical confidence scores or ranges \cite{Cao2024UncertaintyPresentation,Bhattacharya2024EXMOS,Cau2025CuriosityTraits}, or textual/graphical representations \cite{Padilla2020AIConfidenceChartTextual,Prabhudesai2023UncertantyHCI,Zhao2023AIUncertaintyVisualization,Marusich2024AIUncertaintyQuantification}. From an AI perspective, the term \textit{calibration} describes how well an AI's predicted probabilities match the actual chances of its predictions being correct. For example, if an AI gives a confidence score of 0.9 to a certain class, it should be correct about 90\% of the time when reporting that confidence level (perfect calibration).
In this paper, we specifically refer to binary classification tasks, where we present AI-calibrated outputs' probabilities as numerical confidence estimates in percentage. 
Prior work demonstrated that showing AI confidence to users does not always improve human-AI decision-making \cite{Bansal2021FeatureBased,Zhao2023AIUncertaintyVisualization,He2023DunningKruger,Ma2023CorrectnessLikelihoodAIUsersIncome,Cao2024UncertaintyPresentation}, especially when the confidence delivered is not properly calibrated, which can lead to misleading information about the AI's likelihood of correctness for its suggestions \cite{Rechkemmer2022AIPerformanceAndConfidenceEffectsOnUsers,Cao2024UncertaintyPresentation, Ma24UserConfidenceCalibration,Marusich2024AIUncertaintyQuantification,li2024overconfidentunconfidentaihinderARXIV,li2025confidencealigns}. Moreover, an increasing amount of research demonstrates that high AI confidence usually causes users to overrely on its recommendations, while low AI confidence encourages them to dismiss and ignore its advice \cite{Cau2023ExplImageText,Cau2023LogicalReasoningStock,Ma2023CorrectnessLikelihoodAIUsersIncome,Ma24UserConfidenceCalibration,li2024overconfidentunconfidentaihinderARXIV,Cau2025CuriosityTraits,li2025confidencealigns}. 

Parallel to this, the human-AI decision paradigm in which AI assistance is presented to people is another fundamental piece in team decision-making. Current research identifies multiple interaction patterns for
AI-assisted decision making \cite{Gomez2025HumanAICollaboration}. However, here we discuss the four main AI paradigms commonly employed in the HCAI field: one-stage, two-stage, on-demand, and Maximum Confidence Slating (MCS).\footnote{In this work, we do not discuss conversational, actionable, or evaluative AI paradigms \cite{Gomez2025HumanAICollaboration}, as they are typically interactive and often involve the use of AI explanations. Instead, we focus on fixed paradigms that avoid explanations and deliver a suggestion together with an associated confidence score.}

The \textit{one-stage} AI or concurrent paradigm involves providing AI assistance immediately to the human decision-maker \cite{Bucinca2021NFC,Rastogi2022CognitiveBiasAIConfidence,Fogliato2022OneStageTwoStage,Cau2023ExplImageText,Cau2023LogicalReasoningStock,Lu2024DoWeLearnFromEachOtherTwoStage,Swaroop2025AIPersonalizedOverrelianceRate}. While this approach can be convenient for quick or time-critical decisions \cite{Swaroop2024NFCTimePressureOverreliance,Swaroop2025AIPersonalizedOverrelianceRate}, it might introduce an anchoring bias effect \cite{Nourani2021AnchoringBias,Fogliato2022OneStageTwoStage,Ma2023CorrectnessLikelihoodAIUsersIncome,Nattapat2025CognitiveBiases,Romeo2025AutomationBiasReviewXAI} where users remain stuck in the cognitive processes of System 1 thinking (fast and
automatic) without processing the available information, and over-rely on AI advice (which works as an anchor), thereby failing to shift to System 2 thinking (slow and deliberative) \cite{Kahneman2011ThinkingFastSlow}.

Instead, the \textit{two-stage} AI or sequential paradigm is based on the the Judge-Advisor System (JAS) model \cite{SniezekBuckley1989JAS,Yaniv2000JAS,Bonaccio2006JASAdvisor,Pescetelli2021DecisionConfidenceTrust}, where a participant (\enquote{judge})  is typically asked to provide an initial answer to a question and then presented with the opinions of one or more advisors (i.e., an AI) whose advice can then be used to update the initial judgment. 
Prior HCI research on AI-assisted decisions introduced this model as a cognitive forcing function (i.e., cognitive intervention to enhance users' engagement with AI assistance), to allow users to switch from fast decisions to more critical thinking, offering potential benefits in terms of increased accuracy and appropriate reliance on AI advice \cite{Bucinca2021NFC,He2023DunningKruger,He2023AnalogyExplanations,Salimzadeh2024PrognosticVsDiagnosticTasks,Agudo2024HumanError,Morrison2024ImperfectXAI,Cao2024AppropriateRelianceSkinCancer,Kuper2025NFCConfidenceRelianceAITwoStage}. 
However, some studies highlight that the boost in participants' performance may stem from an increased overall overreliance on AI's correct and incorrect predictions, resulting in an improvement due to the increase in incorrect advice alignment, which contradicts the expected purpose of introducing such a cognitive forcing strategy \cite{Lu2024DoWeLearnFromEachOtherTwoStage,Ma24UserConfidenceCalibration,Cao2024AppropriateRelianceSkinCancer}.

Similarly, the \textit{on-demand} AI paradigm is also a form of cognitive intervention, where users can explicitly request AI assistance to complete the assigned tasks \cite{Millecamp2019NFC,Bucinca2021NFC,Martijn2022OnDemand,He2024AnalogiesDemand,bucinca2024optimizinghumancentricobjectivesaiassisted,Cau2025CuriosityTraits,He2025ConversationalXAIOnDemand}. Although it has been studied less extensively than the two‑stage paradigm, on‑demand AI could potentially improve accuracy and mitigate overreliance. However, the current Human-Centered AI (HCAI) literature remains too limited to draw firm conclusions for this type of paradigm, which might tend to lower users' confidence in their decisions when used \cite{Cau2025CuriosityTraits}.

Unlike the previous paradigms, the \textit{MCS} one requires both dyadic parties (i.e., human and AI) to express their confidence in the decision, and then the outcome of the task is determined by selecting the
decision of the more confident member within the human-AI dyad \cite{Bahrami20102heads,Koriat2012MCS,Koriat2015MCS,Bang2014Doesinteractionmatter,Koriat2015MCS,Nguyen2025DiadDecisions,li2025confidencealigns}. This signifies that MCS only became useful based on the calibration similarity of dyad members: the higher the better \cite{Bang2017ConfidenceGroupDecisionMaking,Pescetelli2022Benefitsofspontaneousconfidence,Nguyen2025DiadDecisions,li2025confidencealigns}. 
Since the dyadic decision is dominated by the individual with higher confidence as per definition \cite{Koriat2012MCS}, uncalibrated users (e.g., overconfident) might completely disrupt effective decision-making with this paradigm, leading to suboptimal decisions.

Given the importance of humans' self-confidence calibration as we discussed earlier, previous works suggest that AI assistance should be tailored based on people's self-confidence to improve human-AI collaboration instead of using one fixed human-AI decision paradigm \cite{Ma2023CorrectnessLikelihoodAIUsersIncome,Ma24UserConfidenceCalibration,Steyvers2024AIParadigms,Cau2025CuriosityTraits,li2025confidencealigns}. 
For example, Ma et al. \cite{Ma24UserConfidenceCalibration} explored how different self-calibration mechanisms (i.e., think, bet, and real-time feedback) affect human self-confidence and experience, further examining the effects of self-confidence calibration status on two-stage AI-assisted decision-making. 
The results show that calibration mechanisms improve humans' self-confidence calibration and enhance human-AI team performance (when considering real-time feedback as a calibration method). However, self-confidence calibration was not sufficient to foster appropriate reliance on AI, with the need for targeted interventions based on humans' self-confidence calibration status.
A recent study by Li et al. \cite{li2025confidencealigns} investigated whether the alignment of human self-confidence with AI confidence could influence self-confidence calibration using three different human-AI paradigms (i.e., two-stage, MCS, and supervisor\footnote{In this paradigm, the goal was to measure users' self-confidence, as the AI decision was automatically picked as the final answer.}). 
They found that individuals' self-confidence tends to align with that of AI, and this alignment remains even when the AI assistance is removed. Moreover, the calibration of self-confidence is influenced by this alignment, depending on a person's initial level of self-confidence and how it compares to the AI's confidence and accuracy. 
Overall, miscalibration of users' self-confidence undermines proper reliance by both users and AI and reduces the efficacy of human-AI decision-making, independently of the human-AI paradigm. The authors finally suggest tailoring AI's expression of confidence according to different self-confidence calibration statuses (e.g., overconfident or underconfident users) to achieve better human-AI collaboration.

\subsubsection{Measuring Metacognition}
\label{sec:related_metacognition}
As we discussed earlier, self-assessment of one's abilities using confidence estimates is strictly related to metacognition. For example, local metacognition involves assessing an individual's performance on single instances, whereas global metacognition entails broader self‑evaluations, like rating the overall intellectual abilities over a set of items \cite{Moore2008OverconfidenceOverprecision,Rouault2019GlobalEstimatesFromLocalSelfPerformance,Fleming2024_metacognition}. 
Therefore, the concept of miscalibration refers to altered states of self-assessment, which cause humans' overconfident or underconfident behaviors. 
In this regard, prior work \cite{Moore2008OverconfidenceOverprecision,moore2015OverprecisionInJudgement,Moore2017ThreeOverconfidence,zhang2024confidenceAccuracyCorrelation} distinguishes three forms of overconfidence: overestimation, overplacement, and overprecision.\footnote{The three categories for overconfidence also extend to underconfidence with the same but mirrored definitions, hence underestimation, underplacement, and underprecision \cite{Erev1994simultaneousOverUnderconfidence,Moore2007Underconfidence,Harris2011unrealisticUnderconfidence}. While underestimation and underplacement are common, literature on people exhibiting underprecision by having less confidence than their actual accuracy deserves is minimal \cite{moore2015OverprecisionInJudgement}.} 
Overestimation refers to one's actual ability, performance, level of control, or chance of success (e.g., a student who took a 10-item quiz believes they answered five correctly when they got only three, thus overestimating their score). 
Instead, overplacement happens when someone thinks they’re better than others: for example, a student who believes she had the highest score in class, even though half the class outscored her, has overplaced her performance.
Finally, overprecision refers to the excessive certainty about the accuracy stemming from a single decision on one's beliefs.
In our work, we will focus on overestimation (underestimation), which overlaps with measuring global metacognition. Specifically, we aim to measure people's global metacognition after the main task (retrospective) \cite{Nelson2017JudgmentOFLearningJOL,KATYAL2024_metacognition,Tankelevitch&Kewenig2024MetacognitionGenAI,Lee2025MetacognitiveSensitivity}, considering different types of AI assistance and psychological constructs (see Sec. \ref{sec:related_human_factors}).

On top of this, recent research emphasises that current trends in metacognition focus on first-order cognitive processing that can be verified with objective performance metrics \cite{Tankelevitch&Kewenig2024MetacognitionGenAI,KATYAL2024_metacognition,Mladen2024AffectiveAndMotivationalStatesMetacognition,Cushing2024Metacognitionasawindowintosubjectiveaffectiveexperience}. However, a broad spectrum of human metacognition involves processes that lack an obvious ground truth, such as affective states, where individuals might report feeling excited but sometimes be very sure they are excited, and at other times not so certain. 
As such, prior work \cite{KATYAL2024_metacognition,Cushing2024Metacognitionasawindowintosubjectiveaffectiveexperience}  highlights the need to study links between metacognition and affective experiences. Therefore, affective metacognition can have an important role, for
example, in studying emotion regulation \cite{MCRae2020EmotionRegulation}, or act as a key mechanism mediating metacognitively oriented therapeutic
interventions \cite{Moritz2007AffectiveTherapyinterventions,wells2011metacognitiveAffectiveTherapyinterventions}.

As such, Mladen et al. \cite{Mladen2024AffectiveAndMotivationalStatesMetacognition} propose a Generative AI framework showing that leveraging metacognitive experiences, such as confidence judgments and monitoring, helps users interact more efficiently with Generative AI by lowering overall cognitive load through clearer explanations and customisable interfaces, thus improving users’ accuracy and trust. 
Tankelevitch et al. \cite{Tankelevitch&Kewenig2024MetacognitionGenAI} conducted a lab study with 22 university students on a 45‑minute digital reading and writing task, using wearable EEG and digital trace data to monitor their cognitive, affective, metacognitive, and motivational processes. They found that learners' emotional and motivational states, especially interest, engagement, and excitement, could predict whether they used low‑level, high‑level, or metacognitive strategies and promoted deeper cognitive processing.
Additionally, Scharowski et al. \cite{Scharowski2025TrustDistrustPANAS} measured users' positive and negative emotions during AI interactions, since trust and distrust are believed to cause different emotional responses. They reported that positive affect correlates with greater trust, while negative affect drives distrust, emphasizing the importance of capturing emotional states to predict and interpret changes in human-AI trust dynamics.
Based on this body of research, we aim to incorporate measurements of affective metacognition to evaluate variations before and after a self-confidence calibration phase, considering different types of AI assistance and individual traits.

In summary, past research emphasises the importance of delivering tailored AI assistance based on individuals' self-confidence calibration status, as well as considering their overall confidence and accuracy in comparison to the AI, to enhance human-AI team decision-making.
Consequently, our goal is to explore how human confidence calibration based on real-time feedback impacts decision-making, appropriateness of self-confidence, and metacognitive perceptions considering (i) human decision-maker alone, (ii) two-stage AI-assisted decision making, and (iii) an experimental personalized AI assistance based on human self-confidence calibration status, average confidence, and accuracy, and how they compare with the AI ones.

\subsection{Individual Factors Shaping Self-confidence Calibration}
\label{sec:related_human_factors}

People's characteristics are another key factor influencing human self-confidence calibration.
Considering individuals' expertise \cite{Recchia2019ExpertConfidence,Han2024ExpertConfidence}, Han et al. \cite{Han2024ExpertConfidence} examined whether domain expertise provides metaknowledge across climate science, psychological statistics, and investment by comparing experts' and non-experts' confidence in correct versus incorrect choices and analyzing separation metrics such as Murphy's Resolution\footnote{A high Murphy's Resolution means
that one's confidence is a reliable indicator of whether one answers
a question correctly or not, whereas a low Murphy's
resolution means that one's confidence is not an informative indicator
of correctness.} and Yates' Separation.\footnote{Yates' Separation measures the gap between average
confidence for correct versus incorrect responses. A high Yates'
Separation suggests an overall good confidence alignment.} They found that experts were generally less overconfident and assigned higher confidence to correct answers, yet still showed notable limitations, including low Murphy's Resolution and equal or greater confidence when endorsing wrong answers compared to non-experts. The findings suggest that expertise enhances certainty about what is known while masking awareness of the boundaries of one’s knowledge.
For mental health issues \cite{ROUAULT2018443,Seow2020-ko,Hoven2023Psychopathology,Katyal2025AnxietyDepressionLocalGlobalMetacognition}, Hoven et al. \cite{Hoven2023Psychopathology} examined how local, global, and higher‑order confidence relate to psychopathology in a general population sample. They found that an anxious‑depression dimension was linked to underconfidence, while a compulsive‑intrusive‑thoughts dimension showed overconfidence and a disconnect between self‑beliefs and task confidence, with higher‑order self‑beliefs emerging as the strongest predictor of mental health.
Additionally, Katyal et al. \cite{Katyal2025AnxietyDepressionLocalGlobalMetacognition} investigated how both local confidence in individual tasks and global confidence respond to external feedback in two large general population samples. The findings highlight that global confidence is sensitive to local confidence and the valence of feedback. Instead, individuals with higher anxious-depression symptoms exhibit a specific blunted sensitivity to instances of high local confidence, offering a mechanism for their persistent underconfidence despite otherwise intact feedback processing. 
Regarding cultural influences, Ordin et al. \cite{Ordin2024CulturalInfluenceMetacognition} studied whether cultural values affect metacognition during a mental rotation task across three regions (Saudi Arabia, Portugal, and China). 
The results indicate that metacognitive sensitivity is higher in cultures characterized by low individualism and high uncertainty avoidance, that gender gaps disappear at the metacognitive level, and the cultural dimension of masculinity shapes over‑ or underconfidence independently of actual performance.

In our work, we focus on two psychological constructs that prior studies indicate influence human self-confidence calibration and human-AI decision-making. 
The first is Need for Cognition (NFC) \cite{Cacioppo1984NFC}, a stable personality trait reflecting the propensity to engage in effortful cognitive activities, where high NFC individuals tend to be more curious and higher performing at complex tasks than their low-NFC counterparts \cite{Cazan2014NFC,Carenini2001NFC, gajos2017NFC, Ghai2021NFC,Gajos2022NFC,Bucinca2021NFC}. 
In AI-assisted decision-making, the NFC trait has been studied in various domains, such as nutrition \cite{Bucinca2021NFC,Gajos2022NFC}, explaining music recommendation \cite{Millecamp2019NFC,Millecamp2020NFC}, maze solving \cite{Vasconcelos2023OverrelianceXAI}, intelligent tutoring systems  \cite{Conati2021NFC,BahelConati2024NFC}, robots' domestic service \cite{Kopecka2024NFCHumanRobotsExplanationsPerceptions}, job applications \cite{Cau2025CuriosityTraits}, and exercise recommendation \cite{bucinca2025ContrastiveExplanationsNFCAOT}. 
Nevertheless, these works provide contrasting results on whether high-NFC individuals might benefit from AI assistance and explanations in terms of increased accuracy or overreliance mitigation on AI.
As per NFC effects on self-confidence calibration, Vogt et al. \cite{Vogt2022NonAbilityConfidence} examined non‑ability‑based confidence (i.e., the excess of self‑rated over actual cognitive ability) in 1588 participants aged 7-15. They found that this confidence gap was most strongly linked to the NFC, such that kids who enjoy effortful thinking reported higher confidence regardless of actual performance. 
Zerna et al. \cite{Zerna2024NFCReview} investigated how NFC relates to well‑being in healthy adults, discovering that higher NFC is associated with lower neuroticism, anxiety, negative affect, burnout, public self‑consciousness, and depression, and with higher positive affect, private self‑consciousness, and life satisfaction. 
However, high-NFC individuals experience greater perceived control, which boosts active coping but can also increase overconfidence in their resources, thereby reducing the effectiveness of some health interventions.
Furthermore, Cau and Spano \cite{Cau2025CuriosityTraits} explored how different levels of NFC (low or high) could influence accuracy and overreliance on AI when presented with on-demand multifaceted explanations in an AI-assisted job application context, and found that high-NFC individuals tend to be overconfident in their decisions, despite showing no differences in accuracy compared to low-NFC individuals.

The second construct is Actively Open-minded Thinking (AOT) 
\cite{Baron_1985_AOT}, a style of thinking that includes the tendency to
weigh new evidence against a favored belief, to spend sufficient time on a problem before giving up, and to carefully consider the opinions of others when forming one's own \cite{Ottati2023AOT,Stanovich2023AOTMeasurement}.
Haran et al. \cite{Haran2013AOTinAccuracyAndCalibration} examined whether individual differences in AOT, NFC, grit, and decision‑making style (maximize vs. satisfice) predict performance on an estimation task when participants control how much information they gather. 
The results highlight that AOT reliably predicted information acquisition and performance, considering categorical and quantitative estimates. Additionally, higher AOT led to an increase in persistence in search for information, higher accuracy estimates, and lower overconfidence.
Furthermore, Martin et al. \cite{Martin2024CalibrationFeedbackAOT} tested scalable calibration training, comparing outcome feedback with Practical scoring rule\footnote{The Practical scoring rule is a modification of the logarithmic scoring rule designed to be more intuitive to facilitate learning \cite{greenberg2020calibrationscoringrulespractical}.} performance feedback, and evaluated AOT as a predictor of calibration. The findings indicate that neither feedback approach reduced overconfidence, but individuals with higher AOT consistently demonstrated better calibration overall, even though AOT did not lead to accuracy improvement across training blocks.

Compared to NFC, AOT remains underexplored in the literature on AI-assisted decision making. For example, Swaroop et al. \cite{Swaroop2025AIPersonalizedOverrelianceRate} examined how to quickly assess users' tendency to over‑rely on AI and personalize assistance accordingly using probe questions. They found that a small set of well‑chosen probe questions reliably identifies over‑reliance, and that tailoring AI support based on this hidden trait improves human‑AI team performance. Also, over‑reliers don't differ in NFC or Openness but score higher on Agreeableness and Neuroticism and lower on AOT, suggesting that personality profiles are key for effective adaptation. 
Additionally, Buçinca et al. \cite{bucinca2025ContrastiveExplanationsNFCAOT} studied the benefits of contrastive explanations, which highlight differences between the AI's choice and a likely human choice, compared to \enquote{unilateral} explanations that justify the AI's decision but do not account for users' knowledge and thinking. They found contrastive explanations significantly boost independent decision‑making without harming accuracy, though overreliance on AI still occurs. These benefits are strongest in high-AOT individuals, suggesting that AOT moderates who gain most from contrastive explanations.

Considering the limited existing research on how psychological constructs affect self-confidence calibration in AI-assisted decision-making, results indicate that high-NFC individuals tend to be overconfident in their decisions, whereas high-AOT individuals generally exhibit better calibration.
Despite these findings, there is still a lack of research examining how calibration mechanisms affect individuals with different traits and increase awareness of their actual self-confidence calibration during an AI-assisted task. 
To address this gap, we conducted a user study to explore whether differences exist in the accuracy and appropriateness of self-confidence among individuals with low or high NFC and AOT, and to investigate further
whether they tend to have distinct self-assessment processes related to global and affective metacognition.

\section{Research Questions}
\label{sec:rqs}

Based on the grounding provided in Section \ref{sec:related}, we designed two studies to investigate how calibrating individuals' self-confidence influences their behavior in subsequent interactions, considering different types of AI assistance, as well as individual traits such as the Need for Cognition (NFC) and Actively Open-minded Thinking (AOT). 
Specifically, Study 1 aims to explore how to define \textit{well-calibrated} individuals, setting the groundwork for personalized AI assistance, and examine changes in accuracy and self-confidence appropriateness for people with low and high NFC and AOT during a confidence calibration phase. Conversely, Study 2's objective is to test how different types of AI assistance (no AI, two-stage AI, and personalized AI), along with low and high levels of NFC and AOT, influence humans' accuracy, self-confidence appropriateness, as well as global and affective metacognition after a confidence calibration phase.

As previous work suggests, calibrating humans' self-confidence might introduce benefits in decision-making outcomes, such as increased accuracy, improved appropriateness of confidence calibration, and reliance on AI advice \cite{PULFORD1997OverconfidenceHumanCalibration,Tenney2008HumanConfidenceCredibility,Miller2015HumanConfidenceCalibration,Moore2017HumanConfidenceCalibration,He2023DunningKruger,Ma2023CorrectnessLikelihoodAIUsersIncome,OttenandFischoff2023HumanConfidenceCalibrationScientificReasoningAbility,Benz2024HumanConfidenceCalibration,Ma24UserConfidenceCalibration,FosterAndRenie2024HumanConfidenceCalibrationStudents}. 
Calibration \cite{Camerer1991JudgmentAD,dunlosky2009metacognition} can be defined at the instance level, where people's confidence judgments demonstrate overconfidence (judgments
are more optimistic than actual performance), underconfidence
(judgments are less optimistic than actual performance), or correct alignment\footnote{Correct alignment can be seen as an individual who is \textit{well-calibrated} in a specific instance.} (judgment matches the actual performance) \cite{LichtensteinFischhoffPhillips1982,Koriat2018OutOfFocus,Ma24UserConfidenceCalibration}. Furthermore, calibration can also be attained at a task level by measuring the difference between mean confidence and mean accuracy across items (when both are assessed on the same scale) \cite{Koriat2018OutOfFocus,li2025confidencealigns}. Focusing on the latter, previous HCI research mainly explored overconfident and underconfident disaggregation among users \cite{Ma24UserConfidenceCalibration,li2025confidencealigns}, while attempts to define a \enquote{well-calibrated} user across multiple items are rarely addressed, leaving the criteria for identifying this user category and the strategies for providing personalized AI assistance to them inadequately defined \cite{Ma2023CorrectnessLikelihoodAIUsersIncome,Ma24UserConfidenceCalibration,li2025confidencealigns, Cau2025CuriosityTraits}.
Consequently, to enhance existing findings regarding the role of self-confidence and to provide personalized AI assistance also for well-calibrated individuals in a low-stakes, easy task scenario, we formulated the following research question: 

\begin{itemize}
    \item \textbf{RQ1.1}: 
    What threshold of the gap between a human's average self-confidence and their actual accuracy best identifies well-calibrated individuals?
\end{itemize}

Furthermore, past research highlighted the importance of individual traits in shaping users' decision-making and behavior \cite{BEELER2022IndividualFactorsDisclosure,Khadar2025UserCharacteristicsXAI}. For example, individuals with high NFC might achieve increased accuracy and be more prone to overconfidence in specific tasks when compared to low-NFC individuals \cite{Gajos2022NFC,BahelConati2024NFC,Vogt2022NonAbilityConfidence,Zerna2024NFCReview,Cau2025CuriosityTraits}. Rather, high-AOT individuals can also be more prone to increased accuracy while having lower overconfidence and an overall good self-confidence calibration compared low-AOT individuals \cite{Haran2013AOTinAccuracyAndCalibration,Martin2024CalibrationFeedbackAOT,Swaroop2025AIPersonalizedOverrelianceRate,LiHaleMoore_2025OverconfidenceIndividualDifference}. 
Given that the aspect of confidence calibration is understudied when considered in combination with psychological traits related to self-confidence, we ask the following sub-question: 

\begin{itemize}
    \item \textbf{RQ1.2}: Do humans with different levels of NFC and AOT exhibit differences in accuracy and appropriateness of self-confidence during a confidence calibration session?
\end{itemize}

Along with this, a growing body of research suggests that AI assistance should not only be tailored based on AI-calibrated confidence levels \cite{Cao2024UncertaintyPresentation, Ma24UserConfidenceCalibration,Marusich2024AIUncertaintyQuantification,li2025confidencealigns} and human cognitive biases \cite{Ma2023CorrectnessLikelihoodAIUsersIncome,Nattapat2025CognitiveBiases}, but also on humans' self-confidence status \cite{Pescetelli2021GroupDecisionMetacognition,CHONG2022HumanConfidence,He2023DunningKruger,Ma24UserConfidenceCalibration,Cau2025CuriosityTraits,Nguyen2025DiadDecisions,li2025confidencealigns}. 
For instance, while overconfident users might avoid following or requesting AI assistance due to overestimating their abilities, underconfident ones might completely trust AI assistance, as they doubt their judgment. Additionally, specific dyadic decision-making paradigms, such as Maximum Confidence Slating (MCS), produce accuracy gains that increase with dyadic confidence calibration: the higher the dyad's confidence calibration, the higher their decision accuracy becomes \cite{Bang2014Doesinteractionmatter,Bang2017ConfidenceGroupDecisionMaking,Nguyen2025DiadDecisions}.
This highlights the need to tailor different AI assistance paradigms based on users' self-confidence to optimise human-AI team dyad collaboration, which is still underexplored in the current literature.
Meanwhile, metacognitive research has widely explored the importance of capturing self-confidence judgments regarding individual tasks and specific intellectual abilities (namely, local metacognition and global metacognition) in influencing human decision-making behavior 
\cite{KLEITMAN2007SelfConfidenceMetacognitivePRocesses,Moore2008OverconfidenceOverprecision,moore2015OverprecisionInJudgement,Rouault2019GlobalEstimatesFromLocalSelfPerformance,Fleming2024_metacognition,KATYAL2024_metacognition,Katyal2025AnxietyDepressionLocalGlobalMetacognition}. However, prior research recommends integrating metacognitive perceptions, such as subjective affective experiences (i.e., affective metacognition), which have, to date, been understudied in the HCI field 
\cite{Smith2016MemoryAcuteStress,Tankelevitch&Kewenig2024MetacognitionGenAI,KATYAL2024_metacognition,Cushing2024Metacognitionasawindowintosubjectiveaffectiveexperience,Mladen2024AffectiveAndMotivationalStatesMetacognition}. Exploring these experiences could help capture a broader spectrum of both human decision‐making behavior and how internal emotional states may change during interactions.
Therefore, we aim to bridge these gaps by exploring the potential benefits of tailoring AI assistance according to individuals' self-confidence, while also considering the effects of AI assistance on metacognitive perceptions, such as global and affective metacognition. Thus, we propose the following research question:

\begin{itemize}
    \item \textbf{RQ2}: 
    After calibrating humans' self-confidence (i.e., making them aware of their overconfident, underconfident, or well-calibrated status), how do different types of AI assistance impact their accuracy, appropriateness of self-confidence, and metacognitive perceptions?

    \begin{itemize}
        \item \textbf{RQ2.1}: 
         How do different types of AI assistance (including no AI) affect humans' accuracy and the appropriateness of their self-confidence?

        \item \textbf{RQ2.2}: How do different types of AI assistance (including no AI) affect humans' global and affective metacognition? 
        
    \end{itemize}
\end{itemize}

As mentioned earlier, individual traits also shape people's decision-making and behaviour. While high-NFC individuals appear to achieve higher accuracy, they also exhibit overconfident behaviour. High-AOT individuals likewise seem to achieve higher accuracy while maintaining a good overall calibration of self-confidence.
Nevertheless, most prior research has overlooked how individuals with varying levels of NFC and AOT might respond after a calibration phase, or how calibration might impact their accuracy and the appropriateness of self-confidence in AI-assisted tasks. Moreover, the influence of psychological traits on metacognitive perceptions remains largely unexplored in the current literature.
Finally, our work proposes the third research question:
\begin{itemize}

    \item \textbf{RQ3}: 
     After calibrating humans' self-confidence (i.e., making them aware of their overconfident, underconfident, or well-calibrated status), how do different levels of Need for Cognition (NFC) and Actively Open-minded Thinking (AOT) impact their accuracy, appropriateness of self-confidence, and metacognitive perceptions?
    
    \begin{itemize}
        \item \textbf{RQ3.1}: Do humans with different levels of NFC and AOT exhibit differences in accuracy and appropriateness
        of self-confidence? 
        
        \item \textbf{RQ3.2}: 
        Do humans with different levels of NFC and AOT exhibit differences in global and affective metacognition?

    \end{itemize}
\end{itemize}

\section{Income Prediction Task}
\label{sec:pred_task}
This section describes the shared design for the income prediction task and AI model that served as the experimental testbed for addressing our research questions in Study 1  (Sec. \ref{sec:study1}) and Study 2 (Sec. \ref{sec:study2}).

\subsection{Task}
We employed the \textit{income prediction} task in our study, which has been used by several previous works in AI-assisted decisions \cite{Ribeiro_Singh_Guestrin_2018income,Zhang2020ConfidenceExplanationsAccuracyTrust,Ghai2021income,Ma2023CorrectnessLikelihoodAIUsersIncome,Ma24UserConfidenceCalibration,li2025confidencealigns,Khadar2025UserCharacteristicsXAI}. In this task, participants are requested to predict whether a person's annual income is less/equal or more than \$50K based on some attributes like age, work class, and others.
The data for this task is from the Adult Income dataset of the UCI Machine Learning Repository
\cite{adult1996}, consisting of 48,842 instances and 14 features. 
We selected this specific task since it requires a low domain expertise and limited risks \cite{Salimzadeh2023TaskComplexityDecisionMaking,Salimzadeh2024TaskCharacteristics}, hence suitable for lay users, and has been used by previous HCI research working on the AI-human confidence calibration issue \cite{Ghai2021income,Ma2023CorrectnessLikelihoodAIUsersIncome,Ma24UserConfidenceCalibration,li2025confidencealigns}, so we can faithfully use the same settings and proceed with a fair comparison of the outcomes. Following prior work settings \cite{Zhang2020ConfidenceExplanationsAccuracyTrust,Ma24UserConfidenceCalibration,li2025confidencealigns}, we decided to reduce the features' space from 14 to 8 to reduce participants' task difficulty and cognitive load, keeping the following features: age, work
class, years of education, marital status, occupation, race, gender, and hours worked
per week (see Fig. \ref{fig:calib_phase}-A1).

\subsection{AI Model} 
According to the Adult Income dataset statistics \cite{adult1996}, the best performing models in terms of accuracy are, in order: eXtreme Gradient Boosting (XGB), Random Forest (RF), and Logistic Regression (LR). However, given that we reduced the total number of predictive features to eight, we needed to reassess the models' performance and calibration based on this adjustment. We then trained the models using a 60:20:20 split for the training, calibration, and test sets. After hyperparameter tuning, we had the following values of accuracy (\textit{acc}) and ROC-AUC (\textit{AUC}) for the test set: XGB ($acc=83.76, AUC=.89$), RF ($acc=83.63, AUC=.89$), and LR ($acc=81.43, AUC=.86$). Given the lowest performance, we discarded LR and moved on to calibrating \cite{SilvaFilho2023Calibration,Ma24UserConfidenceCalibration} the XGB and RF models (see Appendix \ref{sec:app_model_calibration} for more details). After the calibration using ensembles of two models, XGB and RF, achieved comparable performance and calibration metrics. However, given the slightly better overall metrics, we decided to use the XGB ensemble as our AI, which results in better-calibrated confidence estimates through probabilities.

\subsection{Selected Instances}
Inspired by Ma et al.'s approach \cite{Ma24UserConfidenceCalibration}, we selected 20 instances for the calibration phase and 20 for the main task, which followed the same selection criteria. Half instances had a confidence below 0.75 (low confidence), with an average confidence of 0.60 and accuracy of 60\% (six out of ten were correct). The remaining half had confidence scores greater than or equal to 0.75 (high confidence), with an average confidence of 0.9 and accuracy of 90\% (nine out of ten were correct). This procedure ensured an optimal AI confidence calibration on the task samples for our study (see Table \ref{tab:final_instances} in Appendix \ref{sec:app_selected_instances}).
We also balanced the binary true classes, having ten instances for incomes less than \$50K and ten for incomes exceeding \$50K.
Finally, we created 300 random permutations for calibration and main task instances to avoid ordering bias \cite{Nourani2021AnchoringBias} and ensure that each participant encountered income profiles uniquely ordered.

\section{Study 1 - Identifying Well-Calibrated Users and How NFC and AOT Shape Self-Confidence and Accuracy}
\label{sec:study1}

In this first study, we aimed to determine how to distinguish well-calibrated users from those who are overconfident or underconfident. This classification would later inform the design of \textit{personalized AI} assistance based on users' self-confidence status after the calibration phase. Additionally, we conducted a preliminary analysis of how the accuracy and appropriateness of self-confidence varied across individuals with low and high levels of NFC and AOT.\footnote{The differentiation between low and high NFC and AOT levels was achieved by using the respective distribution's median as a split.}
Specifically, we measured NFC using the NCS-6 six-item five-point scale
from \cite{LinsDeHolandaCoelho2020NFC6} (see Appendix \ref{sec:app_nfc_scale}), while AOT was assessed using the Actively Open-minded Thinking Scale (seven-item, seven-point) by Haran et al. \cite{Haran2013AOTinAccuracyAndCalibration} (see Appendix \ref{sec:app_aot_scale}).

\begin{figure}
      \centering
    \includegraphics[width=\textwidth]{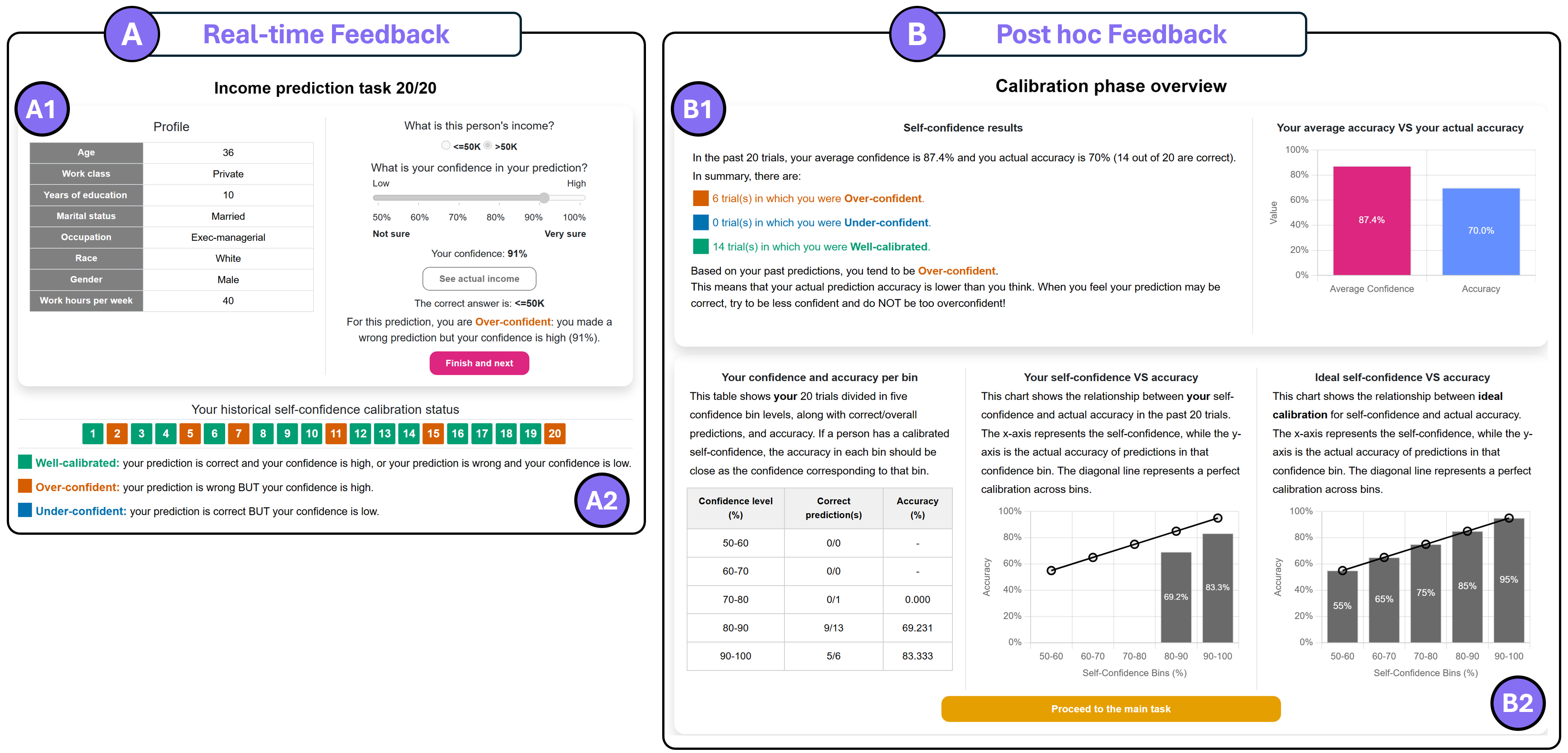}
    \caption{ The interface with which participants interacted in our user study as a part of the calibration phase, inspired by Ma et al. \cite{Ma24UserConfidenceCalibration}. 
(A) Real-time Feedback interface provides participants with correctness and self-confidence calibration feedback for each instance. The bottom of the interface (A2) displays a historical self-confidence status bar for previous choices, color-coded based on self-confidence levels and listed in chronological order of the tasks. 
(B) Post hoc feedback interface shows an overview of participants' calibration status after the calibration phase, summarising the proportion of self-calibration status for well-calibrated, overconfident, and underconfident instances, plus their average confidence and actual accuracy in a bar chart (B1). 
The bottom part of the interface displays their past predictions, divided into five bins based on confidence distribution, along with a chart showing the relationship between average confidence and actual accuracy compared to an ideal self-confidence and accuracy (B2).
    }

    \Description{
     todo.
    }
 
    \label{fig:calib_phase}
\end{figure}

\subsection{Calibration Mechanism}
\label{sec:calibration_mechanism_study1}
To calibrate participants’ self-confidence, we drew on Ma et al.’s \textit{Calibration Status Feedback} \cite{Ma24UserConfidenceCalibration}, which provides immediate feedback on correctness and confidence calibration for each prediction. Specifically, we adopted their instance-level Confidence-Correctness matching, classifying predictions into three categories based on confidence (low or high) and correctness (correct or incorrect). We define \emph{Over-confident} as [High confidence \& Incorrect prediction] and \emph{Under-confident} as [Low confidence \& Correct prediction]. Conversely, both [High confidence \& Correct prediction] and [Low confidence \& Incorrect prediction] are labeled as \emph{Well-calibrated} (C-C Matched). 
In this first study, we only use the \textit{Real-time Feedback} interface, depicted in Figure \ref{fig:calib_phase}-A. 

While Ma et al.'s work gives an indication of which self-confidence status the human is tending to on an instance level (without specific thresholds for instance-based self-confidence classification), the main goal of our first study is to define an overall labeling procedure for users into overconfident, underconfident, or well-calibrated self-confidence status based on the average confidence and accuracy gap.
Specifically, given the perfect calibration line (i.e., average confidence = accuracy), we aim to define a symmetric tolerance interval $\pm \delta$ around this line. Participants whose average confidence deviates from their accuracy by no more than $\delta$ are classified as \textit{well-calibrated}. Namely, we use the following criteria to classify users' self-confidence:

\begin{equation}
\label{eq:calibration_label}
\mathrm{label}(\mathrm{gap}) =
\begin{cases}
\textit{Over\--confident},  & \text{if }\mathrm{gap} > \delta,\\[6pt]
\textit{Under\--confident}, & \text{if }\mathrm{gap} < -\delta,\\[6pt]
\textit{Well\--calibrated}, & \text{if }|\mathrm{gap}|\le \delta.
\end{cases}
\end{equation}

To achieve this, we conducted a sensitivity analysis considering the variability of users' percentage classified as well-calibrated with $\delta \in [0, 25]$. 
We first ran 5000 simulations on 128 synthetic users whose average confidence and accuracy profiles were drawn to mimic plausible \enquote{lay} behavior (see Section \ref{sec:users_simulation} for more details and Table \ref{tab:simulation_result} in Appendix for simulation results). The results led to a candidate maximum gap of $\delta =$ 20, with approximately 50\% of users classified as well-calibrated for that value (practically unfeasible in a real-world scenario where most users are usually overconfident \cite{Frankenberger1997decision,Razmdoost2015UnderconfidentOverconfident,Soll2022OverconfidenceAdvice,Binnendyk2024IndividualDifferencesOverconfidence,li2025confidencealigns,LiHaleMoore_2025OverconfidenceIndividualDifference}). 
Therefore, we set our initial upper bound at $\delta = 20$ and planned to determine the most suitable \(\delta\) threshold for identifying well-calibrated users with three different approaches:

\begin{itemize}
  \item \textit{Quartile‐based}: For $\delta \in [0, 20]$, we computed the percentage $p(\delta)$ of users whose calibration gap falls within that delta. We then look at the distribution of these percentages across all deltas, identify the first quartile $Q_1$ (the value below which 25\% of the percentages lie), and finally choose the delta coinciding with $Q_1$.

  \item \textit{Elbow‐based}: 
  For $\delta \in [0, 20]$, we computed the percentage $p(\delta)$ of users whose calibration gap falls within that delta. We then examine how that percentage grows as $\delta$ increases and look for the \enquote{knee} in the curve (i.e., the point where the rate of increase slows down most sharply). We approximate the curvature by taking the second discrete difference of the percent of well-calibrated values and select the delta at which this absolute second difference is maximal. This \enquote{elbow} identifies the threshold beyond which adding tolerance yields diminishing returns in well‐calibrated coverage.\footnote{Our R implementation approximates the curvature by computing the second discrete difference of the well-calibrated percent users' series and selecting the threshold \(\delta\) that maximizes \(\lvert \Delta^2 p(\delta)\rvert\). This is a lightweight numerical realization of the elbow method as introduced by Thorndike \cite{Thorndike1953whoKnees} and later formalized in the Kneedle algorithm \cite{SatopaaKnee2011}.}  

  \item \textit{ECE-based}:
  For $\delta \in [0, 20]$, we identified users whose calibration gap fell within that delta and computed (i) the percentage of well‑calibrated users and (ii) their mean Expected Calibration Error (ECE, see Equation \ref{eq:ece}). 
  We then selected the optimal tolerance \(\delta\) as the value that minimized mean ECE among those users. 

\end{itemize}

\subsection{Evaluation Metrics}
\label{sec:evaluation_metrics_study1}

Along with exploring how to label a well-calibrated user, we analyzed the differences between low- and high-level NFC and AOT individuals in the calibration phase using the following dependent variables:

\begin{itemize}
    \item \textbf{Accuracy (numerical).}
    Percentage of participants' correct decisions in the 20 calibration phase instances. 
    
    \item \textbf{Appropriateness of human self-confidence (continuous).}
    We measured the appropriateness of human self-confidence using the Expected Calibration Error (ECE) \cite{Ma24UserConfidenceCalibration,li2025confidencealigns}, which computes model calibration by quantifying discrepancies between expected accuracy and self-reported confidence across \( N \) predictions divided into \( M \) bins of equal length (see Eq. \ref{eq:ece}). Specifically, ECE computes the absolute difference between the accuracy \( acc(B_m) \) and average confidence \( conf(B_m) \) for each bin \( B_m \), which is weighted by the number of predictions of each bin \( |B_m| \) over the total prediction number \( N \). 
    A smaller ECE value indicates better human self-confidence calibration, hence reflecting a higher matching between actual accuracy and self-confidence.
    In this paper, we set \( M \) = 5 as the number of bins and \( N \) = 20 as the number of predictions (tasks).
    
    \begin{equation}
    ECE = \sum_{m=1}^{M} \frac{|B_m|}{N} |acc(B_m) - conf(B_m)|
    \label{eq:ece}
    \end{equation}

\end{itemize}

\subsection{Sample Size and Statistical Analysis}
\label{sec:sample_size_stat_study1}

Before recruiting participants, we computed the required sample size considering the investigation of low and high NFC and AOT individuals using
\textit{G*Power} \cite{faul2009statistical} and considering a medium effect size (Cohen's $f$ = 0.25), a desired power of (1 - $\beta$) = 0.8, and a significant threshold $\alpha=.05$. 
Specifically, we conducted two between-subjects ANOVA tests with \textit{accuracy} and \textit{ECE} as dependent variables, examining the main effects of NFC and AOT as independent variables, which resulted in a required sample size of 128 participants. We corrected p-values using Bonferroni's adjustment, accounting for two ANOVA tests.

\subsection{Participants and Procedure}
\label{sec:participants_procedure_study1}

After the user study was approved by the University of Cagliari Ethics Committee,\footnote{Received on 30 April 2025, Prot. 0117403.\label{ethics_approval}} we recruited participants from Prolific\footnote{https://www.prolific.com/\label{prolific_}} with the following selection criteria: balanced participation across genders, U.S. residency, aged 18 or more, high English proficiency, approval rate over 99\%, and Desktop as a mandatory participation device. 
We rewarded participants with \pounds 1.35 for completing the study, considering an average completion time of 9 minutes, paying an average of \pounds 9 per hour. We rewarded an extra \pounds 1 to participants who achieved an ECE below 0.05 to incentivise high-quality work. 
To define the experimental process, we took inspiration from Ma et al.'s self-confidence calibration framework \cite{Ma24UserConfidenceCalibration}. We implemented the user‑study interface on a full \textit{JavaScript} stack: \textit{Node.js} powers the backend, \textit{Bootstrap} drives the frontend, and all user interactions, decisions, and questionnaire responses were registered using \textit{MongoDB}. Only participants who passed all the comprehension checks were included in the statistical analysis.

Upon accepting the informed consent, participants underwent a familiarization tutorial explaining how to use the task interface (see Fig. \ref{fig:calib_phase}-A), where we described each of the eight profile attributes and showed the binary income distribution for each attribute. Participants were then asked to respond to some questions about the tutorial, and only those who provided correct answers qualified to proceed to the next phase.\footnote{We used Comprehension Checks to ensure participants understood critical information that was integral to completing the study successfully, following the good practices of Prolific (see \href{https://researcher-help.prolific.com/en/article/fb63bb}{link}).\label{comprehension_checks}}
Then, participants completed the NFC and AOT questionnaires before beginning the calibration phase. 
Finally, participants engaged in a calibration session (see Sec. \ref{sec:calibration_mechanism_study1}), where they completed 20 tasks stating their prediction and confidence, receiving feedback after each decision about the correctness and their self-confidence status, categorized as well-calibrated, overconfident, or underconfident. 
A historical self-confidence status bar was displayed for all previous choices, color-coded based on self-confidence status and enumerated in chronological order of the tasks (see Fig. \ref{fig:calib_phase}-A2).

\begin{figure} [!t]
      \centering
    \includegraphics[width=\textwidth]{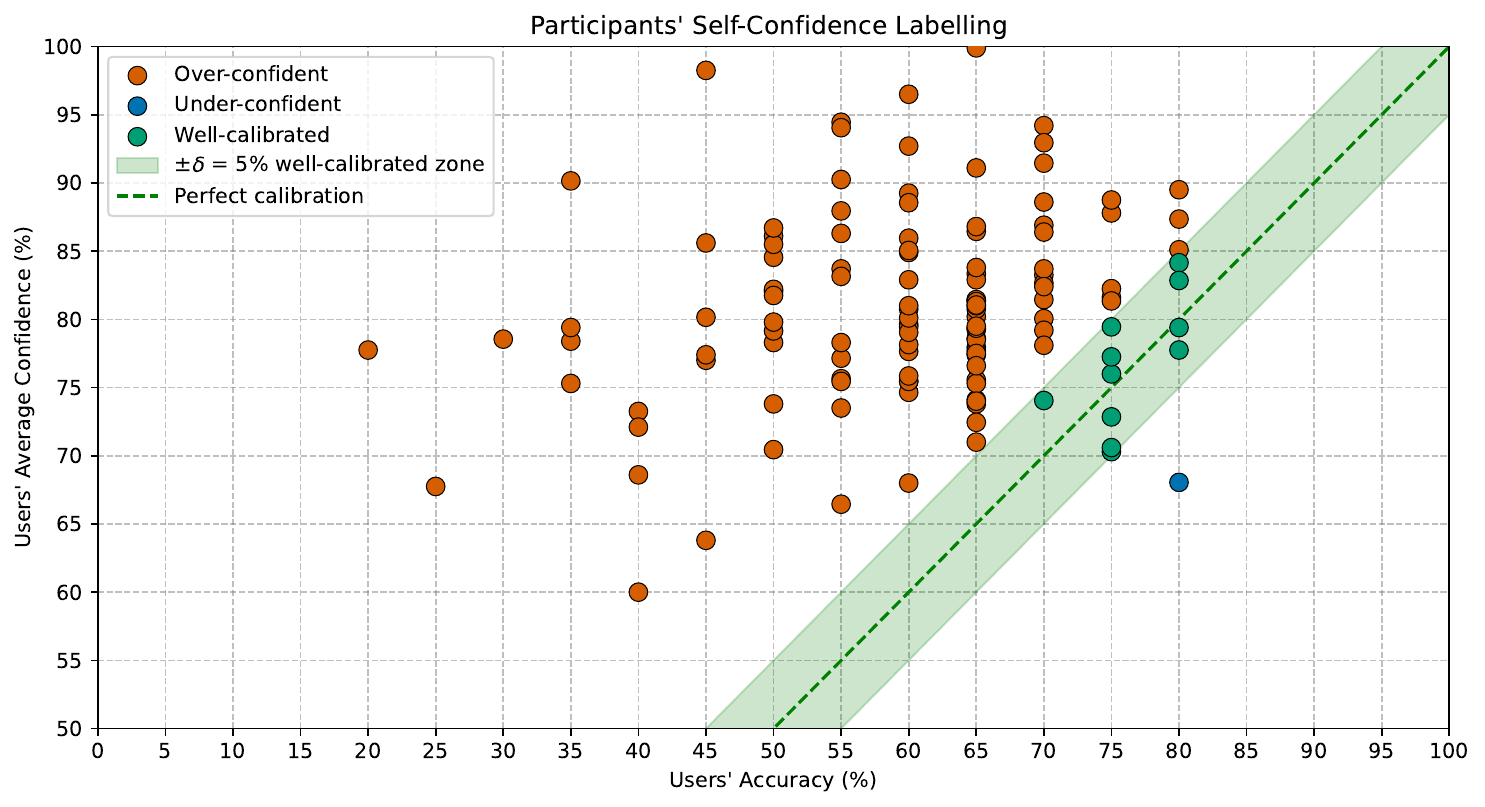}
    \caption{Participants' self-confidence distribution after the calibration phase, disaggregated by over-confident, under-confident, and well-calibrated groups, using a threshold of $\delta$ = 5.
    }

    \Description{
     todo.
    }
 
    \label{fig:study1-well-calibrated-distr}
\end{figure}

\subsection{Results}
\label{sec:results_study1}

\subsubsection{Descriptive Statistics}
Of the 128 participants in our user study, 64 were female and 64 were male, with an average age of $M$ = 36.47 and $SD$ = 13.42. The NFC subdivision into low (56) and high (72) individuals was achieved with a computed median $Mdn$ = 4.00 ($M$ = 3.91 and $SD$ = 0.72). For individuals with AOT low (61) and high (67), we had a median $Mdn$ = 5.14 ($M$ = 5.14 and $SD$ = 1.01). 
Participants achieved a mean accuracy of $M$ = 60.55 ($SD$ = 12.11), in line with previous studies using similar study designs \cite{Ma24UserConfidenceCalibration,li2025confidencealigns}.

\subsubsection{Normality and Data Homogeneity Assumptions for ANOVA Tests}
The Shapiro-Wilk test revealed that both accuracy 
\(\bigl(W = 0.944,\ p = .0001\bigr)\) and ECE 
\(\bigl(W = 0.970,\ p = .0062\bigr)\) distributions 
significantly deviated from normality. 
Instead, the Levene's test assumption was satisfied for both accuracy ($F$ = 1.57, $p$ = 0.2) and ECE ($F$ = 0.81, $p$ = 0.49), as well as random sampling and independence. Consequently, we conducted ANOVA tests to assess the main effects of the NFC and AOT factors.

\subsubsection{RQ1.1 - Defining Well-calibrated Participants}
\begin{figure} [!t]
      \centering
    \includegraphics[width=\textwidth]{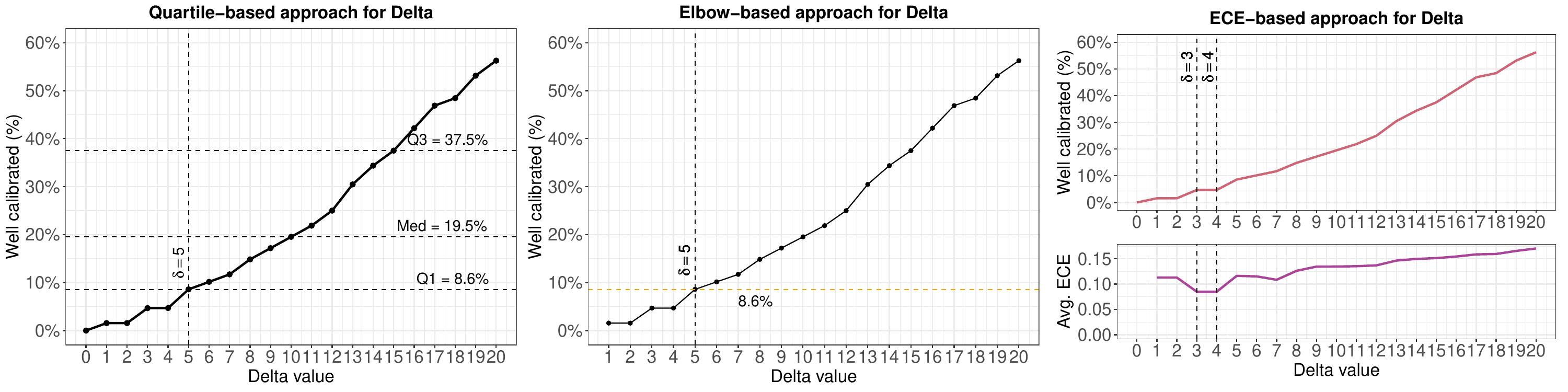}
    \caption{ Threshold results for $\delta$ using three different methods: quartile-based ($\delta$ = 5), elbow-based ($\delta$ = 5), and ECE-based ($\delta$ = 3, and $\delta$ = 4).
    }

    \Description{
     todo.
    }
 
    \label{fig:study1_delta_methods}
\end{figure}

After analyzing the data, we computed the following candidate delta thresholds for classifying well-calibrated users: \textit{Quartile-based} ($Q_1$ = $p(\delta)$ = 8.6\%, $\delta = 5$), \textit{Elbow-based} (elbow = $p(\delta)$ = 8.6\%, $\delta = 5$), and \textit{ECE-based} 
($p(\delta) = 4.7\%$, $\delta_1 = 3$, avg. ECE = 0.085; 
$p(\delta) = 4.7\%$, $\delta_2 = 4$, avg. ECE = 0.085). 
Considering the shared results from Quartile- and Elbow-based approaches lead to a candidate value of $\delta$ = 5, while the ECE-based approach led to two candidates ($\delta \in [3, 4]$), we selected \(\delta = 5\) as the optimal threshold for classifying well-calibrated users (see Figure \ref{fig:study1_delta_methods}).
Figure \ref{fig:study1-well-calibrated-distr} presents participants' self-confidence labelling distribution based on \(\delta = 5\), which consists of: 116 (90.6\%) Over-confident, 1 (0.8\%) Under-confident, and 11 (8.6\%) Well-calibrated participants.

\begin{figure} [!t]
      \centering
    \includegraphics[width=\textwidth]{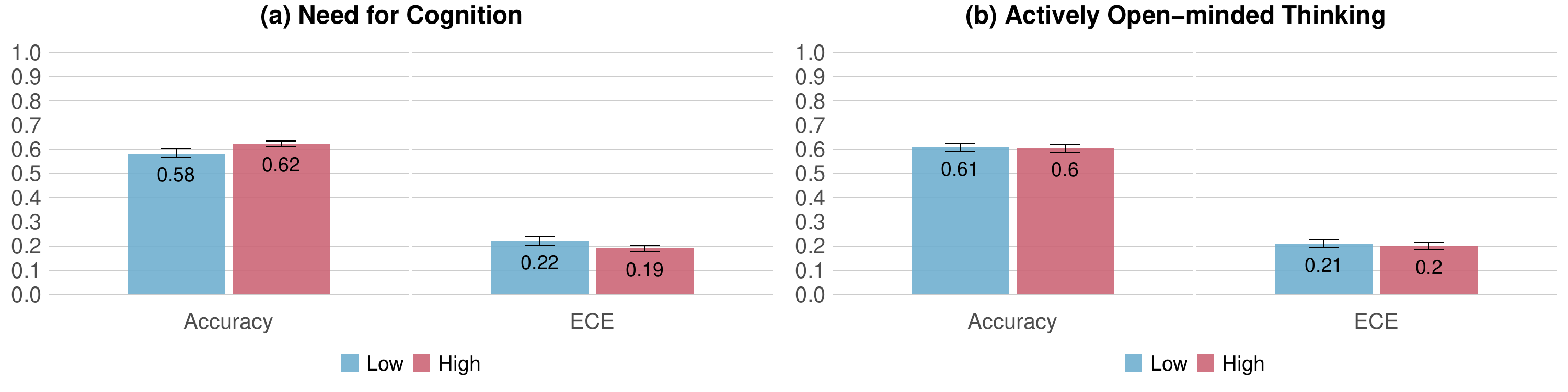}
    \caption{Participants' accuracy and appropriateness of their self-confidence (ECE) in the calibration phase considering low and high (a) Need for Cognition, and (b) Actively Open-minded Thinking.} 

    \Description{
     todo.
    }
 
    \label{fig:study1-traits-categorical}
\end{figure}

\begin{table}[!t]
  \centering
  \footnotesize
  \begin{tabular}{
p{2cm}                           
>{\centering\arraybackslash}p{1.4cm}
>{\centering\arraybackslash}p{1.7cm} 
>{\centering\arraybackslash}p{1.7cm} 
>{\centering\arraybackslash}p{1.7cm} 
>{\centering\arraybackslash}p{1.7cm} 
>{\centering\arraybackslash}p{1.7cm} 
  }
    \toprule
    \multicolumn{7}{c}{\textbf{Accuracy model (ANOVA)}}\\
    Factor & Df & Sum Sq & Mean Sq & F value & $p_{\textnormal{raw}}$ & $p_{\textnormal{adj}}$ \\ 
    \midrule
    NFC category  & 1  & 501    & 501     & 3.47     & .065 & 0.130    \\
    AOT category  & 1  & 44     & 44      & 0.31     & .581 & $>1$     \\
    \addlinespace
    \cmidrule(lr){1-7}
    \addlinespace
    \multicolumn{7}{c}{\textbf{Expected Calibration Error (ECE) model (ANOVA)}}\\
    Factor & Df & Sum Sq & Mean Sq & F value & $p_{\textnormal{raw}}$ & $p_{\textnormal{adj}}$ \\ 
    \midrule
    NFC category  & 1  & 0.009  & 0.0091 & 0.78     & .380 & 0.760    \\
    AOT category  & 1  & 0.0004 & 0.0004 & 0.04     & .850 & $>1$     \\
    \bottomrule
  \end{tabular}
  \caption{ANOVA tests result for Need for Cognition (NFC) and Actively Open-minded Thinking (AOT) on accuracy and ECE in Study 1 with Bonferroni's p-value adjustment ($m$ = 2).}
  \label{tab:study1_anovas}
\end{table}

\subsubsection{RQ1.2 - NFC and AOT Levels Effects on Accuracy and Appropriateness of Human Self-Confidence}
The results from the non-parametric permutation ANOVA, depicted in Figure \ref{fig:study1-traits-categorical} and described in Table \ref{tab:study1_anovas} indicate that individual traits are not significant for both accuracy (NFC: $F = 3.47$, $p_{\textnormal{adj}} = .13$, Cohen's $f$ = 0.17; AOT: $F = 0.31$, $p_{\textnormal{adj}} > 1$, Cohen's $f$ = 0.05) and ECE (NFC: $F = 0.78$, $p_{\textnormal{adj}} = .76$, Cohen's $f$ = 0.08; AOT: $F = 0.04$, $p_{\textnormal{adj}} > 1$, Cohen's $f$ = 0.02).

\subsubsection{Summary of Findings}
The results of Study 1 indicate that, in our settings, $\delta = 5$ is a potential optimal threshold for classifying users' self-confidence into Over-confident, Under-confident, and Well-calibrated groups (\textbf{RQ1.1}). 
As depicted in Figure \ref{fig:study1-well-calibrated-distr}, most participants were Over-confident (90.6\%), with 8.6\% Well‑calibrated and only 0.8\% Under‑confident.
Furthermore, we found no significant differences in accuracy and ECE for both low and high NFC/AOT groups (\textbf{RQ1.2}).

\section{Study 2 - Effects of AI Assistance, NFC, and AOT on Accuracy, Self-Confidence, and Metacognition}
\label{sec:study2}
Our second study investigates how three levels of AI assistance (i.e., no AI, two-stage AI, and personalized AI) and individual differences in Need for Cognition (high vs. low NFC) and Actively Open-minded Thinking (high vs. low AOT) shape users' behavior considering their accuracy, appropriateness of self-confidence, and metacognitive perceptions (i.e., global and affective metacognition).

\subsection{Calibration Mechanism}
\label{sec:calibration_mechanism_study2}
The calibration procedure mirrors that of Study 1 (see Section \ref{sec:calibration_mechanism_study1}), combining \textit{Real-time feedback} with a \textit{Post hoc feedback} (see Figure \ref{fig:calib_phase}). Participants receive immediate feedback during the task, followed by an overall report that includes total self-confidence status, average confidence, accuracy, a table of confidence-accuracy bins, and reliability diagrams comparing their actual confidence-accuracy profile against the ideal. Unlike Ma et al. \cite{Ma24UserConfidenceCalibration}, we display a final self-confidence status label derived from the average confidence and accuracy gap ($\delta = 5$) resulting from Study 1, which highlights participants' self-confidence calibration status (i.e., Over-confident, Under-confident, or Well-calibrated). 

\begin{figure}[t!]
      \centering
    \includegraphics[width=\textwidth]{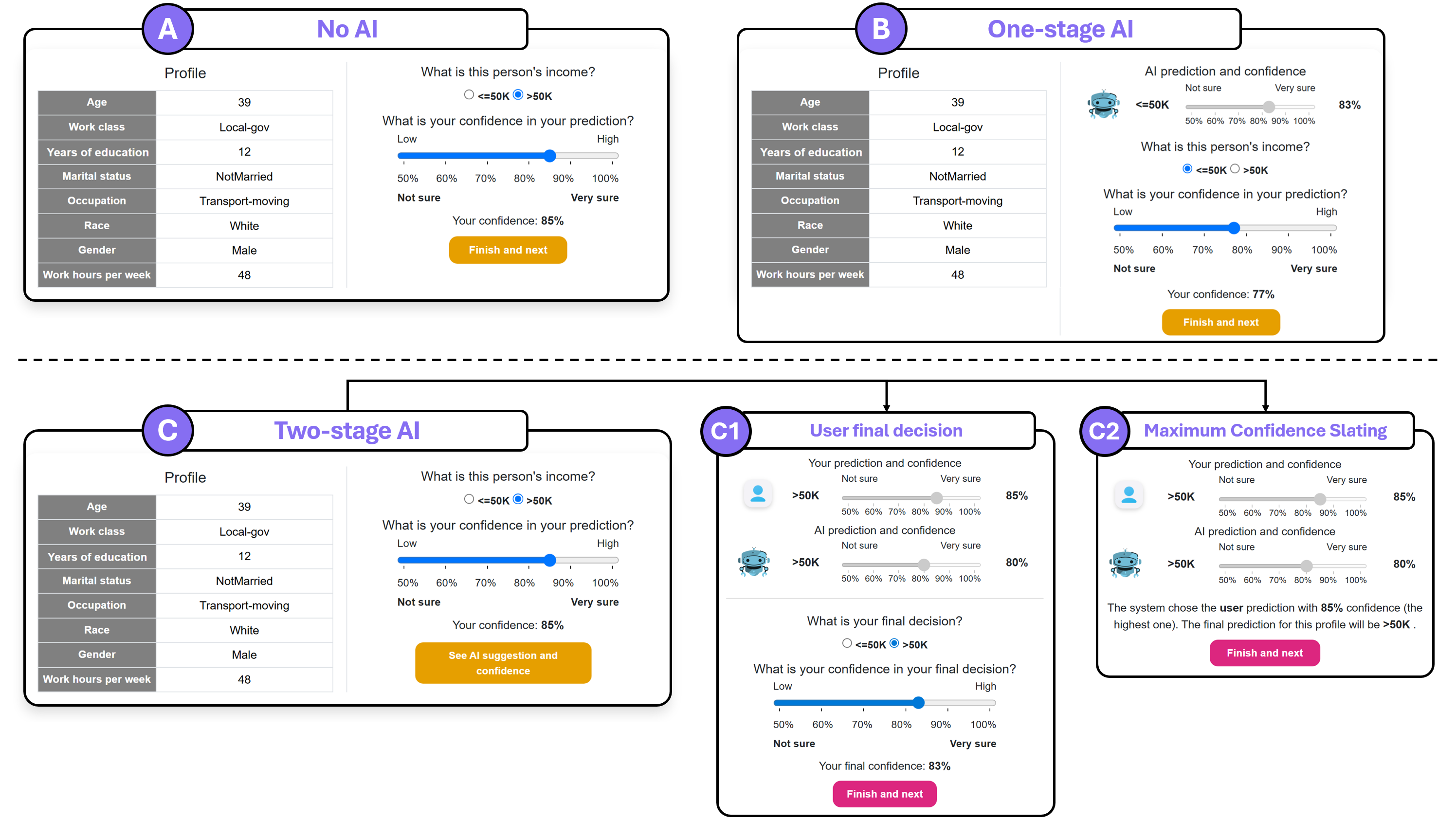}
    \caption{ Different interface conditions for our user studies involve providing users with eight attributes of a profile, and they must decide whether the income is above or below \$50K, while also indicating their confidence in their decisions. Note that personalised AI may utilise all the interactions described below: No AI, One-stage AI, Two-stage AI, and Maximum Confidence Slating (MCS). (A) No AI: Users make all decisions without AI assistance. (B) One-stage AI: Users immediately see the AI suggestion prediction and confidence (available only in the personalized AI condition). (C+C1) Two-stage AI: Users initially decide autonomously about the profile, then see the AI suggestion with prediction and confidence, and make their second and final prediction. (C+C2) Maximum Confidence Slating: Users initially decide autonomously about the profile, then see the AI suggestion with a prediction and confidence, and the system chooses the decision with the highest confidence (available only in the personalized AI condition).
    }

    \Description{
     todo.
    }
 
    \label{fig:ai_paradigms}
\end{figure}

\subsection{Personalized AI Assistance Design}
\label{sec:study2_personalized_ai_design}

As previous work suggests adapting AI assistance based on people's self-confidence to improve human-AI collaboration \cite{Ma2023CorrectnessLikelihoodAIUsersIncome,Ma24UserConfidenceCalibration,Steyvers2024AIParadigms,Cau2025CuriosityTraits,li2025confidencealigns,Rieger2025DSSComplementarity}, we designed the personalized AI condition as a \enquote{facilitator system} \cite{Pescetelli2021HybridAIFacilitator}. This type of system is an AI that adapts to human cognitive strategies to correct biases, whether explicit or implicit. It models its human partner, anticipates strengths and weaknesses, and intervenes when necessary to improve team performance, while also supporting and reducing biases in human decision-makers. 
Specifically, we crafted the personalized AI assistance based on three elements resulting from the calibration phase: (i) users' self-confidence calibration status (i.e., Over-confident, Under-confident, or Well-calibrated), (ii) average confidence, and (iii) accuracy. 

Given the importance of humans' self-confidence for advice-seeking behaviour \cite{Pescetelli2021GroupDecisionMetacognition, Lu2021HumanRelianceAIUserConfidence,Vodrahalli2022HumanTrustAIUserConfidence,CHONG2022HumanConfidence,Pescetelli2022Benefitsofspontaneousconfidence,He2023DunningKruger,li2025confidencealigns}, after calibration, we can compare users’ average confidence and accuracy with the AI’s, to determine the most suitable human-AI interaction type (i.e., no AI, one-stage AI, two-stage AI, or Maximum Confidence Slating) based on AI confidence levels (low or high). Our primary goal is for the resulting AI interface to guide users into the Well-calibrated band (see Figure~\ref{fig:study1-well-calibrated-distr}),  based on their initial self-confidence calibration.
Although the results from Study 1 showed that most users (90.6\%) fall under the \textit{Over-confident} calibration status, we designed the personalized AI assistance to also account for under-represented groups, such as \textit{Under-confident} and \textit{Well-calibrated} users, to cover the entire spectrum of the users' self-confidence status.  
We now describe how we implemented the personalized AI condition, starting from the users' self-confidence calibration status as a macro category.

\paragraph{\textbf{Over-confident users}} This user category was the most prominent in Study 1 (90.6\%), which is in line with previous studies stating that, in general, users tend to be overconfident in their decisions \cite{Frankenberger1997decision,Razmdoost2015UnderconfidentOverconfident,Soll2022OverconfidenceAdvice,Binnendyk2024IndividualDifferencesOverconfidence,li2025confidencealigns,LiHaleMoore_2025OverconfidenceIndividualDifference}.

\begin{itemize}

    \item \textit{Less confident and less accurate than AI} (avg. conf < AI conf \& accuracy <= AI accuracy). In this scenario, users have lower average confidence and accuracy compared to the AI. Considering the common behaviour of overconfident users dismissing AI advice \cite{He2023DunningKruger,Cau2025CuriosityTraits,li2025confidencealigns,Soleimanof2025OverUnderConfidentAI}, we could attempt to address this by using the following approach: for \textbf{high AI confidence}, we can provide AI assistance through a one-stage paradigm, thereby leveraging anchoring bias \cite{Nourani2021AnchoringBias,Rastogi2022CognitiveBiasAIConfidence,Fogliato2022OneStageTwoStage,Ma2023CorrectnessLikelihoodAIUsersIncome,Nattapat2025CognitiveBiases,Romeo2025AutomationBiasReviewXAI} to encourage users to follow AI advice, given the increased likelihood of correctness and well-calibrated confidence by the AI. 
    Conversely, for \textbf{low AI confidence} instances, overconfident users are more likely to dismiss AI advice or, if advice is given, the alignment of their self-confidence with AI confidence might degrade their self-confidence calibration \cite{li2025confidencealigns}. 
    Therefore, we choose not to display AI support and allow users to make their own decisions in such cases, given their natural tendency towards higher self-confidence.

    \item \textit{More confident and less accurate than AI} (avg. conf >= AI conf \& accuracy <= AI accuracy). 
    In this situation, users have higher average confidence than AI but lower accuracy. Again, we can apply a one-stage AI paradigm to instances with \textbf{high AI confidence}, which will preserve alignment with the AI and may enhance their self-confidence calibration \cite{Fogliato2022OneStageTwoStage,li2025confidencealigns}. 
    Furthermore, since this category is more confident than the AI, we use a two-stage AI paradigm with \textbf{low AI confidence} instances trying to reduce users' overconfidence and improve their calibration \cite{Lu2024DoWeLearnFromEachOtherTwoStage,li2025confidencealigns,Soleimanof2025OverUnderConfidentAI}. Specifically, participants with high initial confidence tend to experience more cognitive dissonance when confronted with conflicting AI advice. To protect their self-esteem, they often ignore the AI's suggestions, leaving the dissonance unresolved, which in turn increases their psychological discomfort and ultimately leads to a significant decrease in confidence \cite{Soleimanof2025OverUnderConfidentAI}. 

    \item \textit{More confident and more accurate than AI} (avg. conf >= AI conf \& accuracy > AI accuracy). 
    In this scenario, users have higher average confidence and accuracy compared to AI. A possible solution could be to adopt an on-demand AI paradigm, allowing users to access AI assistance only when they request it. However, overconfident users tend to dismiss AI advice altogether, and without enforced intervention like in the one- or two-stage AI paradigms, there is a high risk that users will bypass the assistance. Additionally, providing AI assistance might negatively impact users' accuracy. For these reasons, we choose not to offer any form of AI assistance to this user group. 
\end{itemize}

\paragraph{\textbf{Under-confident users}} This user category was the least prominent in Study 1 (0.8\%), which aligns with previous studies indicating that, in general, users tend to be over-confident rather than under-confident.
    
    \begin{itemize}

    \item \textit{Less confident and less accurate than AI} (avg. conf < AI conf \& accuracy <= AI accuracy). 
    In this scenario, users exhibit lower average confidence and accuracy compared to the AI. Since underconfident users tend to follow AI advice, we could adjust their self-calibration accordingly. For cases with \textbf{high AI confidence} instances, we can provide assistance through a one-stage AI approach, thereby leveraging anchoring bias and helping users attain high confidence based on the likelihood of correctness and the well-calibrated confidence of the AI. Conversely, for instances with \textbf{low AI confidence}, we could adopt a two-stage AI paradigm to prevent sudden confidence spikes (i.e., turning them into Over-confident users) and maintain proper self-confidence calibration  \cite{li2025confidencealigns,Soleimanof2025OverUnderConfidentAI}. This is because participants with low initial confidence might experience less cognitive dissonance when receiving contradictory advice, as it aligns with their self-doubt and validates their uncertainty. By adjusting their predictions, they reduce dissonance and protect their confidence \cite{Soleimanof2025OverUnderConfidentAI}.

    \item \textit{Less confident and more accurate than AI} (avg. conf < AI conf \& accuracy > AI accuracy).
    In this situation, users have less average confidence than AI but higher accuracy. As we did in the previous condition with \textbf{high AI confidence} instances, we can assist users through a one-stage AI paradigm to boost their confidence while maintaining a higher likelihood of correctness. Instead, for instances with \textbf{low AI confidence}, since underconfident users are more likely to follow AI advice, they could potentially lower their confidence even further. Therefore, we choose not to display AI support and allow users to make their own decisions in such cases, given their natural tendency toward lower self-confidence. 
  
    \item \textit{More confident and more accurate than AI} (avg. conf >= AI conf \& accuracy > AI accuracy). 
    In this scenario, users have higher average confidence and accuracy compared to AI. Providing AI assistance could induce this user category to align with AI confidence, so it may be beneficial to avoid offering any AI assistance to prevent worsening their self-confidence calibration \cite{li2025confidencealigns}.

    \end{itemize}

\paragraph{\textbf{Well-calibrated users}} 
This user category was the second most prominent in Study 1 (8.6\%, but still a large minority compared to Over-confident users), representing those who maintain a good balance between their average confidence and corresponding accuracy based on the predefined threshold $\delta$ = 5 in Study 1. Our goal is to guide users within the well-calibrated band toward its upper-right region, which reflects the most desirable balance of accuracy and self-confidence. Previous research on dyadic interactions \cite{Bahrami20102heads,Koriat2012MCS,Bang2014Doesinteractionmatter,Bang2017ConfidenceGroupDecisionMaking,Pescetelli2022Benefitsofspontaneousconfidence,Nguyen2025DiadDecisions,li2025confidencealigns} that studied confidence calibration and alignment highlights that a dyad's team accuracy is positively related to the self-calibration of each member: the better calibrated their confidence distributions are, the higher the dyad's performance. However, when team members significantly differ in capabilities (i.e., confidence calibration), it results in negative effects on both accuracy and confidence calibration in team decisions. 
Consequently, the Maximum Confidence Slating (MCS) \cite{Bahrami20102heads,Koriat2012MCS,Bang2014Doesinteractionmatter,Bang2017ConfidenceGroupDecisionMaking,Nguyen2025DiadDecisions,li2025confidencealigns} approach seems the most suitable interaction paradigm to improve human-AI team decisions in this scenario, given that user and AI are well-calibrated.\footnote{In case of a self-confidence tie in the human-AI dyad, the final answer will be picked at random, as done in previous work \cite{Bang2014Doesinteractionmatter,li2025confidencealigns}.}

\begin{itemize}

\item \textit{Less confident than AI} (avg. conf < AI conf - $\delta$). In this scenario, users have lower average confidence and accuracy compared to the AI. Given that the human and AI are well-calibrated, we will use the MCS paradigm for both \textbf{low-} and \textbf{high-confidence} instances. 

\item \textit{Aligned with AI confidence} (avg. conf >= AI conf - $\delta$ \& avg. conf <= AI conf + $\delta$). 
In this situation, users have similar average confidence and accuracy compared to the AI, considering our $\delta$ = 5 threshold. As before, given the stronger closeness in confidence calibration of the human-AI dyad, we will use the MCS paradigm for both \textbf{low-} and \textbf{high-confidence} instances.  

\item \textit{More confident than AI} (avg. conf > AI conf). 
In this scenario, users demonstrate higher overall confidence and accuracy than AI, given our $\delta$ = 5 threshold. In our view, this is the only circumstance where the MCS paradigm might potentially degrade human-AI team calibration, given that the AI assistance has worse calibration than the user \cite{Bang2017ConfidenceGroupDecisionMaking,Pescetelli2022Benefitsofspontaneousconfidence,li2025confidencealigns} and loses its role as a \enquote{facilitator} system. Therefore, we decided not to provide any AI assistance in this situation. 

\end{itemize}

\subsection{Study Conditions}
\label{sec:study2_conditions}

To answer our research questions, we conducted a between-subjects user study to assess the effects of AI assistance (no AI, two-stage AI, and personalized AI) and two psychological constructs (NFC and AOT) on humans' accuracy, appropriateness of self-confidence, as well as global and affective metacognition. Hereby, we list the \textit{independent variables} collected:

\begin{itemize}
\item \textbf{AI assistance (between-subjects, categorical).} Following Ma et al.'s work as a basis for no AI and two-stage AI \cite{Ma24UserConfidenceCalibration}, participants saw one of the following conditions in the main task phase:

\begin{itemize}
  \item \textbf{No AI}: In this condition, we did not present participants with AI support (Fig. \ref{fig:ai_paradigms}-A).
  
  \item \textbf{Two-stage AI}: Participants were presented with a two-stage AI assistance in this condition (Fig. \ref{fig:ai_paradigms}-C+C1).

  \item \textbf{Personalized AI}: Participants were provided a personalized AI assistance in this condition, following the premises in Section \ref{sec:study2_personalized_ai_design}.
  
\end{itemize}

\item \textbf{Need for Cognition (categorical).} 
Since individuals with high NFC may exhibit overconfidence in their decisions \cite{Vogt2022NonAbilityConfidence,Zerna2024NFCReview,Cau2025CuriosityTraits}, we investigated whether this tendency changes before and after a self-calibration session and examined its potential effects on accuracy, global metacognition, and affective metacognition.
We measured NFC using the NCS-6 six-item five-point scale
from \cite{LinsDeHolandaCoelho2020NFC6}, assigning participants to low or high levels by comparing their score to the distribution median (see \ref{sec:app_nfc_scale} for details).

\item \textbf{Actively Open-minded Thinking (categorical).} 
Individuals with high AOT already appear well calibrated in their self-confidence \cite{ Haran2013AOTinAccuracyAndCalibration, Martin2024CalibrationFeedbackAOT, Swaroop2025AIPersonalizedOverrelianceRate}, so we explored whether this calibration shifts before and after a self-confidence calibration session and assessed its impacts on accuracy, global metacognition, and affective metacognition.
We measured AOT using the Actively Open-minded Thinking Scale by Haran et al. \cite{Haran2013AOTinAccuracyAndCalibration}, assigning participants to low or high levels by comparing their score to the distribution median (see \ref{sec:app_aot_scale} for details).

\end{itemize}

For a more detailed analysis of our results, we also gathered the following measures:

\begin{itemize}
\item \textit{Timepoint (within-subjects, categorical).} 
We designed this within-subjects independent variable as a marker for specific moments in the study when we calculated accuracy and ECE during the calibration phase and the main task. 
Additionally, this variable also indicates where we asked participants to express their global and affective metacognition, providing a better view of how these states evolved throughout the study. 
Given the subjective nature of metacognitive perceptions, we decided to also measure people's global and affective metacognition before the self-confidence calibration phase, which will serve as a comparison value for subsequent metacognitive measurements after the main task.
Namely, we asked participants to state their global and affective metacognition twice: (i) \textit{before} the calibration phase, and (ii) \textit{after} the main task.\footnote{Considering the two-stage and personalized AI conditions, we aim to measure people's positive and negative affect experienced before calibration and after interacting with the AI, using the no AI condition as a baseline. Prior research found that distrust in AI is linked to more negatively charged emotions, while trust leads to more positive emotional reactions \cite{luhmann1979trust,Scharowski2025TrustDistrustPANAS}.}

\item \textit{AI literacy (continuous).}  
As people have distinct motivations to engage with AI assistance \cite{Long2020AiLiteracy,ChunWeiMing2021AIliteracy,LEICHTMANN2023AILiteracy,FOROUDI2025AISensationAIliteracy,Yurrita2025AIliteracy}, we measured AI literacy by computing the average score of the four items
defined in the Self-Assessed AI Literacy (AILIT) \cite{Schoeffer2022AiLiteracy}, in a 5-point Likert scale (1: Strongly
disagree; 5: Strongly agree).\footnote{Self-Assessed AI Literacy (AILIT) full questions: \url{https://github.com/jakobschoeffer/facct22-130-appendix/blob/main/facct22-130-appendix.pdf}} 

\end{itemize}

\subsection{Evaluation Metrics}
\label{sec:evaluation_metrics_study2}

To examine how AI assistance and individual differences in Need for Cognition (NFC) and Actively Open-minded Thinking (AOT) influence accuracy, calibration of self-confidence, and both global and affective metacognition, we measured the following \textit{dependent variables}:

\begin{itemize}
    \item \textbf{Accuracy (numerical).}
    Percentage of participants' correct decisions in the 20 main task instances.   
    
    \item \textbf{Appropriateness of human self-confidence (continuous).}
    We measured the appropriateness of human self-confidence in the main task using the Expected Calibration Error (ECE) \cite{Ma24UserConfidenceCalibration,li2025confidencealigns} as in Study 1 (see Sec. \ref{sec:evaluation_metrics_study1}). 

    \item \textbf{Global metacognition (numerical).} 
    We asked participants to state how accurate they would think they were in the income prediction task on a 0-100\% scale \cite{KLEITMAN2007SelfConfidenceMetacognitivePRocesses,Rouault2019GlobalEstimatesFromLocalSelfPerformance,Fleming2024_metacognition,KATYAL2024_metacognition,Katyal2025AnxietyDepressionLocalGlobalMetacognition}. Specifically, we measured global metacognition before the calibration phase (prospective) and after the main task (retrospective) asking participants the following questions: \textit{\enquote{How much accuracy (in percentage \%) do you think you will achieve in the upcoming income prediction task of 20 trials?}}, and: \textit{\enquote{How much accuracy (in percentage \%) do you think you achieved in the main task?}}.
    
    \item \textbf{Affective metacognition (continuous).}
    To our knowledge, there are no established affective metacognition measures in the literature \cite{Smith2016MemoryAcuteStress,Tankelevitch&Kewenig2024MetacognitionGenAI,KATYAL2024_metacognition,Mladen2024AffectiveAndMotivationalStatesMetacognition}. 
    For this reason, we adapted an existing metric to capture a person’s predominant positive or negative emotions and their confidence in those emotions at a specific time. 
    Namely, we used the 10-item International Positive and Negative Affect Schedule Short Form (I-PANAS-SF) \cite{WatsonClark1988PanasFULL,Thompson2007ShortPANAS}, which consists of five positive (Determined, Attentive, Alert, Inspired, and Active) and negative  (Afraid, Nervous, Upset, Ashamed, and Hostile) affects. 
    While the original scale assesses emotions over a general timeframe  (i.e.,  \enquote{Thinking about yourself and how you normally feel, to what extent do you generally feel.}) \cite{Thompson2007ShortPANAS}, we adapted it to focus on the present moment by asking participants how they felt \enquote{right now} along with the magnitude of the perceived affect.  
    In the user study, participants completed 20 items in total, including the intensity
    (affective part: \textit{\enquote{How <emotion> do you feel at the moment?}}; \enquote{1 - Very slightly or not at all} to \enquote{5 - Extremely}) and confidence (metacognitive part: \textit{\enquote{How confident are you that this rating reflects your true feelings?}}; \enquote{1 - Not very confident} to \enquote{5 - Extremely confident}) of their emotional states. 

    We named this metric the \textit{Affective Metacognitive Index (AMI)}. First, we computed the total weighted scores for positive and negative affects by summing, over each item \(i\), the product of its affective intensity \(a_i\) and confidence rating 
    \(c_i\)\footnote{The confidence rating \(c_i\) represents participants' metacognitive confidence in their affective judgments \cite{KATYAL2024_metacognition,Mladen2024AffectiveAndMotivationalStatesMetacognition,Tankelevitch&Kewenig2024MetacognitionGenAI}.}. 
    We then normalized these sums by their respective maximum values \(P_{\max}\) and \(N_{\max}\), to map them into \([0,1]\).\footnote{Since each of the five positive and negative affects consists of two questions with a five-point scale, the values of \(P_{\max}\) and \(N_{\max}\) will be equal to 125.} Finally, the AMI is defined as the difference between the normalized positive and normalized negative scores, yielding a value in \([-1,1]\). In Eq. \ref{eq:ami}, \(\mathcal{P}\) and \(\mathcal{N}\) denote the sets of positively and negatively valenced items, \(P_{\max}\) and \(N_{\max}\) are the theoretical maximum sums of \(a_i c_i\) over \(\mathcal{P}\) and \(\mathcal{N}\), respectively, which ensures that \(\mathrm{AMI}\) lies between \(-1\) (predominance of negative meta-affect) and \(+1\) (predominance of positive meta-affect). 
    
            \begin{equation}
            \mathrm{AMI}
            =
            \frac{1}{P_{\max}}
            \sum_{i\in\mathcal{P}} a_i\,c_i
            \;-\;
            \frac{1}{N_{\max}}
            \sum_{i\in\mathcal{N}} a_i\,c_i
            \label{eq:ami}
            \end{equation}

\end{itemize}

Furthermore, we collected exploratory measurements not present in the research questions to get a more nuanced understanding of participants' behavior:

\begin{itemize}
   
    \item \textit{Reliance (continuous).} To complement the accuracy and appropriateness of human self-confidence, we also explored how users' reliance varied considering AI assistance conditions \cite{He2023AnalogyExplanations,Ma24UserConfidenceCalibration,li2025confidencealigns}. Specifically, when we provided AI assistance using the two-stage paradigm, we assessed the following measures: \textit{Agreement fraction}, \textit{Switch fraction}, \textit{Follow high confidence fraction}, \textit{Accuracy with initial disagreement (Accuracy-wid)}, \textit{Over-reliance}, and \textit{Under-reliance}. 
    However, since the personalized AI condition includes multiple assistance types (including no AI), we computed only those metrics relevant to one- and two-stage paradigms: \textit{Agreement fraction}, \textit{Over-reliance}, and \textit{Under-reliance}. 
    
    \begin{align*}
    \text{Agreement fraction} &= \frac{\text{Number of final decisions same as the AI suggestion}}{\text{Total number of decisions}},\\
    \text{Switch fraction}    &= \frac{\text{Number of decisions the user switched to agree with AI suggestion}}{\text{Total number of decisions with initial disagreement}},\\
    \text{Follow high confidence fraction} &= \frac{\text{Number of tasks where user followed the prediction with higher confidence}}{\text{Total number of decisions with initial disagreement}},\\
    \text{Accuracy-wid}       &= \frac{\text{Number of correct final decisions with initial disagreement}}{\text{Total number of decisions with initial disagreement}},\\
    \text{Over-reliance}               &= \frac{\text{Number of incorrect human final decisions with incorrect AI advice}}{\text{Total number of incorrect AI advice}},\\
    \text{Under-reliance}                &= \frac{\text{Number of incorrect human final decisions with correct AI advice}}{\text{Total number of correct AI advice}}
    \end{align*}

\item \textit{User experience (continuous).}
To measure user perceptions as a post-test following prior work \cite{Ma2023CorrectnessLikelihoodAIUsersIncome,Ma24UserConfidenceCalibration,He2025ConversationalXAIOnDemand}, we asked participants to respond about the following qualitative assessments on a seven-point  scale (1: Strongly disagree; 7: Strongly agree): helpfulness of the calibration phase, perceived appropriateness of self-confidence, mental demand, perceived complexity, preference, satisfaction, AI trust and autonomy (see Appendix \ref{sec:user_exp_app} for details).

\end{itemize}

\subsection{Sample Size and Statistical Analysis}
\label{sec:sample_size_stat_study2}
Before recruiting participants, we computed the required sample size considering the scope of our research questions using
\textit{G*Power} \cite{faul2009statistical} and considering a medium effect size (Cohen's $f$ = 0.25), a desired power of (1 - $\beta$) = 0.8, and a significant threshold $\alpha=.05$. 
To investigate RQ2 and RQ3, we conducted four between-subjects ANOVA tests with \textit{accuracy}, \textit{ECE}, \textit{global} and \textit{affective metacognition} as dependent variables, studying the main effects of AI assistance (no AI, two-stage AI, and personalized AI), NFC (low or high), and AOT (low or high) as independent variables, resulting in a required sample size of 158 participants (Df = 2).
We corrected p-values using Bonferroni's adjustment, considering four ANOVA tests ($m$ = 4). 
For post-hoc pairwise comparisons, we used Tukey HSD for contrast tests and p-value adjustments. If normality and homogeneity of variances assumptions were violated, we conducted a fully non‐parametric permutation ANOVA ($p$ = 9999) \cite{Frossard2021PermTests} to test main effects (see Section \ref{sec:study2_posthoc}).

\paragraph{Post hoc analysis}
As already mentioned in Sections \ref{sec:study2_conditions} and \ref{sec:evaluation_metrics_study2}, we extended our analysis of research questions by examining participants' behavioral differences during the calibration phase and main task, reliance on AI, subjective perceptions, and the effect of AI literacy. 
For each analysis, we corrected p-values using Bonferroni's adjustment based on the number of tests (i.e., dependent variables).
To further clarify the \textit{calibration phase} outcomes, we conducted ANOVA tests on accuracy, ECE, global, and affective metacognition dependent variables using the same settings as per RQ2 and RQ3. 
Additionally, to investigate \textit{differences} between the \textit{calibration phase} and the \textit{main task} on the same dependent variables, we introduced a within-subjects timepoint independent variable to conduct Linear Mixed Model (LMM) analyses, for which we approximated the required sample size using mixed-design ANOVA in \textit{G*Power} using AI assistance, NFC, AOT, and timepoint as independent variables.
Specifically, we set a medium effect size (Cohen's $f$ = 0.25), a desired power of (1 - $\beta$) = 0.8, a significant threshold $\alpha=.05$, three groups, and two repeated measurements (calibration phase and main task).\footnote{This resulted in a sample size of 42 participants, which aligns with the previously estimated total sample size of 158 participants for the main analysis, ensuring adequate power for all tests. We complemented the linear mixed model (LMM) analyses by also reporting Type III ANOVA results with Satterthwaite's approximation for degrees of freedom.} 
To assess participants' \textit{reliance on AI}, we conducted ANOVA tests with NFC, AOT, and AI assistance as independent variables, isolating two-stage and personalized AI conditions to measure \textit{agreement on AI}, \textit{over-reliance}, and \textit{under-reliance fractions} as dependent variables. Furthermore, given that recent work \cite{li2025confidencealigns} showed that participants' under AI assistance might align their confidence to that of the AI, we also conducted (i) Pearson's correlations considering ECE and agreement on AI, over-reliance, and under-reliance for both two-stage and personalized AI conditions, and (ii) paired t-test for ECE variations before and after seeing the AI advice only for the two-stage AI condition.
Additionally, we employed Kruskal-Wallis tests to compare participants' \textit{subjective experience} measures across the three levels of AI assistance, while we used Wilcoxon rank-sum tests for the binary categorical variables NFC and AOT.
Finally, we examined the impact of \textit{AI literacy} as a continuous covariate on our dependent variables using ANCOVA tests, building upon our previous RQ2 and RQ3 ANOVA model settings.

\begin{figure} [!t]
      \centering
    \includegraphics[width=\textwidth]{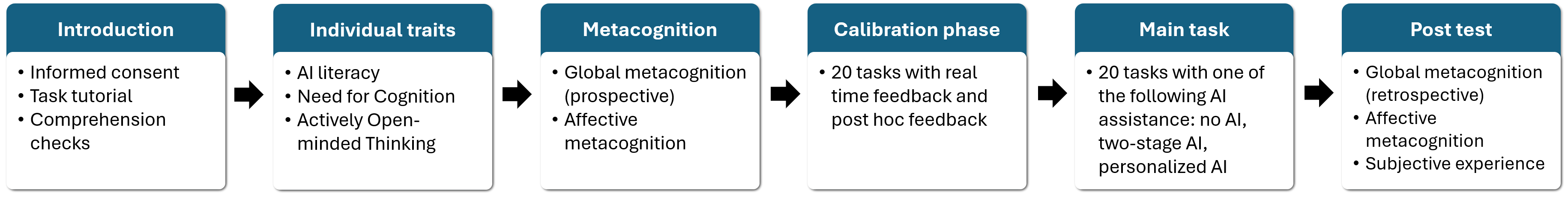}
    \caption{Illustration of the procedure participants engaged in during Study 2.}

    \Description{
     todo.
    }
 
    \label{fig:study2_procedure}
\end{figure}

\subsection{Participants and Procedure}
\label{sec:participants_procedure_study2}

After the user study was approved by the University of Cagliari Ethics Committee\footref{ethics_approval}, we recruited participants from Prolific using the same selection criteria in Study 1 (see Section \ref{sec:participants_procedure_study1}). 
We rewarded participants with \pounds 3.6 for completing the study, considering an average completion time of 24 minutes, paying an average of \pounds 9 per hour. We rewarded an extra \pounds 2 to participants who achieved an accuracy above 85\% in the main task to incentivise high-quality performance. 
The user study procedure is illustrated in Figure \ref{fig:study2_procedure}, and we detail the steps as follows:

\begin{enumerate}
    \item \textbf{Tutorial}: Upon accepting informed consent, participants underwent a familiarization tutorial that explained how to use the task interface for the calibration and main phases. The tutorial described each of the eight profile attributes and showed the binary income distribution for each attribute. Participants were then asked to respond to some questions about the tutorial, and only those who provided correct answers qualified to proceed to the next phase.

    \item \textbf{Individual traits and metacognition}: Participants then filled out the AI literacy, NFC, and AOT questionnaires, and expressed their global (GM) and affective (AM) metacognition before starting with the calibration phase.

    \item \textbf{Calibration phase}: 
    Participants engaged in a calibration session (see Sec. \ref{sec:calibration_mechanism_study1}), where they completed 20 tasks stating their prediction and confidence, receiving feedback after each decision about the correctness and their self-confidence status, categorised as well-calibrated, overconfident, or underconfident. 
    A historical self-confidence status bar was shown for all the previous choices, color-coded based on the self-confidence status and enumerated based on the chronological order of the tasks (see Fig. \ref{fig:calib_phase}-A2). 
    Afterwards, participants were shown an overview of their calibration status,
    summarising the proportion of self-calibration status for well-calibrated, overconfident, and underconfident instances, plus their average confidence and actual accuracy in a bar chart (see Fig. \ref{fig:calib_phase}-B1). The bottom part of the interface also displayed their past predictions, divided into five bins based on confidence distribution, along with a chart showing their relationship of average confidence and actual accuracy vs an ideal self-confidence and accuracy (see Fig. \ref{fig:calib_phase}-B2).

    \item \textbf{Main task}: Participants were then assigned to one of the three conditions in a balanced fashion (i.e., no AI, two-stage AI, and personalized AI; see Section \ref{sec:study2_conditions}), where they completed another 20 tasks by stating their prediction and confidence.

    \item \textbf{Post test}: Finally, participants expressed their global (GM) and affective (AM) once more. Then, we asked participants to rate their subjective experience with the main task interface (see Appendix \ref{sec:user_exp_app} for details).

\end{enumerate}

\subsection{Results}
\label{sec:results_study2}
In this section, we present the results of our study on research questions RQ2 and RQ3, considering the dependent variables collected in the main task. We discuss descriptive statistics, assumptions for ANOVA tests, the findings related to research questions, and exploratory results concerning calibration and main phase differences, reliance behaviors, as well as participants' subjective perceptions and the effects of AI literacy.

\begin{table}[t]
\centering
\footnotesize
\setlength{\tabcolsep}{6.8pt} 
\begin{tabularx}{\textwidth}{
>{\raggedright\arraybackslash}X
>{\centering\arraybackslash}p{0.7cm} >{\centering\arraybackslash}p{0.7cm} 
>{\centering\arraybackslash}p{0.7cm} >{\centering\arraybackslash}p{0.7cm} 
>{\centering\arraybackslash}p{0.7cm} >{\centering\arraybackslash}p{0.7cm} >{\centering\arraybackslash}p{0.7cm} 
>{\centering\arraybackslash}p{0.7cm} >{\centering\arraybackslash}p{0.7cm} >{\centering\arraybackslash}p{0.7cm} 
>{\centering\arraybackslash}p{0.7cm} 
}
\toprule
Condition
& \multicolumn{2}{c}{Accuracy (\%)} 
& \multicolumn{2}{c}{ECE} 
& \multicolumn{3}{c}{Calibration self-confidence} 
& \multicolumn{3}{c}{Main self-confidence} 
& AILIT
\\
\cmidrule(lr){2-3}\cmidrule(lr){4-5}\cmidrule(lr){6-8}\cmidrule(lr){9-11}\cmidrule(lr){12-12}
& Calib & Main & Calib & Main & Over & Under & Well & Over & Under & Well & M
\\
\midrule
No AI
& 62.3 & 60.4
& 0.19 & 0.19
& 42 & 1 & 10
& 44 & 2 & 7
& 3.71 
\\
Two-stage AI
& 62.9 & 70.3
& 0.19 & 0.09
& 49 & 0 & 4
& 25 & 4 & 24
& 3.84 
\\
Personalized AI
& 60.7 & 67.1
& 0.22 & 0.13
& 49 & 0 & 4
& 34 & 2 & 17
& 3.92 
\\
\bottomrule
\end{tabularx}
\caption{Summary of participants' average accuracy, ECE, self-confidence labeling, and AI literacy across AI assistance conditions for the calibration phase and main task. }
\label{tab:study2_summary_ai_assistance}
\end{table}

\begin{table}[!t]
\centering
\footnotesize
\begin{tabular}{p{4.5cm} C{2.7cm} C{2.7cm} C{2.7cm}}
\toprule
\textbf{Calibration label} & \textbf{N. users} & \textbf{AI confidence} & \textbf{AI paradigm} \\
\midrule
\multirow{2}{*}{Over-confident; more confident than AI} & \multirow{2}{*}{46} & High & one-stage AI \\
& & Low  & two-stage AI \\
\midrule
\multirow{2}{*}{Over-confident; less confident than AI} & \multirow{2}{*}{3}  & High & one-stage AI \\
& & Low  & no AI \\
\midrule
\multirow{2}{*}{Well-calibrated; around AI confidence} & \multirow{2}{*}{2} & High & \multirow{2}{*}{MCS} \\
& & Low  & \\
\midrule
\multirow{2}{*}{Well-calibrated; less confident than AI} & \multirow{2}{*}{1} & High & \multirow{2}{*}{MCS} \\
& & Low  & \\
\midrule
\multirow{2}{*}{Well-calibrated; better than AI} & \multirow{2}{*}{1} & High & \multirow{2}{*}{no AI} \\
& & Low  &  \\
\bottomrule
\end{tabular}
\caption{Personalized AI assistance paradigms resulting from participants' self-confidence labeling on the calibration phase.}
\label{tab:study2_personalized_AI}
\end{table}

\subsubsection{Descriptive statistics}
Of the 159 participants gathered for Study 2, 76 were female and 83 were male, with an average age of $M$ = 42.36 and $SD$ = 12.13. 
The NFC subdivision into low (74) and high (85) groups was based on a median $Mdn$ = 3.83 ($M$ = 3.66 and $SD$ = 0.87) and for individuals with AOT, the low (74) and high (85) groups had a median $Mdn$ = 5.43 ($M$ = 5.43 and $SD$ = 0.91). 
Across AI conditions, participants achieved an average accuracy of $M$ = 61.95 ($SD$ = 10.63) and $M$ = 0.24 ($SD$ = 0.10) ECE during the calibration phase. Instead, they achieve an average accuracy of $M$ = 65.91 ($SD$ = 10.48) and $M$ = 0.19 ($SD$ = 0.10) ECE in the main phase. Further, participants reported an average AI literacy level of ($M$ = 3.83 and $SD$ = 0.62). Table \ref{tab:study2_summary_ai_assistance} reports participants' average accuracy, ECE, self-confidence distributions, and AI literacy divided by AI assistance condition and considering calibration phase and main task. Table \ref{tab:study2_personalized_AI} presents the various types of AI paradigms employed in personalised AI assistance, based on the calibration phase results. Instead, Figure \ref{fig:study2_conf_calibration} in the Appendix illustrates the distribution of self-confidence across AI assistance, considering both the calibration phase and the main task.

\subsubsection{Normality and Data Homogeneity Assumptions for ANOVA Tests.}
Shapiro-Wilk tests relative to the main task showed that both accuracy 
\(\bigl(W = 0.926,\ p < .0001\bigr)\) and ECE 
\(\bigl(W = 0.925,\ p < .0001\bigr)\) distributions 
significantly deviated from normality. 
Instead, the Levene's test assumption was satisfied for both accuracy ($F$ = 0.7, $p$ = .74) and ECE ($F$ = 0.94, $p$ = .51).
Similarly, Shapiro-Wilk tests indicated that both global 
\(\bigl(W = 0.958,\ p < .0001\bigr)\) and affective 
\(\bigl(W = 0.978,\ p = .011\bigr)\) metacognition distributions 
significantly deviated from normality. Levene's test assumption was satisfied for both global ($F$ = 1.69, $p$ = .081) and affective ($F$ = 0.97, $p$ = .47) metacognition. 
Given that the other assumptions concerning random sampling and independence were met, we conducted ANOVA tests to assess the main effects of AI assistance, NFC, and AOT factors.

\begin{figure} [!t]
      \centering
    \includegraphics[width=\textwidth]{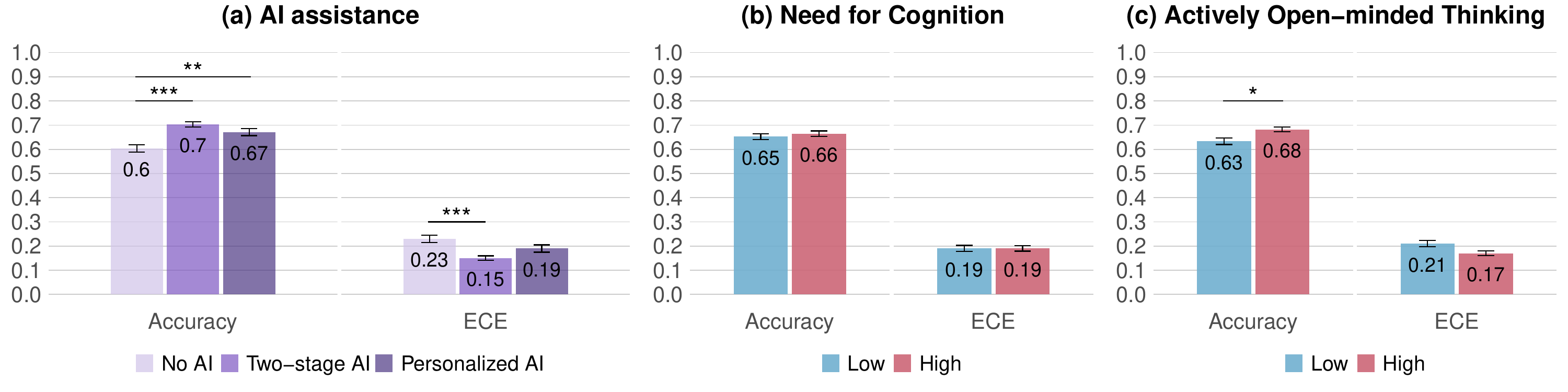}
    \caption{Participants' accuracy and appropriateness of their self-confidence (ECE) in the main task considering: (a) AI assistance, (b) Need for Cognition, and (c) Actively Open-minded Thinking. The asterisks highlight p-value significance strength (*$p$ < .05; **$p$ < .01; ***$p$ < .001).}

    \Description{
     todo.
    }
 
    \label{fig:study2_accuracy_ece}
\end{figure}

\begin{table}[!t]
  \centering
  \footnotesize
  \begin{tabular}{
p{2cm}                           
>{\centering\arraybackslash}p{1.4cm}
>{\centering\arraybackslash}p{1.7cm} 
>{\centering\arraybackslash}p{1.7cm} 
>{\centering\arraybackslash}p{1.7cm} 
>{\centering\arraybackslash}p{1.7cm} 
>{\centering\arraybackslash}p{1.7cm} 
  }
    \toprule
    \multicolumn{7}{c}{\textbf{Accuracy model (ANOVA)}}\\
    Factor & Df & Sum Sq & Mean Sq & F value & $p_{\textnormal{raw}}$ & $p_{\textnormal{adj}}$ \\ 
    \midrule
    AI assistance        & 2  & 2708  & 1354  & 14.86     & $< .0001$ & \textbf{< .0001}     \\
    NFC category           & 1  & 7     & 7     & 0.08      & .784     & $> 1$    \\
    AOT category           & 1  & 592   & 592   & 6.50      & .012     & \textbf{.047}     \\
    \addlinespace
    \cmidrule(lr){1-7}
    \addlinespace
    \multicolumn{7}{c}{\textbf{ECE model (ANOVA)}}\\
    Factor & Df & Sum Sq & Mean Sq & F value & $p_{\textnormal{raw}}$ & $p_{\textnormal{adj}}$ \\
    \midrule
    AI assistance        & 2  & 0.164 & 0.0819 & 8.51     & .0003   & \textbf{.0012}      \\
    NFC category          & 1  & 0.000 & 0.0000 & 0.00     & .974     & $> 1$    \\
    AOT category           & 1  & 0.038 & 0.0381 & 3.96     & .048     & .193     \\
    \bottomrule
  \end{tabular}
  \caption{ANOVA results for AI assistance, Need for Cognition (NFC), and Actively Open-minded Thinking (AOT) on accuracy and ECE in the main task with Bonferroni's correction of $m = 4$.}
  \label{tab:study2_anovas_accuracy_ece}
\end{table}

\begin{table}[!t]
  \centering
  \footnotesize
  \begin{tabular}{
p{4.5cm}
>{\centering\arraybackslash}p{1.7cm}
>{\centering\arraybackslash}p{1.7cm}
>{\centering\arraybackslash}p{1.7cm}
>{\centering\arraybackslash}p{1.7cm}
  }
    \toprule
    \multicolumn{5}{c}{\textbf{Accuracy model (Tukey HSD comparisons)}}\\

    Constrast (AI assistance)                     & diff     & lwr      & upr       & $p_{\textnormal{adj}}$ \\ 
    \midrule
    Personalized AI – no AI            & 6.6981   & 2.3092   & 11.0871   & \textbf{.0012}         \\
    Two-stage AI – no AI            & 9.9057   & 5.5167   & 14.2946   & \textbf{< .0001}       \\
    Two-stage AI – personalized AI      & 3.2075   & –1.1814  & 7.5965    & .1975                  \\
    \addlinespace
    \cmidrule(lr){1-5}
    \addlinespace
    \multicolumn{5}{c}{\textbf{ECE (Tukey HSD comparisons)}} \\

    Constrast (AI assistance)                     & diff       & lwr        & upr        & $p_{\textnormal{adj}}$ \\ 
    \midrule
    Personalized AI – no AI            & –0.0398   & –0.0849   & 0.0053    & .0956                  \\
    Two-stage AI – no AI            & –0.0786   & –0.1237   & –0.0335   & \textbf{.0002}         \\
    Two-stage AI – personalized AI      & –0.0388   & –0.0839   & 0.0063    & .1066                  \\
    \bottomrule
  \end{tabular}
  \caption{Post-hoc pairwise comparisons (Tukey HSD with conf. levels = 95\%) for accuracy and ECE based on AI assistance.}
  \label{tab:study2_accuracy_ece_posthoc}
\end{table}

\subsubsection{Effect of AI assistance on accuracy, appropriateness of self-confidence, global and local metacognition (RQ2)}

For \textbf{RQ2.1}, the results from ANOVA tests (see Table \ref{tab:study2_anovas_accuracy_ece}) demonstrate that AI assistance has a significant effect on accuracy ($F = 14.86$, $p_{\textnormal{adj}} < .0001$, Cohen's $f$ = 0.44) and ECE ($F = 8.51$, $p_{\textnormal{adj}} = .0012$, Cohen's $f$ = 0.39). 
To identify differences across AI assistance conditions, we performed post hoc comparisons using Tukey HSD tests, presented in Table \ref{tab:study2_accuracy_ece_posthoc}.\footnote{Here, we report p-values as \enquote{\textit{p}}, which indicate the adjusted value with Tukey HSD.\label{posthoc_tukey}} 
As for accuracy, the results show that both two-stage AI ($Dif.$ = 9.906, $p$ < .0001) and personalized AI ($Dif.$ = 6.698, $p$ = .0012) conditions outperform the no AI condition. However, no significant difference was found between the personalized AI and two-stage AI conditions ($Dif.$ = 3.208, $p$ = .198).
For appropriateness of self-confidence (ECE), two-stage AI achieved better self-confidence than no AI ($Dif.$ = -0.0786, $p_{\textnormal{adj}}$ = .0002), whilst we found no differences between personalized AI and no AI ($Dif.$ = -0.0399, $p_{\textnormal{adj}}$ = .0956) nor two-stage AI ($Dif.$ = -0.0388, $p_{\textnormal{adj}}$ = .1066). Figure \ref{fig:study2_accuracy_ece}-a summarizes the results.

\begin{figure} [!t]
      \centering
    \includegraphics[width=\textwidth]{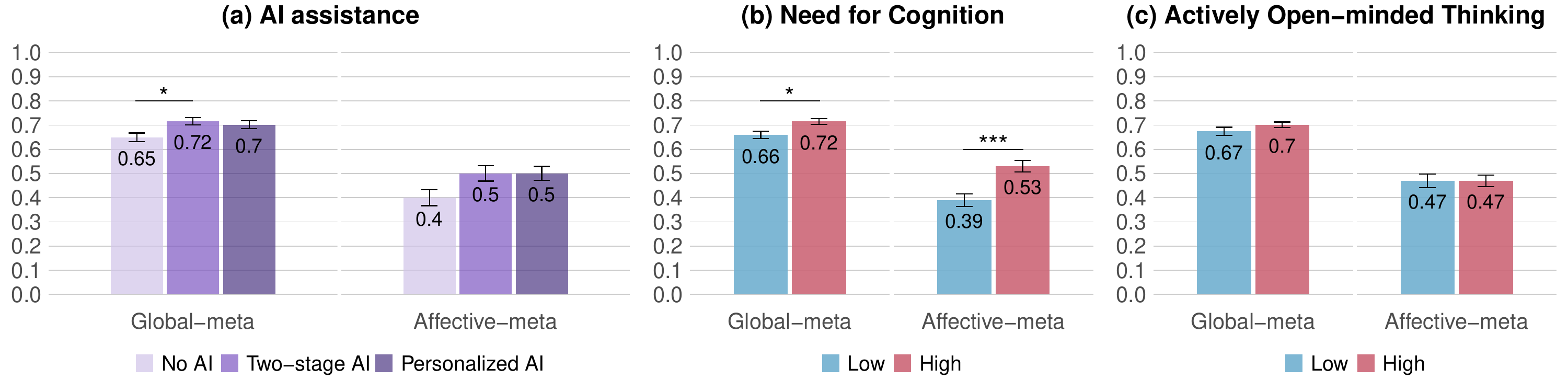}
  \caption{Participants' global and affective metacognition in the main task considering: (a) AI assistance, (b) Need for Cognition, and (c) Actively Open-minded Thinking. The asterisks highlight p-value significance strength (*$p$ < .05; **$p$ < .01; ***$p$ < .001). }

    \Description{
     todo.
    }
 
    \label{fig:study2_global_affective}
\end{figure}

For \textbf{RQ2.2}, ANOVA test results (see Table \ref{tab:study2_metacognition_anovas}) show that AI assistance has a significant effect on global metacognition ($F = 4.70$, $p_{\textnormal{adj}} = .042$, Cohen's $f$ = 0.25), but not on affective metacognition ($F = 3.30$, $p_{\textnormal{adj}} = .157$, Cohen's $f$ = 0.21). 
Post hoc comparisons using Tukey HSD tests (see Table \ref{tab:study2_posthoc_global_metacognition}) reveal that two-stage AI participants reported a greater global metacognition when compared to the no AI condition ($Dif.$ = 6.6415, $p$ = .0115). At the same time, no differences exist across personalized AI and no AI conditions ($Dif.$ = 5.2264, $p$ = .0602) or two-stage and personalized AI conditions ($Dif.$ = 1.415, $p$ = .8094). Figure \ref{fig:study2_global_affective}-a depicts a summarization of these findings.

\begin{table}[t]
  \centering
  \footnotesize
\begin{tabular}{
p{2cm}                           
>{\centering\arraybackslash}p{1.4cm}
>{\centering\arraybackslash}p{1.7cm} 
>{\centering\arraybackslash}p{1.7cm} 
>{\centering\arraybackslash}p{1.7cm} 
>{\centering\arraybackslash}p{1.7cm} 
>{\centering\arraybackslash}p{1.7cm} 
  }
    \toprule
    \multicolumn{7}{c}{\textbf{Global Metacognition model (ANOVA)}}\\
    Factor          & Df & Sum Sq & Mean Sq & F value & $p_{\textnormal{raw}}$ & $p_{\textnormal{adj}}$ \\ 
    \midrule
    AI assistance          & 2  & 1297    & 649     & 4.70     & .0105   & \ \textbf{.042} \\
    NFC category   & 1  & 1010    & 1010    & 7.32     & .0076   & \ \textbf{.03} \\
    AOT category    & 1  & 129     & 129     & 0.93     & .3357   & \ > 1 \\
    \addlinespace
    \cmidrule(lr){1-7}
    \addlinespace
    \multicolumn{7}{c}{\textbf{Affective Metacognition model (ANOVA)}}\\
    Factor          & Df & Sum Sq & Mean Sq & F value & $p_{\textnormal{raw}}$ & $p_{\textnormal{adj}}$ \\
    \midrule
    AI assistance          & 2  & 0.31    & 0.156   & 3.30     & .0393   & \ .157 \\
    NFC category    & 1  & 0.74    & 0.742   & 15.72    & .0001  & \ \textbf{.0004} \\
    AOT category    & 1  & 0.01    & 0.010   & 0.22     & .6426   & \ > 1 \\
    \bottomrule
  \end{tabular}
  \caption{ANOVA results for AI assistance, Need for Cognition (NFC), and Actively Open-minded Thinking (AOT) on global and affective metacognition in the main task with Bonferroni's correction of $m = 4$.}
  \label{tab:study2_metacognition_anovas}
\end{table}

\begin{table}[t]
  \centering
  \footnotesize
  \begin{tabular}{
p{4.5cm}
>{\centering\arraybackslash}p{1.3cm}
>{\centering\arraybackslash}p{1.3cm}
>{\centering\arraybackslash}p{1.3cm}
>{\centering\arraybackslash}p{1.3cm}
}
    \toprule
    \addlinespace
    \multicolumn{5}{c}{\textbf{Global Metacognition model (Tukey HSD comparisons)}}\\
    Contrast (AI assistance)                & diff     & lwr       & upr       & $p_{\mathrm{adj}}$ \\ 
    \midrule
    Personalized AI – no AI                 & 5.2264   & –0.1749   & 10.6277   & .0602              \\
    Two-stage AI – no AI                    & 6.6415   & 1.2402    & 12.0428   & \textbf{.0115}     \\
    Two-stage AI – personalized AI          & 1.4151   & –3.9862   & 6.8164    & .8094              \\
    \bottomrule
  \end{tabular}
  \caption{Post-hoc pairwise comparisons (Tukey HSD, 95\% CI) for Global Metacognition model based on AI assistance.}
  \label{tab:study2_posthoc_global_metacognition}
\end{table}

\subsubsection{Effect of low and high levels of NFC and AOT on accuracy, appropriateness of self-confidence, global and affective metacognition (RQ3)}

Considering \textbf{RQ3.1}, ANOVA tests presented in Table \ref{tab:study2_anovas_accuracy_ece} showed that high-AOT individuals achieved greater accuracy ($F = 6.50$, $p_{\textnormal{adj}} = .047$, Cohen's $f$ = 0.21) compared to low-AOT ones, although we found no differences considering ECE ($F = 3.96$, $p_{\textnormal{adj}} = .193$, Cohen's $f$ = 0.17). Also, there were no significant differences for low and high NFC individuals (accuracy: $F = 0.08$, $p_{\textnormal{adj}} > 1$, Cohen's $f$ = 0.02; ECE: $F = 0.00$, $p_{\textnormal{adj}} > 1$, Cohen's $f$ = 0.08). A summary of these findings is shown in Figure \ref{fig:study2_accuracy_ece}-b and \ref{fig:study2_accuracy_ece}-c.

For \textbf{RQ3.2}, ANOVA tests highlight that high-NFC individuals reported greater global ($F = 7.32$, $p_{\textnormal{adj}} = .03$, Cohen's $f$ = 0.22) and affective ($F = 15.72$, $p_{\textnormal{adj}} = .0004$, Cohen's $f$ = 0.32) metacognition compared to low-NFC ones. Instead, no significant differences for low or high AOT individuals (global metacognition: $F = 0.93$, $p_{\textnormal{adj}} > 1$, Cohen's $f$ = 0.08; affective metacognition: $F = 0.22$, $p_{\textnormal{adj}} > 1$, Cohen's $f$ = 0.04). Figure \ref{fig:study2_global_affective}-b and \ref{fig:study2_global_affective}-c outline these results.

\begin{table}[!t]
  \centering
  \footnotesize
  \begin{tabular}{
p{2cm}                           
>{\centering\arraybackslash}p{1.4cm}
>{\centering\arraybackslash}p{1.7cm} 
>{\centering\arraybackslash}p{1.7cm} 
>{\centering\arraybackslash}p{1.7cm} 
>{\centering\arraybackslash}p{1.7cm} 
>{\centering\arraybackslash}p{1.7cm} 
  }
    \toprule
    \multicolumn{7}{c}{\textbf{Accuracy model (ANOVA)}}\\
    Factor & Df & Sum Sq & Mean Sq & F value & $p_{\textnormal{raw}}$ & $p_{\textnormal{adj}}$ \\ 
    \midrule
    AI assistance        & 2  & 144   & 72    & 0.65      & .524     & $> 1$     \\
    NFC category         & 1  & 6     & 6     & 0.05      & .816     & $> 1$     \\
    AOT category         & 1  & 646   & 646   & 5.84      & .017   & .068      \\
    \addlinespace
    \cmidrule(lr){1-7}
    \addlinespace
    \multicolumn{7}{c}{\textbf{ECE model (ANOVA)}}\\
    Factor & Df & Sum Sq & Mean Sq & F value & $p_{\textnormal{raw}}$ & $p_{\textnormal{adj}}$ \\
    \midrule
    AI assistance        & 2  & 0.033 & 0.0166 & 1.18     & .310     & $> 1$     \\
    NFC category         & 1  & 0.014 & 0.0143 & 1.02     & .314     & $> 1$     \\
    AOT category         & 1  & 0.052 & 0.0522 & 3.72     & .056 \,  & .224      \\
        \addlinespace
    \cmidrule(lr){1-7}
    \addlinespace
\multicolumn{7}{c}{\textbf{Global Metacognition model (ANOVA)}}\\
    Factor          & Df & Sum Sq & Mean Sq & F value & $p_{\textnormal{raw}}$ & $p_{\textnormal{adj}}$ \\ 
    \midrule
    AI assistance          & 2  & 1710    & 855     & 2.86     & .060 \,  & .240    \\
    NFC category    & 1  & 1966    & 1966    & 6.57     & .011  & \ \textbf{.044} \\
    AOT category    & 1  & 138     & 138     & 0.46     & .498    & \ > 1 \\
    \addlinespace
    \cmidrule(lr){1-7}
    \addlinespace
    \multicolumn{7}{c}{\textbf{Affective Metacognition model (ANOVA)}}\\
    Factor          & Df & Sum Sq & Mean Sq & F value & $p_{\textnormal{raw}}$ & $p_{\textnormal{adj}}$ \\
    \midrule
    AI assistance          & 2  & 0.19    & 0.095   & 2.44     & .091 \,  & .364    \\
    NFC category    & 1  & 0.72    & 0.719   & 18.40    & < .0001  & \ \textbf{.0001} \\
    AOT category    & 1  & 0.01    & 0.008   & 0.19     & .661    & \ > 1 \\
    \bottomrule
  \end{tabular}
  \caption{ANOVA results for AI assistance, Need for Cognition (NFC), and Actively Open-minded Thinking (AOT) on accuracy, ECE, global and affective metacognition in the calibration phase with Bonferroni's correction of $m = 4$.}
  \label{tab:study2_anovas_calibration}
\end{table}

\subsubsection{Post Hoc Analysis}
\label{sec:study2_posthoc}

As already mentioned in Section \ref{sec:sample_size_stat_study2}, we extended our analysis of research questions by examining participants'  behavioral differences during the calibration phase and main task (timepoint),\footnote{Please note that we include the AI-assistance factor in models of the calibration phase for completeness; results are robust to its exclusion, and all statistically significant findings remain unchanged.} reliance on AI, subjective perceptions, and the effect of AI literacy. Further details for statistical models' assumptions and full results are presented in Section \ref{sec:app_results_study2} in the Appendix. 
For the calibration phase, we conducted ANOVA tests, as most of the assumptions were met. Furthermore, to examine differences between the calibration phase and the main task on the same dependent variables, most assumptions for linear mixed models were also satisfied, making them suitable for this analysis.

For reliance on AI, we compared two-stage and personalized AI conditions as follows: we filtered participants who did not receive AI assistance in the personalized condition, as well as well-calibrated ones from both conditions, since those in the personalized condition consisted of no AI or MCS paradigms, which would prevent computing some of the metrics. After this filtering, the two-stage AI condition had 49 participants, while personalized AI had 46, consisting only of those who were overconfident. Then, we were able to calculate the agreement fraction, as well as the over- and under-reliance fractions.\footnote{A sample size of N=95 is insufficient for 80\% power at Cohen's $f = 0.25$ considering two group categories, but adequate for detecting an effect size of approximately $f = 0.29$.}
Given agreement fraction $(W = 0.86,\ p < .0001$; $F$ = 1.93, $p$ = .074), over-reliance $(W = 0.816,\ p < .0001$; $F$ = 2.92, $p$ = .0087), and under-reliance ($W = 0.851,\ p < .0001$; $F$ = 1.08, $p$ = 0.38) distributions violate normality, and the majority also violate homogeneity assumptions, we conducted a permutation ANOVA to assess the main effects of AI assistance, NFC, and AOT factors (see Section \ref{sec:sample_size_stat_study2}).
To study ECE correlations with agreement on AI, over-reliance, and under-reliance, we employed Spearman's rank correlation analysis, as the distributions of these variables did not meet the normality assumption, as mentioned earlier.
Finally, we investigated the effects of AI literacy as a continuous covariate on our dependent variables by conducting ANCOVA tests, extending our previous ANOVA models (see Section \ref{sec:app_results_study2} in the Appendix for assumption checks).

\begin{figure} [!t]
      \centering
    \includegraphics[width=\textwidth]{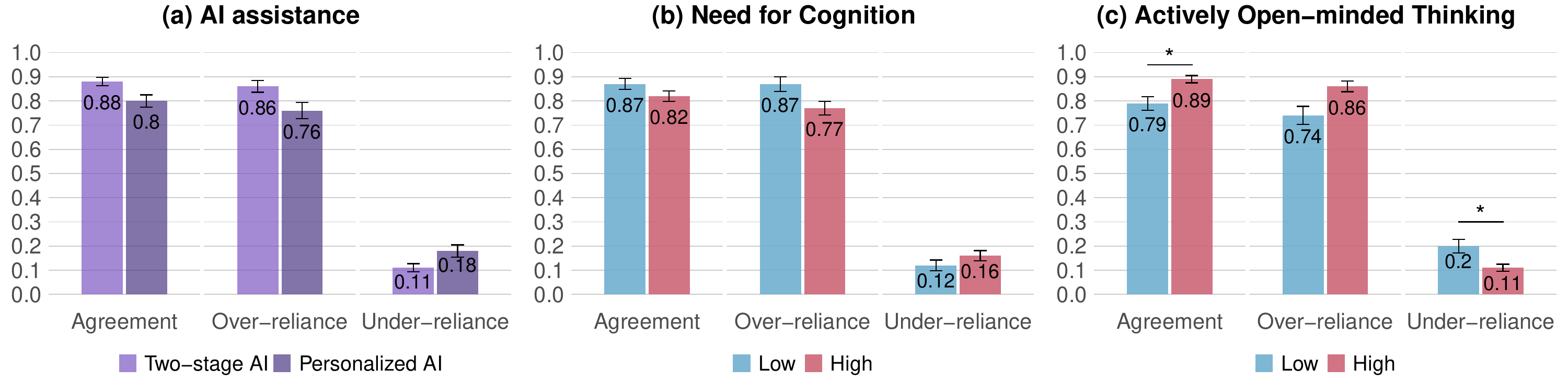}
  \caption{Participants' agreement fraction, over-reliance, and under-reliance on AI in the main task considering: (a) AI assistance, (b) Need for Cognition, and (c) Actively Open-minded Thinking. The asterisks highlight p-value significance strength (*$p$ < .05). }

    \Description{
     todo.
    }
 
    \label{fig:study2_reliance}
\end{figure}

\begin{table}[t]
  \centering
  \footnotesize
\begin{tabular}{
p{2cm}                           
>{\centering\arraybackslash}p{1.4cm}
>{\centering\arraybackslash}p{1.7cm} 
>{\centering\arraybackslash}p{1.7cm} 
>{\centering\arraybackslash}p{1.7cm} 
>{\centering\arraybackslash}p{1.7cm} 
>{\centering\arraybackslash}p{1.7cm} 
  }
    \toprule
    \multicolumn{7}{c}{\textbf{Agreement fraction model (Permutation ANOVA)}} \\
    Factor          & Df & Sum Sq & F value & $p_{\textnormal{parametric}}$ & $p_{\textnormal{resampled}}$ & $p_{\textnormal{adj}}$\\ 
    \midrule
    AI assistance          & 1  & 0.09510 & 4.640   & .0339  & .0338  & .1014 \\
    NFC category    & 1  & 0.0378 & 1.848   & .1774  & .1809  & .5427 \\
    AOT category    & 1  & 0.15717 & 7.669   & .0068  & .0066  & \textbf{.0198} \\
    \\
     \cmidrule(lr){1-7}
    \addlinespace
    \multicolumn{7}{c} {\textbf{Over-reliance model (Permutation ANOVA)}} \\
    Factor          & Df & Sum Sq & F value & $p_{\textnormal{parametric}}$ & $p_{\textnormal{resampled}}$ & $p_{\textnormal{adj}}$\\ 
    \midrule
    AI assistance          & 1  & 0.1625 & 4.410   & .0385  & .0396  & .1188 \\
    NFC category    & 1  & 0.1613 & 4.376   & .0392  & .0416  & .1248 \\
    AOT category    & 1  & 0.2243 & 6.085   & .0155  & .0176  & .0528 \\
     \cmidrule(lr){1-7}
    \addlinespace
    \multicolumn{7}{c}
{\textbf{Under-reliance model (Permutation ANOVA)}} \\
    Factor          & Df & Sum Sq & F value & $p_{\textnormal{parametric}}$ & $p_{\textnormal{resampled}}$ & $p_{\textnormal{adj}}$\\ 
    \midrule
    AI assistance          & 1  & 0.0766 & 3.7554  & .0557  & .0548  & .1644 \\
    NFC category    & 1  & 0.0158 & 0.7731  & .3816  & .3865  & >1 \\
    AOT category    & 1  & 0.1375 & 6.7366  & .0110  & .0122  & \textbf{.0366} \\

    \bottomrule
  \end{tabular}
  \caption{Permutation ANOVA results for AI assistance (two-stage, N=49; and personalized AI, N=46), Need for Cognition (NFC), and Actively Open-minded Thinking (AOT) on agreement fraction, over-reliance, and under-reliance on AI   in the main task with Bonferroni's correction of $m = 3$.}
  \label{tab:reliance_permutation_anova_full}
\end{table}

\begin{table}[t]
\centering
\footnotesize
\label{tab:metrics_ai}
\begin{tabular}{
p{1.7cm}                           
>{\centering\arraybackslash}p{1.4cm}
>{\centering\arraybackslash}p{1.7cm} 
>{\centering\arraybackslash}p{2.0cm} 
>{\centering\arraybackslash}p{1.7cm} 
>{\centering\arraybackslash}p{1.7cm} 
>{\centering\arraybackslash}p{1.7cm} 
  }
\toprule
\multicolumn{7}{c}{\textbf{Reliance metrics}} \\
AI assistance & Agreement & Switch & Follow High Conf. & Accuracy-wid & Over-Reliance & Under-Reliance \\
\midrule
Two-stage AI    & 0.88 & 0.59 & 0.57 & 0.61 & 0.86 & 0.11 \\
Personalized AI & 0.80 & --   & --   & --   & 0.76 & 0.18 \\
\bottomrule
  \end{tabular}
  \caption{Comparison of reliance-related metrics between two-stage (N=49) and personalized AI (N=46) conditions considering overconfident users only across Agreement fraction, Switch fraction, Follow High Confidence fraction, Accuracy-wid, Over-Reliance, and Under-Reliance. }
  \label{tab:reliance_full_metrics}
\end{table}

\begin{table}[t]
\centering
\small
\setlength{\tabcolsep}{6pt} 
\begin{tabularx}{\textwidth}{
>{\raggedright\arraybackslash}X 
>{\centering\arraybackslash}p{0.9cm} >{\centering\arraybackslash}p{0.9cm} >{\centering\arraybackslash}p{0.9cm} 
>{\centering\arraybackslash}p{0.9cm} >{\centering\arraybackslash}p{0.9cm} >{\centering\arraybackslash}p{0.9cm} 
>{\centering\arraybackslash}p{0.9cm} >{\centering\arraybackslash}p{0.9cm} >{\centering\arraybackslash}p{0.9cm} 
}
\toprule
Factor 
& \multicolumn{3}{c}{Source}
& \multicolumn{3}{c}{NFC}
& \multicolumn{3}{c}{AOT}
\\
\cmidrule(lr){2-4}\cmidrule(lr){5-7}\cmidrule(lr){8-10}
Metric & $\chi^2$ & $p_{\textnormal{raw}}$ & $p_{\textnormal{adj}}$
& $W$ & $p_{\textnormal{raw}}$ & $p_{\textnormal{adj}}$
& $W$ & $p_{\textnormal{raw}}$ & $p_{\textnormal{adj}}$
\\
\midrule
Calibration helpfulness
& 5.887 & .053 & > 1
& 3699.5 & .041 & > 1
& 3132.0 & .962 & > 1
\\
Self confidence
& 5.887 & .053 & > 1
& 3699.5 & .041 & > 1
& 3132.0 & .962 & > 1
\\
Mental demand
& 3.004 & .223 & > 1
& 3315.0 & .550 & > 1
& 3358.5 & .453 & > 1
\\
Complexity
& 5.646 & .059 & > 1
& 3104.5 & .887 & > 1
& 2541.0 & .033 & > 1
\\
Preference
& 1.083 & .582 & > 1
& 3717.5 & .037 & > 1
& 3413.0 & .329 & > 1
\\
Satisfaction
& 3.277 & .194 & > 1
& 3725.0 & .034 & > 1
& 3660.0 & .060 & > 1
\\
Trust in AI
& 0.179 & .672 & > 1
& 1496.0 & .412 & > 1
& 1445.5 & .665 & > 1
\\
Autonomy
& 0.116 & .733 & > 1
& 1427.0 & .715 & > 1
& 2030.5 & $<$ .0001 & \textbf{.0002}
\\
\bottomrule
\end{tabularx}

\caption{Kruskal-Wallis (for AI assistance; Bonferroni's correction of $m = 8$) and Wilcoxon (NFC and AOT; Bonferroni's correction of $m = 16$) comparison results considering post-test participants' reported subjective metrics: calibration helpfulness, perceived self-confidence, mental demand, complexity, preference, satisfaction, trust in AI, and autonomy.  }
\label{tab:study2_subjective_measures}
\end{table}

\paragraph{\textbf{Participants' differences in calibration phase vs main task}}
Overall, ANOVA tests (see Table \ref{tab:study2_anovas_calibration}) from the calibration phase only highlight significant differences when considering high-NFC individuals reporting higher global ($F = 6.57$, $p_{\textnormal{adj}} = .044$, Cohen's $f$ = 0.21) and affective ($F = 18.40$, $p_{\textnormal{adj}} = .0001$, Cohen's $f$ = 0.35) metacognition when compared to low-NFC individuals. 
Instead, ANOVA results considering variations on calibration phase and main task on \textbf{accuracy} (see Table \ref{tab:accuracy_lmm} in the Appendix) indicated a significant main effect of timepoint ($F=17.61$, $p_{\textnormal{adj}}=.0002$) and a significant timepoint $\times$ AI assistance interaction ($F=9.50$, $p_{\textnormal{adj}}=.0005$), showing that accuracy increased more from calibration to the main task in the AI-assisted conditions than in the user-only condition. 
The overall effect of AOT was also significant ($F=9.82$, $p_{\textnormal{adj}}=.0082$). At the linear mixed model coefficient level, the contrast for low versus high AOT confirmed lower accuracy for low-AOT individuals ($\beta$ = -0.0410, SE = 0.0159, $t$ = -2.57, $p_{\textnormal{adj}}=.0425$). 
The timepoint $\times$ AI assistance contrasts were likewise significant (personalized AI: $\beta$ = 0.0818, SE = 0.0231, $t=3.55$, $p_{\textnormal{adj}}=.0020$; two-stage AI: $\beta$ = 0.0920, SE = 0.0232, $t=3.97$, $p_{\textnormal{adj}}=.0004$). 
The ANOVA on \textbf{ECE} (see Table \ref{tab:ece_lmm} in the Appendix) showed a significant effect on timepoint ($F=28.49$, $p_{\textnormal{adj}}<.0001$) and timepoint $\times$ AI assistance interaction ($F=8.89$, $p_{\textnormal{adj}}=.0009$), indicating larger reductions in ECE from calibration to main task for the AI-assisted conditions. 
ANOVA results also indicated an effect of AOT ($F=6.68$, $p_{\textnormal{adj}}=.0425$), but the estimated coefficient for low versus high AOT did not remain significant after LMM coefficient-level adjustment ($\beta$ = 0.0368, SE = 0.0182, $t=2.02$, $p_{\textnormal{adj}}=.1752$). 
The timepoint $\times$ AI assistance contrasts were significant (personalized AI: $\beta$ = -0.0893, SE = 0.0266, $t$=-3.35, $p_{\textnormal{adj}}=.004$; two-stage AI: $\beta$ = -0.1041, SE = 0.0267, $t$=-3.89, $p_{\textnormal{adj}}=.0006$). 
For \textbf{global metacognition} (see Table \ref{tab:gmeta_lmm} in the Appendix), ANOVA findings highlight that NFC had a significant main effect ($F=9.66$, $p_{\textnormal{adj}}=.0088$). The LMM coefficient comparing low versus high NFC confirmed lower global metacognition for the low-NFC group ($\beta$ = -6.9895, SE = 2.3360, $t$=-2.99, $p_{\textnormal{adj}}=.0120$). 
The ANOVA for \textbf{affective metacognition} (see Table \ref{tab:ami_lmm} in the Appendix)  showed that NFC had a significant main effect ($F=20.33$, $p_{\textnormal{adj}}<.0001$), and the LMM coefficient for low versus high NFC indicated significantly lower affective metacognition in the low-NFC group ($\beta$ = -0.1368, SE = 0.0328, $t$=-4.17, $p_{\textnormal{adj}}=.0120$).

\paragraph{\textbf{Participants' reliance behaviors}}
\label{sec:study2_reliance_behaviors}
Considering users' reliance on AI behaviors in the main task, permutation ANOVA results highlight that high-AOT individuals tend to have an increased agreement fraction ($F = 7.67$, $p_{\textnormal{adj}} = .0198$) and a trend for increased over-reliance ($F = 6.09$, $p_{\textnormal{adj}} = .0528$) on AI compared to low-AOT ones. Instead, only low-AOT individuals seem to under-rely more on AI compared to high-AOT ones ($F = 6.74$, $p_{\textnormal{adj}} = .0366$). 
The results are presented in Figure \ref{fig:study2_reliance}, while Table \ref{tab:reliance_permutation_anova_full} displays the analysis test results, and Table \ref{tab:reliance_full_metrics} provides the complete reliance metrics for two-stage and personalized AI. 
Spearman's correlation results (see Table \ref{tab:correlations_ece_effect} in the Appendix) for two-stage AI highlight a significant moderate negative correlation between ECE and agreement fraction ($\rho$ = -0.536, $p < .001$) and a significant moderate positive correlation between ECE and under-reliance fraction ($\rho$ = 0.604, $p < .001$). For the personalized AI, Spearman's correlation analysis revealed a significant, moderate positive correlation between ECE and the under-reliance fraction ($\rho$ = 0.391, $p = .0072$). 
As regards the variations between ECE before and after seeing the AI advice (i.e., $\Delta\mathrm{ECE}$), the distribution of paired differences was non-normal (Shapiro-Wilk $W = 0.913$, $p = .0009$), so we used a Wilcoxon signed-rank test, which confirmed a significant reduction in ECE (pre: $M = 0.1629$, post: $M = 0.0882$; $\Delta\mathrm{ECE} = 0.075$, $V = 1146$, $p < .0001$). The paired effect size was medium-to-large (Cohen's $d = 0.79$, 95\% CI $[0.45, 1.12]$).

\paragraph{\textbf{Participants' subjective perceptions and AI literacy}}
For subjective metrics relative to main task interactions, the only significant difference was for high-AOT individuals in perceiving higher levels of autonomy ($M$ = 6.20 and $SD$ = 0.97) than low-AOT ones ($M$ = 5.09 and $SD$ = 1.53).
Full results are shown in Table \ref{tab:study2_subjective_measures}
ANCOVA tests on AI literacy revealed only significant findings related to metacognitive perceptions. Participants with higher AI literacy demonstrated increased metacognitive perceptions across both prospective (before calibration) and retrospective (after the main task) measures. Full results are shown in Tables \ref{tab:ancova_gmeta} and \ref{tab:ancova_ameta}.
For global metacognition, AI literacy positively predicted scores both calibration phase ($B = 11.36$, $SE = 2.23$, $t = 5.09$, $p_{\textnormal{adj}} < .0001$) and main task ($B = 4.96$, $SE = 1.59$, $t = 3.12$, $p_{\textnormal{adj}} = .0088$).
For affective metacognition, AI literacy was associated with higher affective metacognition ($F = 41.81$, $p_{\textnormal{adj}} < .001$, $\eta^{2} =.07$) in the calibration phase. The same holds in the main task ($F = 32.59$, $p_{\textnormal{adj}} < .001$, $\eta^{2} =.12$).

Additionally, there was a significant interaction between AI literacy and NFC on affective metacognition in both the main task ( $F = 8.23$, $p_{\textnormal{adj}} = .0188$, $\eta^{2} =.05$) and calibration phase ($F = 14.65$, $p_{\textnormal{adj}} = .00076$, $\eta^{2} =.09$). 
Post hoc analyses on the significant NFC $\times$ AI literacy interaction revealed consistent patterns across both phases. 
In the \textit{main task}, adjusted means indicated that participants with high NFC reported higher affective metacognition ($M = 0.489$, $SE = 0.024$) than those with low NFC ($M = 0.403$, $SE = 0.025$; $p_{\text{adj}} < .001$). Simple slope analyses showed that AI literacy positively predicted affective metacognition among high NFC individuals ($B = 0.212$, $SE = 0.041$, $p_{\text{adj}} < .001$), but not among low NFC individuals ($B = 0.049$, $SE = 0.040$, $p_{\text{adj}} = .23$).
Similarly, in the \textit{calibration phase}, high NFC participants again exhibited higher affective metacognition ($M = 0.508$, $SE = 0.021$) than low NFC participants ($M = 0.426$, $SE = 0.022$; $p_{\text{adj}} < .001$). AI literacy was positively associated with affective metacognition for high NFC individuals ($B = 0.228$, $SE = 0.036$, $p_{\text{adj}} < .001$), but not for low NFC individuals ($B = 0.042$, $SE = 0.035$, $p_{\text{adj}} = .23$). 

Together, these results suggest that while high NFC participants consistently reported higher affective metacognition, the positive link between AI literacy and affective metacognition was mainly driven by individuals with high NFC, both during calibration and in the main task.
To further investigate this mechanism, a causal mediation analysis was conducted to test whether AI literacy mediated the relationship between NFC and affective metacognition. Results (5,000 bootstrap simulations) revealed a significant indirect effect (ACME = 0.055, 95\% CI [0.027, 0.084], $p < .001$), indicating that higher NFC predicted higher AI literacy, which in turn was associated with greater affective metacognition. However, the direct effect of NFC remained significant (ADE = 0.084, 95\% CI [0.015, 0.158], $p = .015$), consistent with partial mediation. Approximately 39\% of the total effect of NFC on affective metacognition was mediated through AI literacy (95\% CI [0.176, 0.818]).

\subsubsection{Summary of findings}
Study 2 results highlight that AI assistance had a significant effect on accuracy, with both two-stage and personalized AI outperforming the no-AI condition. However, only two-stage AI led to improved appropriateness of self-confidence (ECE) compared to the no-AI condition. The post hoc analysis confirmed that the changes in accuracy and ECE between the calibration phase and main task were driven by the introduction of two-stage and personalised AI assistance (\textbf{RQ2.1}). Additionally, descriptive statistics and correlation analyses highlight that the improvement in ECE for two-stage AI was probably due to a pronounced agreement on AI and confidence alignment. For metacognitive measurements, only the two-stage AI resulted in increased global metacognition compared to the no AI condition (\textbf{RQ2.2}).
For \textbf{RQ3.1} and \textbf{RQ3.2}, we found that high-NFC individuals reported greater global and affective metacognition compared to low-NFC individuals in the main task, with post hoc analysis confirming these results also for the calibration phase. 
Instead, high-AOT individuals achieved higher accuracy than low-AOT individuals in the main task. Additionally, the high-AOT group showed increased agreement and reduced under-reliance on AI, further indicating higher perceived autonomy compared to low-AOT individuals.
Finally, post hoc results emphasize that the higher an individual’s AI literacy, the greater their global and affective metacognition during both the calibration phase and the main task; however, this relationship was evident only among participants with high NFC for affective metacognition. A causal mediation analysis further confirmed that AI literacy partially mediated the effect of NFC on affective metacognition, accounting for about 39\% of the total effect, while a direct influence of NFC remained.


\section{Discussion}
\label{sec:discussion}
In this work, we investigated (\textbf{RQ1}) how to categorize individuals into overconfident, underconfident, and well-calibrated self-confidence groups, and evaluated the accuracy and appropriateness of self-confidence differences across individuals with low and high NFC and AOT during a calibration phase. We also examined how different types of AI assistance (\textbf{RQ2}) and psychological constructs, such as NFC and AOT (\textbf{RQ3}), influenced people's accuracy, appropriateness of self-confidence, and both global and affective metacognition. This section summarizes the results from Studies 1 and 2, highlighting the shared findings and differences with the current literature. Additionally, we offer guidance for HCI research regarding the implications of our findings.

\subsection{Study 1}
\label{sec:study1_discussion}
The results from Study 1 indicate that during a real-time feedback calibration phase, participants are predominantly overconfident in their decisions when completing a low-stakes, simple loan prediction task, while underconfident users are relatively rare. 
Instead, well-calibrated people are slightly more common than underconfident ones, consisting of individuals who are close to the perfect-calibration line within the $\delta$ distance threshold.
This also holds for the calibration phase conducted in Study 2 (see Table \ref{tab:study2_summary_ai_assistance}).
We identified a $\delta$ threshold by applying three different approaches (quartile, elbow, and ECE-based). Two of them provided the same value (5), while the latter was more restrictive (between 3 and 4).  
Overall, given the unbalanced distribution of the three calibration groups in this specific task, we considered it fair to use the higher value,  as lower values would have made the underconfident or even the well-calibrated category vanish. 
The goal of the threshold was to find a plausible categorization across individuals' self-confidence groups while establishing a well-calibrated category to later inform a personalized AI approach based on self-confidence status. However, results show that introducing a well-calibrated category proved to be challenging. 
Still, its identification is the basis for making informed design decisions on the AI personalization (e.g., which paradigm to use for showing AI suggestions and confidence, the policy for making the final decision between AI and user, etc.). Therefore, it is necessary to compute and test this $\delta$ threshold in other comparable, low-stakes, and simple task scenarios to validate the replicability of the results and their usefulness for the potential identification of well-calibrated individuals.

As per the prevalence of overconfident users, it may be due to the general low performance for this type of task, considering our results on calibration phases (Study 1: $M$ = 60.55; Study 2: $M$ = 61.95): earlier studies with similar designs revealed that achieving high accuracy on this specific task is generally challenging without AI assistance, suggesting that the majority of participants are expected to be overconfident \cite{Ma24UserConfidenceCalibration,li2025confidencealigns}. 
This also aligns with previous research considering relatively straightforward tasks, where most people tend to be overconfident \cite{Soll2022OverconfidenceAdvice,Binnendyk2024IndividualDifferencesOverconfidence,li2025confidencealigns,LiHaleMoore_2025OverconfidenceIndividualDifference}. 
We expect a different distribution for more difficult tasks, which are more likely to elicit underconfidence \cite{Razmdoost2015UnderconfidentOverconfident,Frankenberger1997decision}. 
In any case, genuinely underconfident individuals are rare in judgment and decision-making contexts \cite{moore2015OverprecisionInJudgement}. 
Taken together, these results underscore the importance of designing targeted interventions for overconfident individuals, a topic we discuss further in Section \ref{sec:implications}.

Neither low nor high levels of NFC and AOT appear to exhibit differences in the accuracy and appropriateness of self-confidence during the calibration phase.
Regarding NFC, these results align with previous work \cite{Bucinca2021NFC,Gajos2022NFC,Millecamp2019NFC,Millecamp2020NFC,Vasconcelos2023OverrelianceXAI,Conati2021NFC,BahelConati2024NFC,Kopecka2024NFCHumanRobotsExplanationsPerceptions}, suggesting that differences in objective and subjective metrics between low- and high-NFC individuals tend to emerge when people engage with AI or explanations. By contrast, recent research reports the absence of significant differences between low- and high-NFC groups even when AI or explanations are provided \cite{bucinca2024optimizinghumancentricobjectivesaiassisted,bucinca2025ContrastiveExplanationsNFCAOT,Cau2025CuriosityTraits}. 
We can make a similar argument for AOT: low versus high levels may interact with AI explanations, but evidence from AI-assisted studies is limited \cite{Swaroop2025AIPersonalizedOverrelianceRate,bucinca2025ContrastiveExplanationsNFCAOT}, and other work finds that high-AOT individuals show greater accuracy \cite{Haran2013AOTinAccuracyAndCalibration} or better calibration \cite{Martin2024CalibrationFeedbackAOT} even without AI assistance.
Nonetheless, we will explore the role of NFC and AOT individual traits in more detail in the following sections through the results of Study 2.

\subsection{Study 2}
\label{sec:study2_discussion}

\subsubsection{AI assistance}

Study 2 results confirm what we found in Study 1 regarding individuals' self-confidence levels: the vast majority were overconfident. Participants showed no improvements in accuracy or appropriate self-confidence in the no AI assistance condition. Post hoc analysis confirmed that improvements only appeared when AI assistance was provided, showing that \enquote{Calibration Status Feedback} mechanism \cite{Ma24UserConfidenceCalibration} alone was insufficient to improve people's accuracy and self-confidence calibration in the main task without AI support.  
This also highlights that, for this particular task, an accuracy gap exists between users and AI, showing that AI can be helpful and potentially fill this gap.

In our case, the two-stage AI condition enhanced individuals' accuracy, appropriateness of self-confidence, and global metacognition compared to the no AI condition. Although the main task measures suggest improvements in global metacognition for the two-stage condition, post hoc analysis revealed that these gains disappeared when both calibration and main phases were considered. Thus, we cannot conclude that the two-stage intervention produced a stable change in global metacognition.
Conversely, the personalized AI led to improvements only considering accuracy when compared to no AI: most participants (46 out of 53) were overconfident and more confident than the AI, so they were assigned to a two-stage paradigm for instances with low AI confidence and to a one-stage paradigm for instances with high AI confidence. 
As somewhat expected from the results of Study 1, participants' self-confidence distribution was unbalanced towards overconfident users, especially those who were more confident than and less accurate than the AI, thus reducing the personalised AI spectrum to predominantly one macro sub-category. This finding raises multiple issues for tailoring AI assistance given our study settings, which we will discuss in the implications section.
Furthermore, a general tendency towards overconfidence was also acknowledged in high-NFC participants. 
We will discuss the NFC outcomes in more detail in the next paragraph.

Although the two-stage and personalized AI conditions did not differ significantly in terms of agreement or over-reliance on AI, further analyses revealed an interesting pattern. The improvement in ECE in the two-stage AI condition appears to be driven almost entirely by users' high agreement with the AI: the higher the agreement, the lower the ECE, which suggests an overreliance effect. While our findings are consistent with prior work showing that two-stage decision designs can produce gains in accuracy or promote appropriate reliance on AI \cite{Bucinca2021NFC,Ma2023CorrectnessLikelihoodAIUsersIncome,Salimzadeh2024PrognosticVsDiagnosticTasks,Agudo2024HumanError,Morrison2024ImperfectXAI,Cao2024AppropriateRelianceSkinCancer,Kuper2025NFCConfidenceRelianceAITwoStage}, some of the same studies and other works report an increase of decision-making accuracy as a result of agreement with incorrect advice more than with correct advice, a pattern that we also observe in our findings \cite{Lu2024DoWeLearnFromEachOtherTwoStage,Cao2024AppropriateRelianceSkinCancer,Ma24UserConfidenceCalibration,Kuper2025NFCConfidenceRelianceAITwoStage}.
Complementary to this, the difference we observed in ECE before and after viewing the advice further supports the identification of this overreliance behavior, suggesting that improvements in self-confidence calibration are due to a confidence alignment effect. 
This finding is consistent with prior work on confidence alignment within human-human and human-AI dyads \cite{Bang2014Doesinteractionmatter,Bang2017ConfidenceGroupDecisionMaking,Pescetelli2021GroupDecisionMetacognition,CHONG2022HumanConfidence,Pescetelli2022Benefitsofspontaneousconfidence,li2025confidencealigns}. Namely, Li et al. \cite{li2025confidencealigns} showed that, under two-stage AI assistance, people's confidence in their final joint decisions more closely matches the AI's confidence than their initial self-confidence. However, our post-hoc observations require confirmation through adequately powered studies to distinguish appropriate reliance on AI from problematic overreliance and confidence alignment.

\subsubsection{Need for Cognition}
Our findings showed that individuals with high NFC reported greater perceived decision-making abilities (global metacognition) and more positive emotions (affective metacognition) than those with low NFC. 
However, this does not translate to actual improvements in accuracy or self-confidence appropriateness, resulting in an overestimation of one's abilities. 
For high-NFC individuals, this behavior is commonly known as the Dunning-Krueger effect \cite{He2023DunningKruger}, a metacognitive bias causing individuals to overestimate their competence and performance. 
This applies to both the calibration phase and the main task, although most participants were classified as overconfident regardless of their individual trait categories, complementing the results from Study 1. Therefore, this outcome shows that NFC influences self-assessment measures of confidence, with high-NFC individuals displaying inflated self-assessments that lead to overconfidence behaviors, in line with previous research on human confidence \cite{Barden2008NFCAffectConfidence,DeMarree2020NFCDispositionalAttitudeCertainty,Vogt2022NonAbilityConfidence,Zerna2024NFCReview,Cau2025CuriosityTraits}. 
This pattern also holds in previous AI-assisted research, where no differences were observed between low and high NFC in decision-making evaluation metrics, whether with AI assistance alone or with AI plus explanation(s) \cite{Vasconcelos2023OverrelianceXAI,cau2025exploringimpactexplainableaiNFCARXIV,bucinca2024optimizinghumancentricobjectivesaiassisted,Kuper2024NFCRelianceAIAssistanceArtPeriod,Cau2025CuriosityTraits,bucinca2025ContrastiveExplanationsNFCAOT}.
Moreover, higher AI literacy predicted greater global and affective metacognition in both the calibration phase and the main task. However, the positive association with affective metacognition was observed only among high-NFC participants: NFC had a direct positive effect on affective metacognition, while AI literacy partially mediated that effect. Further research is needed to study AI literacy and its interplay with NFC on metacognitive perceptions.

\subsubsection{Actively Open-minded Thinking}
Our results reveal interesting findings for high-AOT individuals: in the main task, they achieved higher accuracy, greater agreement with AI, less under-reliance on AI, a trend toward over-reliance on AI, and a higher perceived level of decision-making autonomy compared to low-AOT individuals. 
On the one hand, these results align with previous work without AI assistance, where high-AOT individuals generally achieve increased decision-making accuracy \cite{Haran2013AOTinAccuracyAndCalibration,Martin2024CalibrationFeedbackAOT}, but not better calibration outcomes.
On the other hand, the increase in accuracy also seems to stem from the introduction of AI assistance during the main phase, which aligns with the increased agreement and less under-reliance on AI they exhibited. 
This finding that high-AOT individuals exhibit excessive agreement with AI might contrast with recent results for AOT in AI-assisted decisions \cite{Swaroop2025AIPersonalizedOverrelianceRate}, where they discovered that overreliers actually have lower AOT scores. 
Due to the limited research on the AOT construct in AI-assisted decisions, further studies are necessary to assess decision-making and metacognitive perceptions among individuals with varying levels of AOT.

\subsection{Implications for HCI Studies} 
\label{sec:implications}

\subsubsection{Complementing self-confidence calibration with other design patterns}

Our results show that calibrating users' self-confidence using the \enquote{Calibration Status Feedback} mechanism by Ma et al. \cite{Ma24UserConfidenceCalibration} did not improve participants' decision-making accuracy or appropriateness of self-confidence without AI support. 
Given our study settings and that most participants were overconfident, these findings suggest that further testing in similar contexts is needed, as well as experimenting with different confidence calibration techniques and incorporating other existing design patterns that might improve participants' decision-making. 
For example, the \enquote{Calibration Status Feedback} mechanism could be complemented by introducing specific \textit{nudging} techniques (i.e., soft interventions that steer users toward better choices, like setting defaults, adding or reducing friction in required actions) \cite{TABATABAI2005Nudges,Caraban2019_23Nudges,Bach2023BiasMitigationBackfire,Nattapat2025CognitiveBiases} after the calibration phase, specifically targeted for the most prominent self-confidence group. 
Integrating nudges when users decide autonomously or with AI assistance could help them to mitigate biased behaviors: for example, we could use nudges to gently remind users of their self-confidence calibration status after the calibration phase, or suggest adjusting their self-confidence by a certain percentage based on their calibration status, using targeted warning labels or dialogs which have proven to be successful in debiasing users' behavior in domains like web search and phishing detection 
\cite{Rieger2024NudgesConfirmationBiasSERP,Desolda2025Warnings,cau2025largelanguagemodelsimprovewarningARXIV}.

\subsubsection{Pitfalls of two-stage AI and challenges of self-confidence-based AI personalization}

As recent work in AI-assisted decisions suggests \cite{Lu2024DoWeLearnFromEachOtherTwoStage,Cao2024AppropriateRelianceSkinCancer,Ma24UserConfidenceCalibration,Kuper2025NFCConfidenceRelianceAITwoStage}, the two-stage AI  paradigm seems to act as a double-edged sword effect also in our study: on the one hand, it increased users' decision-making accuracy, improved self-confidence, appropriateness, and global metacognition (i.e., prospective/retrospective decision-making accuracy in the task). 
On the other hand, it increased users' tendency to agree with the AI's predictions and to match their confidence levels to those reported by the AI, which indicates a pronounced overreliance effect. 
Considering our study, the findings underscore that the improvements observed under the two-stage AI paradigm are largely driven by users' excessive alignment with the AI in both predictions and confidence scores. Notably, users did not distinguish correct from incorrect AI suggestions and accepted advice indiscriminately, a behavior contrary to the intended effect of this cognitive forcing strategy \cite{Green2019Loan,Gajos2022NFC,Lu2024DoWeLearnFromEachOtherTwoStage,Cao2024AppropriateRelianceSkinCancer,Ma24UserConfidenceCalibration,Kuper2025NFCConfidenceRelianceAITwoStage,li2025confidencealigns}. This observation suggests exercising caution regarding the outcomes produced by the widely used two-stage AI paradigm in AI-assisted decision-making \cite{Bucinca2021NFC,He2023DunningKruger,He2023AnalogyExplanations,Salimzadeh2024PrognosticVsDiagnosticTasks,Agudo2024HumanError,Morrison2024ImperfectXAI,Cao2024AppropriateRelianceSkinCancer,Kuper2025NFCConfidenceRelianceAITwoStage}. 
As discussed earlier, one potential approach to reducing incorrect alignment with the two-stage AI is to introduce subtle interface interventions that encourage users' appropriate reliance, thereby complementing cognitive forcing with nudges.

Personalized AI could also be improved by introducing nudges, but it first requires several refinements before being complemented with other strategies. In our study, we adopted a user-centric approach that delivered AI assistance by leveraging people's self-confidence after a calibration phase. This strategy compared their average confidence and accuracy with that of AI, with the ultimate goal of leveraging their cognitive biases to strengthen human-AI team decision-making. 
Nevertheless, further studies with similar conditions are necessary to establish whether overconfident users still dominate. If so, research should focus on reducing overconfidence and adjusting personalized AI accordingly. 
Another viable solution to implement or complement this user-centric personalized AI would be to add Reinforcement Learning strategies \cite{Swaroop2025AIPersonalizedOverrelianceRate} to find the optimal AI assistance type based on the actual context (e.g., task stakes and difficulty, people's expertise, instance-level complexity, etc.). This could result in a framework that combines multiple tasks with different stakes and user traits, and that is adaptable to a wider range of contexts and people's profiles.

\subsubsection{Leveraging psychological constructs to improve personalized AI assistance} 
Our findings on the Need for Cognition and Actively Open-minded Thinking indicate that high-NFC individuals tend to be overconfident in their decisions, which stems from their inflated knowledge about the task. However, this does not lead to an actual increase in task accuracy compared to low-NFC individuals.
Conversely, high-AOT individuals perform better in the task but seem to rely too much on AI advice. Still, they feel more autonomous during the task compared to individuals with low-AOT. 
Our work suggests that we should use NFC and AOT combinations as signals for when to provide targeted interventions, helping to develop personalized AI assistance tailored to specific psychological constructs \cite{Cau2025CuriosityTraits,Zschech2025PersonalizingIML}. For example, individuals with high NFC and high AOT might show both overconfidence in themselves and overreliance on AI. Additionally, individuals with high NFC but low AOT may signal overconfidence (and so forth). Prior work has adopted this strategy to deliver personalized explanations considering low or high NFC and Conscientiousness when evaluating students in learning environments \cite{BahelConati2024NFC,Yanez2024NFCAdaptiveVisualizations}. 
In a broader sense, NFC and AOT psychological constructs can also be integrated into our personalized AI as additional indications, further leading to the development of a more comprehensive framework, as we previously discussed.


\subsection{Limitations}
\label{sec:limitations_fw}
Our work has several limitations, which we list as follows. 
The first concerns the income prediction decision-making task, which can be seen as a low-stakes, simple task scenario. This involves all the settings related to the task, including the classification model, AI confidence estimation, selected criteria for instances, confidence intervals, and population sample. We aimed to replicate previous study settings with the same data \cite{Ma24UserConfidenceCalibration,li2025confidencealigns} to determine whether differences might arise when introducing a personalised AI based on self-confidence calibration. Furthermore, the generalisability of results also needs to be acknowledged for other stakes and task complexities. For example, high-stakes and difficult tasks might require more domain-specific knowledge, which could likely influence people's decision-making and behaviours.
The second limitation is related to the choice of the real-time and post hoc feedback calibration mechanism we used in our study. Our goal was to replicate Ma et al.'s \cite{Ma24UserConfidenceCalibration} approach for the same task and compare our findings. However, we recognize that different calibration mechanisms exist, and further research should focus on developing, testing, and crafting new solutions based on the context at hand.
The third limitation concerns the choice of the $\delta$ threshold to classify individuals as overconfident, underconfident, or well-calibrated. Although we used three established methods to achieve this, we recognize and expect variability across different domains, tasks and users characteristics. Additionally, this classification is only applicable for binary-confidence levels, while strategies involving multi-level confidence levels could be explored in future research.
The fourth limitation relates to the implementation rules of the personalized AI condition. We designed it to incorporate and model known human cognitive strategies and biases to enhance human-AI collaboration. However, there is room for many improvements, such as considering a broader range of human biases and complementing them with Reinforcement Learning techniques 
\cite{Swaroop2025AIPersonalizedOverrelianceRate} to account for more nuanced user traits, which could also be tailored based on task characteristics, different AI paradigms, or refined to prioritize either user or AI autonomy depending on the situation and stakes.
The last limitation involves measuring individuals' affective metacognition. Although we adapted a well-known, stable scale (I-PANAS-SF) to measure positive and negative affects for our affective metacognition assessments at specific points in time, it remains unknown whether the same results would be observed under the same or different conditions, given that affective states can vary and no ground truth exists for them. In this regard, we encourage further research to test the Affective Metacognition Index (AMI) in other settings and advocate for the development of more reliable and nuanced scales to measure affective metacognition, which is currently lacking.

\section{Conclusion and Future Work}
\label{sec:conclusion}

This work examined a strategy to define a well-calibrated user category other than under-confident and over-confident as a result of a calibration phase, and the effects of how different types of AI assistance (no AI, two-stage AI, and personalized AI) and individual trait levels (low or high) of NFC and AOT impact users' accuracy, appropriateness of self-confidence, and meta-perceptions (global and affective metacognition) considering an income prediction task.
Our results showed that defining a well-calibrated user category is problematic because most participants tended to be overconfident, which reduces the value of personalizing AI for well-calibrated users. Real-time calibration alone did not improve decision accuracy or the appropriateness of self-confidence when users acted without AI. In fact, improvements emerged only with AI-assisted decisions: the two-stage AI increased task accuracy, appropriateness of human self-confidence, and global metacognition compared to no AI, but at the cost of excessive overreliance and confidence alignment with AI. Instead, personalized AI improved task accuracy compared with no AI. For psychological constructs, the work showed that high-NFC participants tended to be intrinsically overconfident, while high-AOT participants exhibited increased task accuracy at the cost of excessive agreement on the AI. These results highlight the importance of strengthening and exploring calibration mechanisms across different user self-confidence categories and contexts, and underscore the need to adopt a user-centric approach to deliver personalized AI assistance that accounts for individual traits such as cognitive biases and psychological constructs.

Based on these findings, we propose multiple directions for future research. 
For example, there is a need to explore our user categorization with different $\delta$ thresholds and tasks of multiple natures, such as low-stakes hard tasks and both easy and hard high-stakes tasks, to observe how users' distributions change across different contexts. Additionally, developing and testing new calibration mechanisms and pairing them with post-calibration nudges tailored to users' bias profiles might help to categorize which calibration techniques work better in specific scenarios. Finally, multiple tests are necessary for personalizing AI across various stakes and task difficulties, either by explicitly incorporating psychological traits like NFC and AOT into the personalization logic or by complementing personalization strategies using reinforcement learning approaches.

\begin{acks}
This research is partially funded by the Italian Ministry of University and Research (MUR) and by the European Union - NextGenerationEU, Mission 4, Component 2, Investment 1.1, under grant
PRIN 2022 PNRR \enquote{DAMOCLES: Detection And Mitigation Of Cyber
attacks that exploit human vuLnerabilitiES} (Grant P2022FXP5B) —
CUP: F53D23009220001.
\end{acks}

\bibliographystyle{ACM-Reference-Format}
\bibliography{sample-base}

\appendix
\section{Appendix}
\label{sec:appendix}

\subsection{Model Calibration}
\label{sec:app_model_calibration}
We calibrated XGB and RF models
using the following methods: Isotonic Regression \cite{Zadrozny2001Isotonic}, Platt Scaling \cite{Platt2000Calib}, and Venn-Abers \cite{Vovk2014VennAbers,Vovk2015PRoba,Manokhin2017VennAbers}. During the calibration procedure, we converted the models into \textit{ensembles of two} and evaluated their performance taking into consideration the following metrics: accuracy, Brier loss \cite{Brier1950VERIFICATIONOF}, Log loss \cite{Domingos1999LogLoss}, ROC-AUC \cite{Fawcett2004ROCAUC}, and Expected Calibration Error (ECE) \cite{Chuan2017ECE}. The Isotonic Regression method brought the best improvements, considering calibration metrics like Brier loss and ECE.
Although all the metrics are very similar between models, the XGB ensemble has slightly better metrics than the RF ensemble (see Table \ref{tab:model_calibration}). Consequently, we decided to use the XGB ensemble for the income prediction tasks since its probabilities appear well-calibrated and can be directly interpreted as confidence levels. The full implementation details are publicly available in this repository.\footref{open_repo}

\begin{table}[!ht]
    \centering
    \small
    \setlength{\tabcolsep}{5pt}
    \caption{Summary of the eXtreme Gradient Boosting (XGB) and Random Forest (RF) calibration results using the following metrics: accuracy, Brier loss, Log loss, ECE, and ROC-AUC.}
    \begin{tabular}{ccccccc}
        \toprule
        \textbf{Model} & 
        \textbf{Calibration method} &
        \textbf{Accuracy} & \textbf{Brier loss} & \textbf{Log loss} &
        \textbf{ECE} &
        \textbf{ROC-AUC}\\
        
        \midrule

        XGB ensemble &
        Isotonic regression &
        83.54\% &	\textbf{.1111}	& \textbf{.3472} &	\textbf{.0079} &	\textbf{.8921} \\

        RF ensemble &
        Isotonic regression &
        \textbf{83.58}\% &	
        .1116 &	
        .3483 & 
        .0083 &	
        .8912 \\  

        \bottomrule
    \end{tabular}
    \label{tab:model_calibration}
\end{table}

\subsection{Selected Instances}
\label{sec:app_selected_instances}

\begin{table}[htbp]
\centering
\scriptsize
\begin{tabular}{%
>{\centering\arraybackslash}C{0.05\linewidth} %
>{\centering\arraybackslash}C{0.15\linewidth} %
>{\centering\arraybackslash}C{0.15\linewidth} %
>{\centering\arraybackslash}C{0.15\linewidth} %
>{\centering\arraybackslash}C{0.15\linewidth} %
>{\centering\arraybackslash}C{0.15\linewidth} %
}
\toprule
\multicolumn{6}{c}{\textbf{Calibration phase instances}} \\ 
\midrule
\textbf{ID} & \textbf{True prediction} & \textbf{AI prediction} & \textbf{AI confidence level} & \textbf{AI correctness} & \textbf{AI confidence} \\

\midrule
1  & >50K      & >50K    & high           & correct      & 0.8981 \\
2  & >50K      & >50K    & high           & correct      & 0.8557 \\
3  & >50K      & >50K    & high           & correct      & 0.8003 \\
4  & >50K      & >50K    & high           & correct      & 0.8927 \\
5  & >50K      & >50K    & high           & correct      & 0.8557 \\
6  & $\leq$50K & $\leq$50K & high        & correct      & 0.7876 \\
7  & $\leq$50K & $\leq$50K & high        & correct      & 0.9207 \\
8  & $\leq$50K & $\leq$50K & high        & correct      & 1.0000 \\
9  & $\leq$50K & $\leq$50K & high        & correct      & 0.9883 \\
10 & $\leq$50K & >50K    & high           & incorrect    & 0.7705 \\
11 & >50K      & >50K    & low            & correct      & 0.7155 \\
12 & >50K      & >50K    & low            & correct      & 0.6415 \\
13 & >50K      & >50K    & low            & correct      & 0.7450 \\
14 & $\leq$50K & $\leq$50K & low         & correct      & 0.5420 \\
15 & $\leq$50K & $\leq$50K & low         & correct      & 0.5957 \\
16 & $\leq$50K & $\leq$50K & low         & correct      & 0.6125 \\
17 & $\leq$50K & >50K    & low            & incorrect    & 0.6511 \\
18 & $\leq$50K & >50K    & low            & incorrect    & 0.5769 \\
19 & >50K      & $\leq$50K & low         & incorrect    & 0.5361 \\
20 & >50K      & $\leq$50K & low         & incorrect    & 0.6190 \\

\midrule
\multicolumn{6}{c}
{\textbf{Main task instances}} \\ 
\midrule
\textbf{ID} & \textbf{True prediction} & \textbf{AI prediction} & \textbf{AI confidence level} & \textbf{AI correctness} & \textbf{AI confidence} \\

\midrule
1  & >50K      & >50K    & high           & correct      & 0.8003 \\
2  & >50K      & >50K    & high           & correct      & 0.7745 \\
3  & >50K      & >50K    & high           & correct      & 0.7708 \\
4  & >50K      & >50K    & high           & correct      & 0.7548 \\
5  & $\leq$50K & $\leq$50K & high        & correct      & 0.9529 \\
6  & $\leq$50K & $\leq$50K & high        & correct      & 0.8280 \\
7  & $\leq$50K & $\leq$50K & high        & correct      & 0.8280 \\
8  & $\leq$50K & $\leq$50K & high        & correct      & 0.9603 \\
9  & $\leq$50K & $\leq$50K & high        & correct      & 0.9969 \\
10 & >50K      & $\leq$50K & high        & incorrect    & 0.8894 \\
11 & >50K      & >50K    & low            & correct      & 0.6301 \\
12 & >50K      & >50K    & low            & correct      & 0.5296 \\
13 & >50K      & >50K    & low            & correct      & 0.7155 \\
14 & $\leq$50K & $\leq$50K & low         & correct      & 0.6190 \\
15 & $\leq$50K & $\leq$50K & low         & correct      & 0.6930 \\
16 & $\leq$50K & $\leq$50K & low         & correct      & 0.6607 \\
17 & $\leq$50K & >50K    & low            & incorrect    & 0.7374 \\
18 & $\leq$50K & >50K    & low            & incorrect    & 0.5733 \\
19 & >50K      & $\leq$50K & low         & incorrect    & 0.6125 \\
20 & >50K      & $\leq$50K & low         & incorrect    & 0.6530 \\
\bottomrule
\end{tabular}
\caption{Calibration phase and main task instances.}
\label{tab:final_instances}
\end{table}

\subsection{Need for Cognition Scale}
\label{sec:app_nfc_scale}

We measured participants' Need for Cognition (NFC) with the Need for Cognition Scale (NCS-6) by Lins de Holanda Coelho et al. \cite{LinsDeHolandaCoelho2020NFC6}, consisting of a five-point scale (1 = extremely uncharacteristic of
me; 5 = extremely characteristic of me). We will sum up all six item scores and then compute the \textit{median} to split participants into low and high NFC.

We used the following six items to compute NFC\footnote{note: (R) = reversed items \label{ft:reverse}}:

\begin{enumerate}
    \item I would prefer complex to simple problems.
    \item I like to have the responsibility of handling a situation that requires a lot of thinking.
    \item Thinking is not my idea of fun. (R)
    \item I would rather do something that requires little thought than something that is sure to challenge my thinking abilities. (R)
    \item I really enjoy a task that involves coming up with new solutions to problems. 
    \item  I would prefer a task that is intellectual, difficult, and important to one that is somewhat important. 
\end{enumerate}

\subsection{Actively Open-minded Thinking Scale}
\label{sec:app_aot_scale}
We measured participants' Actively Open-minded Thinking (AOT) with the Actively Open-minded Thinking Scale by Haran et al. \cite{Haran2013AOTinAccuracyAndCalibration}, consisting of a seven-point scale (1 = completely Disagree; 7 = completely Agree). We will sum up all the seven-item scores and then compute the \textit{median} to split participants into low and high AOT.

We used the following seven items to compute AOT\footref{ft:reverse}: 

\begin{enumerate}
    \item Allowing oneself to be convinced by an opposing argument is a sign of good character.
    
    \item People should take into consideration evidence that
goes against their beliefs.

\item People should revise their beliefs in response to new
information or evidence.

\item Changing your mind is a sign of weakness. (R)

\item Intuition is the best guide in making decisions. (R)

\item It is important to persevere in your beliefs even
when evidence is brought to bear against them. (R)

\item One should disregard evidence that conflicts with
one’s established beliefs. (R)

\end{enumerate}

\subsection{Study 1 Details}
\label{sec:appendix_study1}

\subsubsection{Synthetic users simulation}
\label{sec:users_simulation}

To determine an initial maximum range value for $\delta$ around the perfect calibration line, we conducted 5,000 simulations on 128 synthetic users using the following procedure.
We implemented a user-level simulator that generates synthetic subjects who complete 20 binary trials. For each user we sample a base correctness probability \(p\sim\mathcal{U}(0.20,0.85)\),\footnote{This stems from previous study results for minimum and maximum users' accuracy thresholds \cite{Ma24UserConfidenceCalibration,li2025confidencealigns}.} a confidence bias \(b\sim\mathcal{U}(-50,50)\), and add Gaussian trial noise (\(\sigma=7\)); per-trial confidences are clamped to \([50,100]\). Reported metrics are accuracy, mean confidence, and ECE (computed on five equal-width bins over \([50,100]\)), plus a composite calibration index that combines an exponentially transformed normalized gap \(\lvert\text{conf}-\text{accuracy}\rvert/50\) with the fraction of trials classified as well calibrated. 
To stabilize the population, we apply caps (accuracy clipped to \([20,85]\), mean confidence floored at \(65\%\)). 
The simulation results are shown in Table \ref{tab:simulation_result}.

\begin{table}[t]
\centering
\footnotesize
\begin{tabular}{%
>{\centering\arraybackslash}C{0.10\linewidth} 
>{\centering\arraybackslash}C{0.20\linewidth} 
>{\centering\arraybackslash}C{0.15\linewidth} 
}
\toprule
$\delta$ & Mean well-calibrated (\%) & SD well-calibrated \\
\midrule
0  &  3.723 &  1.688 \\
1  &  4.348 &  1.814 \\
2  &  5.001 &  1.938 \\
3  &  5.658 &  2.077 \\
4  &  6.288 &  2.193 \\
5  & 14.380 &  3.128 \\
6  & 15.028 &  3.187 \\
7  & 15.689 &  3.245 \\
8  & 16.346 &  3.311 \\
9  & 17.011 &  3.362 \\
10 & 25.065 &  3.887 \\
11 & 25.773 &  3.930 \\
12 & 26.488 &  3.959 \\
13 & 27.253 &  3.994 \\
14 & 28.098 &  4.028 \\
15 & 37.802 &  4.314 \\
16 & 38.425 &  4.322 \\
17 & 39.066 &  4.331 \\
18 & 39.724 &  4.334 \\
19 & 40.434 &  4.342 \\
20 & 50.333 &  4.439 \\
21 & 50.882 &  4.455 \\
22 & 51.439 &  4.447 \\
23 & 52.017 &  4.448 \\
24 & 52.666 &  4.441 \\
25 & 59.603 &  4.328 \\
\bottomrule
\end{tabular}
\caption{Sensitivity analysis results from 5000 simulations considering the variability of N = 128 users' percentage classified as well-calibrated with $\delta \in [0, 25]$. }
\label{tab:simulation_result}
\end{table}

\begin{figure} [t]
      \centering
    \includegraphics[width=\textwidth]{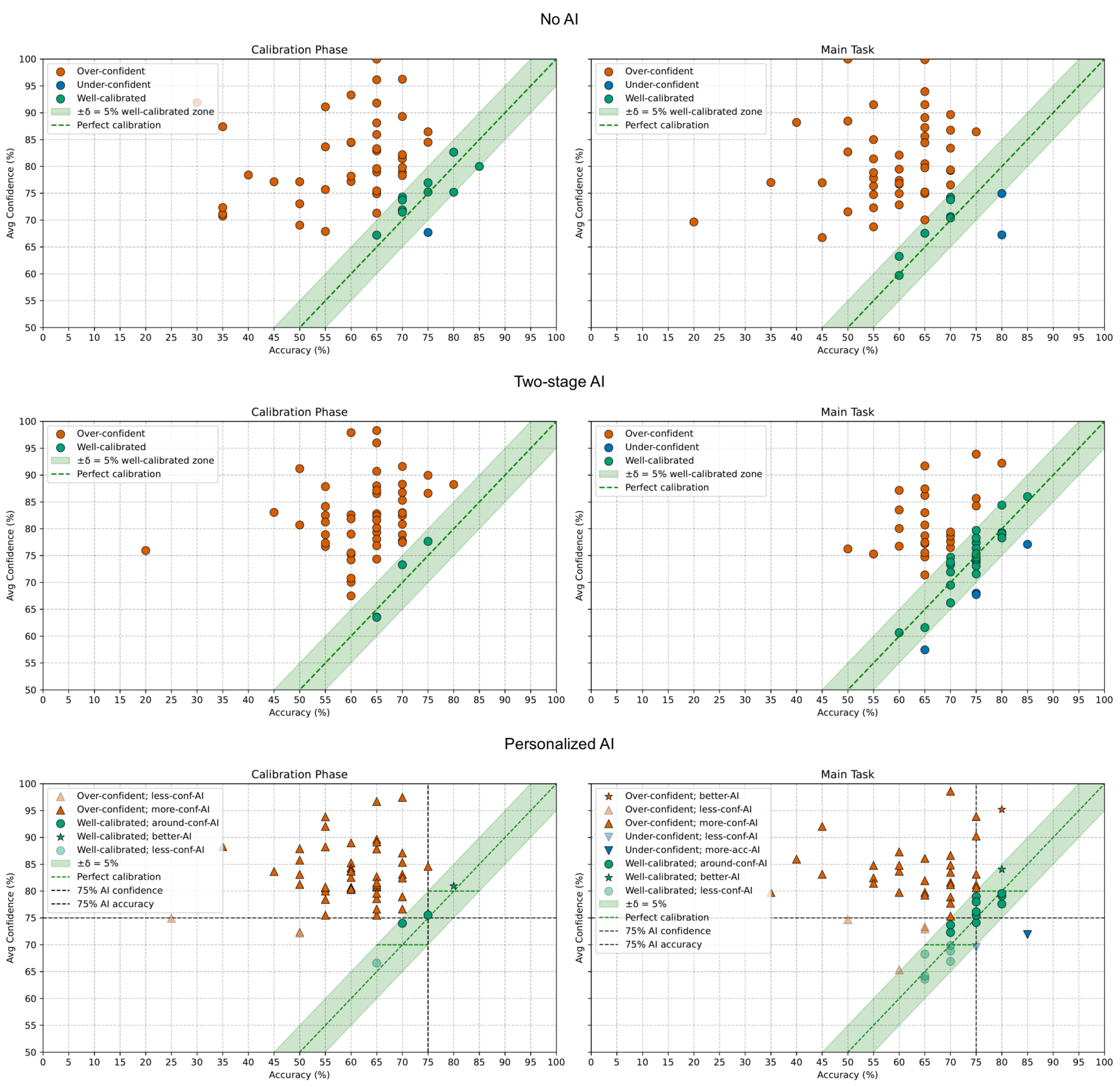}
    \caption{Participants' self-confidence distribution after the calibration phase and after the main task across AI assistance conditions (no AI, two-stage, and personalized AI).
    }

    \Description{
     todo.
    }
 
    \label{fig:study2_conf_calibration}
\end{figure}

\subsection{Study 2 Details}
\label{sec:appendix_study2}

\subsubsection{User experience questions}
\label{sec:user_exp_app}

As a post-test, we asked participants the following questions about their decision-making experience:

\begin{itemize}
    \item  \textbf{No AI assistance condition:}
    \begin{itemize}

    \item Helpfulness of the calibration phase: The calibration phase (20 tasks with feedback) helped me calibrate my confidence to better match my actual performance in the main task. 
    
    \item  Perceived appropriateness of self-confidence: I think my self-confidence was appropriate (being able to reflect the actual correctness likelihood of my predictions accurately).
    
    \item Mental demand: I found this task mentally demanding.

    \item Perceived complexity: The decision-making process with the interface was complex.
    
    \item Preference: I liked the decision-making process with this interface.
    
    \item Satisfaction: I was satisfied with the decision-making process.
\end{itemize}

    \item \textbf{Two-stage and personalized AI assistance conditions:} 
    \begin{itemize}

   \item Helpfulness of the calibration phase: The calibration phase (20 tasks with feedback) helped me calibrate my confidence to better match my actual performance in the main task.

    \item  Perceived appropriateness of self-confidence: I think my self-confidence was appropriate (being able to reflect the actual correctness likelihood of my predictions accurately).
    
    \item Mental demand: I found this task mentally demanding.

    \item Perceived complexity: The decision-making process with the interface was complex.
    
    \item Preference: I liked the decision-making process with this interface.
    
    \item Satisfaction: I was satisfied with the decision-making process collaborating with the AI.

        \item Trust: I trusted the AI’s recommendations in this experiment.
    
    \item Autonomy: I thought I had autonomy in the decision-making process.

    \end{itemize}
    
\end{itemize}

\subsubsection{Post Hoc Result Details}
\label{sec:app_results_study2}

\renewcommand{\arraystretch}{1.2}
\begin{table}[t]
\centering
\footnotesize
\begin{tabular}{p{3cm} C{3.0cm} C{2.1cm} C{2.1cm} C{2.1cm}}
\toprule
Condition & Variable & $\rho$ & Effect & $p$ \\
\hline
\multirow{3}{*}{Two-stage AI} 
 & Agreement fraction & -0.536 & moderate & $<$ \textbf{.001} \\
 & Over-reliance fraction & -0.169 & weak & .25 \\
 & Under-reliance fraction & 0.604 & moderate & $<$ \textbf{.001} \\
\hline
\multirow{3}{*}{Personalized AI} 
 & Agreement fraction & -0.281 & weak & .058 \\
 & Over-reliance fraction & -0.062 & weak & .68 \\
 & Under-reliance fraction & 0.391 & moderate & \textbf{.0072} \\
\bottomrule
\end{tabular}
\caption{Spearman correlations ($\rho$) between reliance measures and ECE for two-stage AI and personalized AI. For the interpretation of the Spearman's correlation coefficients (i.e., Effect column), see Akoglu \cite{Akoglu2018Correlation} and Dancey et al. \cite{dancey2007statistics}.}

\label{tab:correlations_ece_effect}
\end{table}

\paragraph{Assumption checks for ANOVA tests on calibration phase}
\label{sec:app_assumptions_calib_anova}
For the calibration phase, Shapiro-Wilk showed that the accuracy 
\(\bigl(W = 0.902,\ p < .0001\bigr)\), ECE 
\(\bigl(W = 0.958,\ p < .0001\bigr)\) and global metacognition \(\bigl(W = 0.958,\ p < .0001\bigr)\) significantly deviated from normality. Instead
affective metacognition \(\bigl(W = 0.994,\ p = .068\bigr)\) had a normal distribution
Instead, the Levene's test assumption was satisfied for accuracy ($F$ = 1.13, $p$ = .34), ECE ($F$ = 1.71, $p$ = .076), global metacognition ($F$ = 0.95, $p$ = .49), and affective metacognition ($F$ = 0.89, $p$ = .56). Then, we proceeded with ANOVA tests to assess the main effects of AI assistance, NFC, and AOT factors.

\begin{table}[t]
\centering
\footnotesize
\begin{tabular}{lccccccc}
\toprule
\multicolumn{8}{c}{\textbf{Accuracy models (Type III ANOVA Satterthwaite)}} \\
Term & Sum Sq & Mean Sq & NumDF & DenDF & $F$ value & $p$ & $p_{\mathrm{adj}}$ \\
\midrule
Timepoint & 0.1223 & 0.1223 & 1 & 159 & 17.61 & < .0001 & \textbf{.0002} \\
AI assistance & 0.0611 & 0.0305 & 2 & 159 & 4.40  & .0138  & .0553 \\
NFC category & 0.0002 & 0.0002 & 1 & 159 & 0.03  & .8561  & >1 \\
AOT category & 0.0682 & 0.0682 & 1 & 159 & 9.82  & .0021  & \textbf{.0082} \\
Timepoint * AI assistance & 0.1319 & 0.0660 & 2 & 159 & 9.50  & .0001  & \textbf{.0005} \\
Timepoint * NFC category & 0.0013 & 0.0013 & 1 & 159 & 0.19  & .6645  & >1 \\
Timepoint * AOT category & 0.0001 & 0.0001 & 1 & 159 & 0.01  & .9267  & >1 \\
    \addlinespace
    \cmidrule(lr){1-8}
    \addlinespace
\multicolumn{8}{c}{\textbf{Fixed effects (LMM)}} \\
Term & Estimate ($\beta$) & Std. Error & df & $t$ value & $p$ & $p_{\mathrm{adj}}$ \\
\midrule
Timepoint (main) & -0.0153 & 0.0214 & 159.0 & -0.72 & .4755 & >1 \\
AI assistance *  personalized AI & -0.0158 & 0.0194 & 293.0 & -0.82 & .4138 & >1 \\
AI assistance *  two-stage AI & 0.0003 & 0.0194 & 293.0 & 0.01 & .9896 & >1 \\
NFC category (low) & 0.0064 & 0.0159 & 293.0 & 0.40 & .6859 & >1 \\
AOT category (low) & -0.0410 & 0.0159 & 293.0 & -2.57 & .0106 & \textbf{.0425} \\
Timepoint * AI assistance (personalized AI) & 0.0818 & 0.0231 & 159.0 & 3.55 & .0005 & \textbf{.002} \\
Timepoint * AI assistance (two-stage AI) & 0.0920 & 0.0232 & 159.0 & 3.97 & .0001 & \textbf{.0004} \\
Timepoint * NFC category (low) & -0.0082 & 0.0189 & 159.0 & -0.43 & .6645 & >1 \\
Timepoint * AOT category (low) & 0.0018 & 0.0190 & 159.0 & 0.09 & .9267 & >1 \\
\bottomrule
\end{tabular}

\caption{ANOVA (Type III, Satterthwaite) and fixed-effect estimates from the LMM results for \textit{accuracy} across calibration phase and main task with Bonferroni's correction of $m = 4$.}
\label{tab:accuracy_lmm}
\end{table}

\begin{table}[t]
\centering
\footnotesize
\begin{tabular}{lccccccc}
\toprule
\multicolumn{8}{c}{\textbf{ECE models (Type III ANOVA Satterthwaite)}} \\

Term & Sum Sq & Mean Sq & NumDF & DenDF & $F$ value & $p$ & $p_{\mathrm{adj}}$ \\
\midrule
Timepoint & 0.2634 & 0.2634 & 1 & 159 & 28.4900 & <.0001 & \textbf{< .0001} \\
AI assistance & 0.0688 & 0.0344 & 2 & 159 & 3.7200 & .0264 & .1055 \\
NFC category & 0.0183 & 0.0183 & 1 & 159 & 1.9800 & .1612 & .6449 \\
AOT category & 0.0618 & 0.0618 & 1 & 159 & 6.6800 & .0106 & \textbf{.0425} \\
Timepoint * AI assistance & 0.1644 & 0.0822 & 2 & 159 & 8.8900 & .0002 & \textbf{.0009} \\
Timepoint * NFC category & 0.0001 & 0.0001 & 1 & 159 & 0.0100 & .9219 & >1 \\
Timepoint * AOT category & 0.0000 & 0.0000 & 1 & 159 & 0.0000 & .9495 & >1 \\
    \addlinespace
    \cmidrule(lr){1-8}
    \addlinespace
\multicolumn{8}{c}{\textbf{Fixed effects (LMM)}} \\
Term & Estimate ($\beta$) & Std. Error & df & $t$ value & $p$ & $p_{\mathrm{adj}}$ \\
\midrule
Timepoint (main) & 0.0049 & 0.0247 & 159.0000 & 0.2000 & .8442 & >1 \\
AI assistance *  personalized AI & 0.0283 & 0.0221 & 295.8730 & 1.2800 & .2019 & .8074 \\
AI assistance *  two-stage AI & 0.0045 & 0.0222 & 295.8730 & 0.2000 & .8386 & >1 \\
NFC category (low) & -0.0214 & 0.0181 & 295.8730 & -1.1800 & .2380 & .9521 \\
AOT category (low) & 0.0368 & 0.0182 & 295.8730 & 2.0200 & .0438 & .1752 \\
Timepoint * AI assistance (personalized AI) & -0.0893 & 0.0266 & 159.0000 & -3.3500 & .0010 & \textbf{.0040} \\
Timepoint * AI assistance (two-stage AI) & -0.1041 & 0.0267 & 159.0000 & -3.8900 & .0002 & \textbf{.0006} \\
Timepoint * NFC category (low) & 0.0021 & 0.0218 & 159.0000 & 0.1000 & .9219 & >1 \\
Timepoint * AOT category (low) & 0.0014 & 0.0219 & 159.0000 & 0.0600 & .9495 & >1 \\
\bottomrule
\end{tabular}

\caption{ANOVA (Type III, Satterthwaite) and fixed-effect estimates from the LMM results for \textit{ECE} across calibration phase and main task with Bonferroni's correction of $m = 4$.}
\label{tab:ece_lmm}
\end{table}

\begin{table}[t]
\centering
\footnotesize
\begin{tabular}{lccccccc}
\toprule
\multicolumn{8}{c}{\textbf{Global metacognition models (Type III ANOVA Satterthwaite)}} \\

Term & Sum Sq & Mean Sq & NumDF & DenDF & $F$ value & $p$ & $p_{\mathrm{adj}}$ \\
\midrule
Timepoint               & 270 & 270 & 1 & 159 & 1.9900 & .1602  & .6408  \\
AI assistance             & 988 & 494 & 2 & 159 & 3.6400 & .0284  & .1136  \\
NFC category      & 1310 & 1310 & 1 & 159 & 9.6600 & .0022  & \textbf{.0088} \\
AOT category      & 126 & 126 & 1 & 159 & 0.9300 & .3371  & >1     \\
Timepoint * AI assistance        & 86  & 43  & 2 & 159 & 0.3200 & .7277  & >1     \\
Timepoint * NFC category & 78  & 78  & 1 & 159 & 0.5800 & .4489  & >1     \\
Timepoint * AOT category & 0   & 0   & 1 & 159 & 0.0000 & .9804  & >1     \\
    \addlinespace
    \cmidrule(lr){1-8}
    \addlinespace
\multicolumn{8}{c}{\textbf{Fixed effects (LMM)}} \\
Term & Estimate ($\beta$) & Std. Error & df & $t$ value & $p$ & $p_{\mathrm{adj}}$ \\
\midrule
Timepoint (main)                             & 1.2820  & 2.9885 & 159.0000 & 0.4300  & .6690    & >1 \\
AI assistance *  personalized AI                     & 6.3621  & 2.8484 & 281.6480 & 2.2300  & .0260    & .1040 \\
AI assistance *  two-stage AI                    & 5.3397  & 2.8610 & 281.6480 & 1.8700  & .0630    & .2520 \\
NFC category (low)                      & -6.9895 & 2.3360 & 281.6480 & -2.9900 & .0030    & \textbf{.0120} \\
AOT category (low)                      & -1.8942 & 2.3451 & 281.6480 & -0.8100 & .4200    & >1 \\
Timepoint * AI assistance (personalized AI)               & -1.9221 & 3.2245 & 159.0000 & -0.6000 & .5520    & >1 \\
Timepoint * AI assistance (two-stage AI)              & 0.5213  & 3.2388 & 159.0000 & 0.1600  & .8720    & >1 \\
Timepoint * NFC category (low)                 & 2.0076  & 2.6444 & 159.0000 & 0.7600  & .4490    & >1 \\
Timepoint * AOT category (low)                 & 0.0654  & 2.6547 & 159.0000 & 0.0200  & .9800    & >1 \\
\bottomrule
\end{tabular}

\caption{ANOVA (Type III, Satterthwaite) and fixed-effect estimates from the LMM results for \textit{global metacognition} across calibration phase and main task with Bonferroni's correction of $m = 4$.}
\label{tab:gmeta_lmm}
\end{table}

\begin{table}[t]
\centering
\footnotesize
\begin{tabular}{lccccccc}
\toprule
\multicolumn{8}{c}{\textbf{Affective metacognition models (Type III ANOVA Satterthwaite)}} \\
Term & Sum Sq & Mean Sq & NumDF & DenDF & $F$ value & $p$ & $p_{\mathrm{adj}}$ \\
\midrule
Timepoint               & 0.0517   & 0.0517   & 1 & 159     & 4.7100  & .0320  & .1280 \\
AI assistance             & 0.0484   & 0.0242   & 2 & 159     & 2.2000  & .1140  & .4560 \\
NFC category      & 0.2232   & 0.2232   & 1 & 159     & 9.6600  & .0000  & \textbf{<.0001} \\
AOT category      & 0.0027   & 0.0027   & 1 & 159     & 0.2400  & .6230  & >1    \\
Timepoint * AI assistance        & 0.0089   & 0.0045   & 2 & 159     & 0.4100  & .6670  & >1    \\
Timepoint * NFC category & 0.0001   & 0.0001   & 1 & 159     & 0.5800  & .4489  & >1    \\
Timepoint * AOT category & 0.0001   & 0.0001   & 1 & 159     & 0.0100  & .9804  & >1    \\
    \addlinespace
    \cmidrule(lr){1-8}
    \addlinespace
\multicolumn{8}{c}{\textbf{Fixed effects (LMM)}} \\
Term & Estimate ($\beta$) & Std. Error & df & $t$ value & $p$ & $p_{\mathrm{adj}}$ \\
\midrule
Timepoint (main)                            & 1.2820   & 2.9885  & 159.0000 & 0.4300  & .6690    & >1    \\
AI assistance *  personalized AI                    & 6.3621   & 2.8484  & 281.6480 & 2.2300  & .0260    & .1040 \\
AI assistance *  two-stage AI                   & 5.3397   & 2.8610  & 281.6480 & 1.8700  & .0630    & .2520 \\
NFC category (low)                     & -6.9895  & 2.3360  & 281.6480 & -2.9900 & .0030    & \textbf{.0120} \\
AOT category (low)                     & -1.8942  & 2.3451  & 281.6480 & -0.8100 & .4200    & >1    \\
Timepoint * AI assistance (personalized AI)              & -1.9221  & 3.2245  & 159.0000 & -0.6000 & .5520    & >1    \\
Timepoint * AI assistance (two-stage AI)             & 0.5213   & 3.2388  & 159.0000 & 0.1600  & .8720    & >1    \\
Timepoint * NFC category (low)                & 2.0076   & 2.6444  & 159.0000 & 0.7600  & .4490    & >1    \\
Timepoint * AOT category (low)                & 0.0654   & 2.6547  & 159.0000 & 0.0200  & .9800    & >1    \\
\bottomrule
\end{tabular}

\caption{ANOVA (Type III, Satterthwaite) and fixed-effect estimates from the LMM results for \textit{affective metacognition} across calibration phase and main task with Bonferroni's correction of $m = 4$.}
\label{tab:ami_lmm}
\end{table}

\paragraph{Assumption checks for Linear Mixed Models on calibration phase vs main task}
\label{sec:app_assumptions_lmm}
For \textit{accuracy}, the model converged successfully without singularity. The Shapiro-Wilk test indicated a deviation from normality of residuals ($W = 0.962$, $p < .001$), although LMMs are known to be relatively robust to such violations. Homogeneity of variance was satisfied (Levene $F = 1.02$, $p = 0.44$), and the dispersion test confirmed no evidence of overdispersion (DHARMa dispersion = 1, $p = 0.94$). A small number of marginal outliers was detected (3 cases, binomial test $p = 0.027$). Multicollinearity was moderate, with VIF values of 5.23 for \textit{timepoint} and 5.72 for \textit{timepoint * AI assistance} interaction, which remains acceptable.  
For \textit{expected calibration error (ECE)}, the model also converged without singularity. Normality of residuals was again violated (Shapiro-Wilk $W = 0.973$, $p < .001$), but variance homogeneity was respected (Levene $F = 1.24$, $p = 0.21$). Dispersion was not problematic (dispersion = 1, $p = 0.94$), and no outliers were detected ($p = 1$). Multicollinearity was comparable to the accuracy model, with VIF values of 5.23 for \textit{timepoint} and 5.83 for \textit{timepoint * AI assistance} interaction, which remains acceptable.   

For \textit{global metacognition}, the model converged and showed no singularity. Normality was again violated (Shapiro-Wilk $W = 0.973$, $p < .001$). Unlike the previous outcomes, homogeneity of variance was not satisfied (Levene $F = 1.86$, $p = 0.011$). The dispersion test was acceptable (dispersion = 1, $p = 0.96$), but several marginal outliers were identified (5 cases, $p < .001$). Multicollinearity was again moderate with VIF values of 5.23 for \textit{timepoint} and 5.31 for \textit{timepoint * AI assistance} interaction, which remains acceptable. 
Finally, for \textit{affective metacognition}, the model converged successfully with no singularity issues. The strongest deviation from normality was observed here (Shapiro-Wilk $W = 0.903$, $p < .001$). Nevertheless, homogeneity of variance was satisfied (Levene $F = 0.85$, $p = 0.67$), and dispersion was not problematic (dispersion = 1, $p = 0.95$). No outliers were detected ($p = 1$). Multicollinearity was moderate and comparable to the other models with a VIF value of 5.23 for \textit{timepoint}.  

Taken together, these diagnostics suggest that while deviations from normality and, in the case of global metacognition, homogeneity of variance exist, the use of linear mixed models remains appropriate.

\paragraph{Assumption checks for ANCOVA tests on AI literacy}
\label{sec:app_assumptions_ai_literacy}
For metacognitive measurements on the main task, global metacognition residuals showed a deviation from normality ($W = 0.98, p = .023$), homoscedasticity was satisfied (BP = 8.74, $p = .12$), and no autocorrelation was detected (DW = 2.10, $p = .69$). Additionally, there were no interaction effects of other independent variables with AI literacy. Regarding affective metacognition, we identified an interaction between NFC and AI literacy and used this model for subsequent assessments. The normality assumption was met ($W = 0.991, p = .46$), homoscedasticity was confirmed (BP = 5.36, $p = .80$), and there was no autocorrelation (DW = 2.27, $p = .94$). All results are shown in Table \ref{tab:ancova_gmeta}.

For metacognitive measurements during the calibration phase, global metacognition residuals indicated a deviation from normality ($W = 0.974, p = .0045$), homoscedasticity was supported (BP = 3.75, $p = .59$), and no autocorrelation was observed (DW = 2.02, $p = .49$). Additionally, there were no interaction effects of other independent variables with AI literacy. For affective metacognition, we observed an interaction between NFC and AI literacy and subsequently employed this model for further assessments. The residuals were normally distributed ($W = 0.989, p = .28$), homoscedasticity was supported (BP = 5.49, $p = .79$), and no autocorrelation was detected (DW = 2.24, $p = .92$). All results are shown in Table \ref{tab:ancova_ameta}.

\begin{table}[t]
  \centering
  \footnotesize
  \begin{tabular}{
    p{3.5cm}
    >{\centering\arraybackslash}p{1.7cm}
    >{\centering\arraybackslash}p{1.7cm}
    >{\centering\arraybackslash}p{1.7cm}
    >{\centering\arraybackslash}p{1.7cm}
    >{\centering\arraybackslash}p{1.7cm}
  }
    \toprule
    \multicolumn{6}{c}{\textbf{Global metacognition model (ANCOVA) - Main task}} \\
    Predictor & Estimate & Std. Error & t value & $p_{\textnormal{raw}}$ & $p_{\textnormal{adj}}$ \\ 
    \midrule
    AI assistance: personalized AI & 3.718   & 2.250   & 1.65   & .1004    & .4016 \\
    AI assistance: two-stage AI    & 5.514   & 2.250   & 2.45   & .0154    & .0616 \\
    NFC category: low              & 1.426   & 0.979   & 1.46   & .1475    & .5900 \\
    AOT category: low              & 0.668   & 0.924   & 0.72   & .4709    & > 1 \\
    AI literacy (covariate)                   & 4.958   & 1.591   & 3.12   & .0022    & \textbf{.0088} \\
    \addlinespace
    \cmidrule(lr){1-6}
    \addlinespace
    \multicolumn{6}{c}{\textbf{Global metacognition (ANOVA, Type II) - Main task}} \\
    \end{tabular}

  \begin{tabular}{
    p{3.5cm}
    >{\centering\arraybackslash}p{1.4cm}
    >{\centering\arraybackslash}p{1.1cm}
    >{\centering\arraybackslash}p{1.4cm}
    >{\centering\arraybackslash}p{1.4cm}
    >{\centering\arraybackslash}p{1.6cm}
    >{\centering\arraybackslash}p{1.4cm}
  }
    Factor & Sum Sq & Df & $F$ value & $p_{\textnormal{raw}}$ & $\eta^{2}$ & $p_{\textnormal{adj}}$ \\
    \midrule
    AI assistance   & 816   & 2   & 3.12  & .0469 & 0.04   & .1876 \\
    NFC category             & 277   & 1   & 2.12  & .1475 & 0.01   & .5900 \\
    AOT category             & 68    & 1   & 0.52  & .4709 & 0.0034 & 1.0000 \\
    AI literacy (covariate)  & 1269  & 1   & 9.71  & .0022 & 0.06   & \textbf{.0088} \\
  \end{tabular}

   \vspace{4pt}

   \begin{tabular}{
    p{3.5cm}
    >{\centering\arraybackslash}p{1.7cm}
    >{\centering\arraybackslash}p{1.7cm}
    >{\centering\arraybackslash}p{1.7cm}
    >{\centering\arraybackslash}p{1.7cm}
    >{\centering\arraybackslash}p{1.7cm}
  }
        \toprule
    \multicolumn{6}{c}{\textbf{Global metacognition model (ANCOVA) - Calibration phase}} \\
    Predictor & Estimate & Std. Error & t value & $p_{\textnormal{raw}}$ & $p_{\textnormal{adj}}$ \\ 
    \midrule
    AI assistance: personalized AI & 4.709   & 3.158   & 1.49   & .1380    & .5520 \\
    AI assistance: two-stage AI    & 4.545   & 3.159   & 1.44   & .1520    & .6080 \\
    NFC category: low              & 1.053   & 1.375   & 0.77   & .4450    & > 1 \\
    AOT category: low              & 0.383   & 1.298   & 0.30   & .7680    & > 1 \\
    AI literacy  (covariate)                 & 11.362  & 2.234   & 5.09   & $< .0001$ & \textbf{< .0001} \\
    \addlinespace
    \cmidrule(lr){1-6}
    \addlinespace
    \multicolumn{6}{c}{\textbf{Global metacognition (ANOVA, Type II) - Calibration phase}} \\
    \end{tabular}

  \begin{tabular}{
    p{3.5cm}
    >{\centering\arraybackslash}p{1.4cm}
    >{\centering\arraybackslash}p{1.1cm}
    >{\centering\arraybackslash}p{1.4cm}
    >{\centering\arraybackslash}p{1.4cm}
    >{\centering\arraybackslash}p{1.6cm}
    >{\centering\arraybackslash}p{1.4cm}
  }
    Factor & Sum Sq & Df & $F$ value & $p_{\textnormal{raw}}$ &  $\eta^{2}$  & $p_{\textnormal{adj}}$ \\
    \midrule
AI assistance   & 736    & 2   & 1.4291  & .2427   & 0.02   & .9708 \\
    NFC category             & 151    & 1   & 0.5870  & .4448   & 0.0038 & > 1\\
    AOT category             & 22     & 1   & 0.0871  & .7683   & 0.0006 & > 1 \\
    AI literacy (covariate)  & 6663   & 1   & 25.8706 & < .0001 & 0.14 & \textbf{< .0001} \\
    \bottomrule
  \end{tabular}

  \caption{ANCOVA / Type II ANOVA results for global metacognition in the main task (top), and in the calibration phase (bottom) considering AI literacy as a covariate with Bonferroni's correction of $m = 4$.}
  \label{tab:ancova_gmeta}
\end{table}

\begin{table}[t]
  \centering
  \footnotesize
  \begin{tabular}{
    p{3cm}
    >{\centering\arraybackslash}p{1.4cm}
    >{\centering\arraybackslash}p{1.4cm}
    >{\centering\arraybackslash}p{1.4cm}
    >{\centering\arraybackslash}p{1.4cm}
    >{\centering\arraybackslash}p{1.4cm}
    >{\centering\arraybackslash}p{1.4cm}
  }
    \toprule
    \multicolumn{7}{c}{\textbf{Affective metacognition (ANCOVA) - Main task}} \\
    Factor & Sum Sq & Df & $F$ value & $\eta^{2}$ & $p_{\textnormal{raw}}$ & $p_{\textnormal{adj}}$ \\ 
    \midrule
    AI assistance               & 0.3120 & 2   & 3.83  & 0.03 & .0238  & .0952 \\
    AI literacy (covariate)     & 1.3262 & 1   & 32.59 & 0.12 & $<$ .001 & \textbf{$<$ .001} \\
    NFC category                & 0.2365 & 1   & 5.81  & 0.04 & .0171  & .0684 \\
    AOT category                & 0.0321 & 1   & 0.79  & 0.003 & .3761  & 1.0000 \\
    AI literacy * AI assistance & 0.0267 & 2   & 0.33  & 0.002 & .7209  & 1.0000 \\
    AI literacy * NFC           & 0.3348 & 1   & 8.23  & 0.05 & .0047  & \textbf{.0188} \\
    AI literacy * AOT           & 0.0022 & 1   & 0.05  & $<$ .001 & .8171  & 1.0000 \\
    \addlinespace
    \cmidrule(lr){1-7}
    \addlinespace
    
  \end{tabular}

    \begin{tabular}{
    p{3cm}
    >{\centering\arraybackslash}p{1.4cm}
    >{\centering\arraybackslash}p{1.4cm}
    >{\centering\arraybackslash}p{1.4cm}
    >{\centering\arraybackslash}p{1.4cm}
    >{\centering\arraybackslash}p{1.4cm}
    >{\centering\arraybackslash}p{1.4cm}
  }
    \multicolumn{7}{c}{\textbf{Affective metacognition (ANCOVA) - Calibration phase}} \\
    Factor & Sum Sq & Df & $F$ value & $\eta^{2}$ & $p_{\textnormal{raw}}$ & $p_{\textnormal{adj}}$ \\ 
    \midrule
    AI assistance               & 0.1905 & 2   & 3.0446  & 0.02   & .0506  & .2024 \\
    AI literacy (covariate)     & 1.3079 & 1   & 41.8141 & 0.07   & $<$ .001 & \textbf{$<$ .001} \\
    NFC category                & 0.2257 & 1   & 7.2152  & 0.05   & .0080  & \textbf{.0320} \\
    AOT category                & 0.0271 & 1   & 0.8656  & 0.003  & .354   & 1.0000 \\
    AI literacy * AI assistance & 0.0367 & 2   & 0.5864  & 0.003  & .558   & 1.0000 \\
    AI literacy * NFC           & 0.4581 & 1   & 14.6463 & 0.09   & .00019 & \textbf{.00076} \\
    AI literacy * AOT           & 0.0296 & 1   & 0.9454  & 0.006  & .332   & 1.0000 \\
    \bottomrule
  \end{tabular}

  \caption{ANCOVA results for affective metacognition in the main task and calibration phase considering AI literacy as a covariate with Bonferroni correction of $m = 4$.}
  \label{tab:ancova_ameta}
\end{table}

\end{document}